\documentclass[fleqn,usenatbib]{mnras}
\usepackage[utf8]{inputenc}
\usepackage{newtxtext,newtxmath}
\usepackage[T1]{fontenc}
\usepackage{ae,aecompl}
\usepackage{natbib}
\usepackage{graphicx}	

\newcommand\nata{Nature Astronomy}

\def\fesc{\ifmmode f_{\rm esc} \else $f_{\rm esc}$\fi}

\title[Low-mass LyC leaking galaxies]{Lyman continuum leakage from low-mass
galaxies with $M_\star$~$<$~10$^8$~M$_\odot$}

\author[Y. I. Izotov et al.]{
Y. I. Izotov$^{1}$\thanks{E-mail: yizotov@bitp.kiev.ua},
G. Worseck$^{2}$, 
D. Schaerer$^{3,4}$, 
N. G. Guseva$^{1}$,
J. Chisholm$^{5}$, 
\newauthor 
~T. X. Thuan$^{6}$, K. J. Fricke$^{7}$ \& A. Verhamme$^{3}$ 
\\
$^{1}$Bogolyubov Institute for Theoretical Physics,
National Academy of Sciences of Ukraine, 14-b Metrolohichna str., Kyiv,
03143, Ukraine,\\
$^{2}$ Institut f\"ur Physik und Astronomie, Universit\"at Potsdam, Karl-Liebknecht-Str. 24/25, D-14476 Potsdam, Germany,\\
$^{3}$Observatoire de Gen\`eve, Universit\'e de Gen\`eve, 
51 Ch. des Maillettes, 1290, Versoix, Switzerland,\\
$^{4}$IRAP/CNRS, 14, Av. E. Belin, 31400 Toulouse, France\\
$^{5}$Astronomy Department, University of Texas at Austin,
2515 Speedway, Stop C1400 Austin, TX 78712-1205, USA, \\
$^{6}$Astronomy Department, University of Virginia, P.O. Box 400325, 
Charlottesville, VA 22904-4325, USA,\\
$^{7}$Institut f\"ur Astrophysik, G\"ottingen Universit\"at, 
Friedrich-Hund-Platz 1, D-37077 G\"ottingen, Germany.\\
}

\date{Accepted XXX. Received YYY; in original form ZZZ}

\pubyear{2021}

\begin{document}
\label{firstpage}
\pagerange{\pageref{firstpage}--\pageref{lastpage}}
\maketitle

\begin{abstract}
We present observations with the Cosmic Origins Spectrograph onboard the 
{\sl Hubble Space Telescope} of nine low-mass star-forming galaxies at 
redshifts, $z$, in the range 0.3179 -- 0.4524, with stellar masses 
$M_\star$ $<$ 10$^8$ M$_\odot$ and very high specific star-formation rates
sSFR $\sim$ 150 -- 630 Gyr$^{-1}$, aiming to study the dependence of leaking 
Lyman continuum (LyC) emission on stellar mass and some other characteristics 
of the galaxy. We detect LyC emission in four out of nine galaxies with escape 
fractions, $f_{\rm esc}$(LyC), in the range of 11 -- 35 per cent, and establish 
upper limits for $f_{\rm esc}$(LyC) in the remaining five galaxies. We observe a 
narrow Ly$\alpha$ emission line with two peaks in seven galaxies and likely more
complex Ly$\alpha$ profiles in the two remaining galaxies. The velocity 
separation between the peaks $V_{\rm sep}$ varies in the range from 
$\sim$~229~km~s$^{-1}$ to $\sim$~512~km~s$^{-1}$. Our additional data on low-mass
galaxies confirm and strengthen the tight anti-correlation between 
$f_{\rm esc}$(LyC) and $V_{\rm sep}$ found for previous low-redshift galaxy samples 
with higher stellar masses. $V_{\rm sep}$ remains the best indirect indicator of 
LyC leakage. It is better than O$_{32}$ on which $f_{\rm esc}$(LyC) depends 
weakly, with a large scatter. Finally, contrary to expectations, we find no 
increase of $f_{\rm esc}$(LyC) with decreasing galaxy stellar mass $M_\star$.
\end{abstract}




\section{Introduction}\label{intro}

The nature of the main contributors to the reionisation of the Universe
at redshift $z$ $\ga$ 6 is still unknown. Two main types of objects 
have been proposed, active galactic nuclei \citep{Madau15} and
star-forming galaxies (SFGs) \citep[e.g. ][and references therein]{St16}. 
Several recent studies \citep*{H18,M13,M18,Mats18,Ku19,W19,S20} have found 
that the contribution of AGNs 
is small. It has been generally thought that, at high redshifts,  numerous 
faint low-mass SFGs were responsible for the bulk
of the ionising radiation \citep*{O09,WC09,M13,Y11,B15a,Fi19,Le20}, whereas
the number of relatively bright SFGs is insufficient to fully ionise the 
Universe \citep*{S01,C09,Iw09,R13}. However, recently \citet{N20},
based on an empirical model of reionisation have found that 5 per cent of
bright galaxies with $M_{\rm FUV}$~$<$~$-$18~mag and stellar masses 
$M_\star$/M$_\odot$~$>$~10$^8$ contribute
$\ga$~80 per cent of the ionising luminosity, whereas lower-mass galaxies 
have a minor effect. 
On the other hand, \citet{Me20}, based on the correlation of foreground 
galaxies with the intergalactic medium (IGM) transmission toward background QSO at the 
end of reionisation, found that a contribution from faint galaxies is
necessary to reproduce the observed decreasing IGM opacity and that reionisation
might be driven by different sub-populations of Lyman-break galaxies (LBGs) and Lyman-alpha emitters (LAEs) at $z \sim 6$.

Additional conditions for galaxies to reionise the Universe  
require that the escape fraction of their Lyman continuum (LyC) is of the 
order of 10 -- 20~per cent or higher 
\citep[e.g. ][]{O09,R13,D15,Robertson15,K16} and that the ionising photon
production is $\xi_{\rm ion}$~$\sim$~10$^{25.2}$~Hz~erg$^{-1}$ \citep[e.g. ][]{R13}.
One of the differences between these studies and the one by \citet{N20} is that 
the former authors assume a constant $f_{\rm esc}$(LyC), whereas the latter 
allowed $f_{\rm esc}$(LyC) to scale with various galaxy properties, e.g. SFR 
surface density and stellar mass. 
\citet{Fi19} looked at how variations in the galaxy properties can translate 
into different reionisation histories. Specifically, they found that variations 
in the $\xi_{\rm ion}$ strongly favour low mass galaxies as the source of 
reionisation with $f_{\rm esc}$(LyC) values as low as 5 per cent. 
Thus, scaling relations with galaxy properties are 
highly uncertain and adopting different ones can completely change the 
reionisation history of the Universe. 
Studying these limitations, e.g. how the escape fraction of the ionising radiation depends on the galaxy's stellar mass, is among the main
motivations of the present study.

The most reliable LyC leakers
detected at high redshifts thus far appear to fulfill the condition 
that $f_{\rm esc}$(LyC)~$\ga$~10~--~20 per cent. They are 
the objects {\em Ion2} \citep[$z$~=~3.212, ][]{Va15,B16} with a
relative escape fraction $f^{\rm rel}_{\rm esc}$(LyC)~=~64 per cent, 
Q1549-C25 \citep[$z$~=~3.212, ][]{Sh16} with $f_{\rm esc}$(LyC)~$>$~51 per cent,
A2218-Flanking \citep[$z\approx$~2.5, ][]{B17} with 
$f_{\rm esc}$(LyC)~$>$~28~--~57 per cent, {\em Ion3} 
\citep[$z\approx$~4.0, ][]{Va18} with 
$f_{\rm esc}$(LyC)~$\sim$~60 per cent, {\em Sunburst Arc} 
\citep[$z\approx$~2.37, ][]{RT19} with 
$f_{\rm esc}$(LyC)~$\sim$~20 per cent, a $z$~$\sim$~1.4 galaxy with 
$f_{\rm esc}$(LyC)~$>$~20 per cent \citep{Sa20}, two $z$~$\ge$~3.5 galaxies 
with $f_{\rm esc}$(LyC)~$\sim$~5~--~73 per cent \citep{Mes20} and two GRB 
galaxies at $z$ $\sim$ 3 with $f_{\rm esc}$(LyC)~$\sim$~35 and 8 per cent 
\citep{Vi20}. Recently, \citet{Fl19} have shown that 
fifteen $z$~$\sim$~3.1 LyC leakers have escape fractions ranging 
from 2 per cent to 82 per cent. 
\citet{Ma17,Ma18} and \citet{St18} report
LyC detection in stacked spectra of $z$~$\sim$~3 spectra, after careful removal
of possible interlopers with the help of {\sl HST} imaging.

Thus, a small sampling of galaxies near the peak of cosmic star formation 
history empirically shows that galaxies can emit sufficient ionising photons 
to reionise the high-redshift Universe.

  \begin{table*}
  \caption{Coordinates, redshifts, distances, oxygen abundances and O$_{32}$ 
ratios of selected galaxies
\label{tab1}}
\begin{tabular}{lrrccrcr} \hline
Name&R.A.(2000.0)&Dec.(2000.0)&$z$&$D_L$$^{\rm a}$&\multicolumn{1}{c}{$D_A$$^{\rm b}$}&12+logO/H$^{\rm c}$&O$_{32}$$^{\rm c}$ \\ \hline                                                                            
J0232$-$0426&02:32:16.09&$-$04:26:26.72&0.45236&2539&1204&7.88&14.2\\ 
J0919$+$4906&09:19:55.78&$+$49:06:08.75&0.40512&2227&1128&7.77&11.5\\ 
J1046$+$5827&10:46:01.98&$+$58:27:56.95&0.39677&2172&1114&8.01& 4.7\\ 
J1121$+$3806&11:21:18.22&$+$38:06:42.80&0.31788&1675& 965&7.96& 7.3\\ 
J1127$+$4610&11:27:21.00&$+$46:10:42.49&0.32230&1702& 974&7.84& 6.0\\ 
J1233$+$4959&12:33:30.78&$+$49:59:49.45&0.42194&2337&1156&8.11&11.0\\ 
J1349$+$5631&13:49:55.10&$+$56:31:10.90&0.36366&1960&1054&7.91& 4.0\\ 
J1355$+$1457&13:55:53.46&$+$14:57:01.48&0.36513&1970&1057&7.77& 6.1\\ 
J1455$+$6107&14:55:59.57&$+$61:07:19.70&0.36793&1987&1062&7.91& 4.1\\ 
\hline
\end{tabular}

\hbox{$^{\rm a}$Luminosity distance in Mpc \citep[NED, ][]{W06}.}

\hbox{$^{\rm b}$Angular size distance in Mpc \citep[NED, ][]{W06}.}

\hbox{$^{\rm c}$Derived in this paper from the SDSS spectrum.}

  \end{table*}

  \begin{table*}
  \caption{Apparent AB magnitudes with errors in parentheses compiled
from the SDSS and {\sl GALEX} databases and apparent Vega magnitudes from
the {\sl WISE} database
\label{tab2}}
\begin{tabular}{lccccccccccccc} \hline
Name&\multicolumn{5}{c}{SDSS}
&&\multicolumn{2}{c}{\sl GALEX}&&\multicolumn{4}{c}{{\sl WISE}} \\ 
    &\multicolumn{1}{c}{$u$}&\multicolumn{1}{c}{$g$}&\multicolumn{1}{c}{$r$}&\multicolumn{1}{c}{$i$}&\multicolumn{1}{c}{$z$}&&FUV&NUV&&\multicolumn{1}{c}{$W1$}&\multicolumn{1}{c}{$W2$}&\multicolumn{1}{c}{$W3$}&\multicolumn{1}{c}{$W4$}
\\
    &\multicolumn{1}{c}{(err)}&\multicolumn{1}{c}{(err)}&\multicolumn{1}{c}{(err)}&\multicolumn{1}{c}{(err)}&\multicolumn{1}{c}{(err)}&&(err)&(err)&&\multicolumn{1}{c}{(err)}&\multicolumn{1}{c}{(err)}&\multicolumn{1}{c}{(err)}&\multicolumn{1}{c}{(err)} \\
\hline
J0232$-$0426& 23.38& 22.13& 21.93& 20.85& 21.59&& ... & 21.26&&\multicolumn{1}{c}{ ... }&\multicolumn{1}{c}{ ... }&\multicolumn{1}{c}{ ... }&\multicolumn{1}{c}{ ... } \\
            &(0.73)&(0.11)&(0.11)&(0.06)&(0.45)&&(...)&(0.09)&&\multicolumn{1}{c}{(...)}&\multicolumn{1}{c}{(...)}&\multicolumn{1}{c}{(...)}&\multicolumn{1}{c}{(...)} \\
J0919$+$4906& 22.33& 21.97& 21.95& 20.78& 22.32&& 22.54& 21.92&&\multicolumn{1}{c}{ ... }&\multicolumn{1}{c}{ ... }&\multicolumn{1}{c}{ ... }&\multicolumn{1}{c}{ ... } \\
            &(0.27)&(0.08)&(0.10)&(0.06)&(0.62)&&(0.21)&(0.12)&&\multicolumn{1}{c}{(...)}&\multicolumn{1}{c}{(...)}&\multicolumn{1}{c}{(...)}&\multicolumn{1}{c}{(...)} \\
J1046$+$5827& 21.20& 21.07& 21.23& 20.55& 20.57&& 21.30& 21.02&&\multicolumn{1}{c}{17.70}&\multicolumn{1}{c}{ ... }&\multicolumn{1}{c}{ ... }&\multicolumn{1}{c}{ ... } \\
            &(0.09)&(0.04)&(0.06)&(0.05)&(0.16)&&(0.03)&(0.02)&&\multicolumn{1}{c}{(0.18)}&\multicolumn{1}{c}{(...)}&\multicolumn{1}{c}{(...)}&\multicolumn{1}{c}{(...)} \\
J1121$+$3806& 22.18& 22.05& 21.19& 22.75& 21.19&&  ... &  ... &&\multicolumn{1}{c}{ ... }&\multicolumn{1}{c}{ ... }&\multicolumn{1}{c}{ ... }&\multicolumn{1}{c}{ ... } \\
            &(0.17)&(0.07)&(0.05)&(0.28)&(0.26)&& (...)& (...)&&\multicolumn{1}{c}{(...)}&\multicolumn{1}{c}{(...)}&\multicolumn{1}{c}{(...)}&\multicolumn{1}{c}{(...)} \\
J1127$+$4610& 21.97& 22.31& 21.63& 23.31& 21.63&&  ... & 22.22&&\multicolumn{1}{c}{ ... }&\multicolumn{1}{c}{ ... }&\multicolumn{1}{c}{ ... }&\multicolumn{1}{c}{ ... } \\
            &(0.21)&(0.11)&(0.11)&(0.61)&(0.52)&& (...)&(0.39)&&\multicolumn{1}{c}{(...)}&\multicolumn{1}{c}{(...)}&\multicolumn{1}{c}{(...)}&\multicolumn{1}{c}{(...)} \\
J1233$+$4959& 21.70& 21.92& 21.90& 20.80& 21.48&&  ... & 21.91&&\multicolumn{1}{c}{ ... }&\multicolumn{1}{c}{ ... }&\multicolumn{1}{c}{ ... }&\multicolumn{1}{c}{ ... } \\
            &(0.16)&(0.08)&(0.11)&(0.06)&(0.41)&& (...)&(0.12)&&\multicolumn{1}{c}{(...)}&\multicolumn{1}{c}{(...)}&\multicolumn{1}{c}{(...)}&\multicolumn{1}{c}{(...)} \\
J1349$+$5631& 22.48& 22.45& 22.08& 23.00& 22.25&& 22.24& 22.14&& 17.96& 16.89&\multicolumn{1}{c}{ ... }&\multicolumn{1}{c}{ ... } \\
            &(0.29)&(0.11)&(0.11)&(0.38)&(0.61)&&(0.36)&(0.32)&&\multicolumn{1}{c}{(0.19)}&\multicolumn{1}{c}{(0.27)}&\multicolumn{1}{c}{(...)}&\multicolumn{1}{c}{(...)} \\
J1355$+$1457& 22.05& 21.62& 21.40& 21.66& 20.73&&  ... & 21.13&&\multicolumn{1}{c}{ ... }&\multicolumn{1}{c}{ ... }&\multicolumn{1}{c}{ ... }&\multicolumn{1}{c}{ ... } \\
            &(0.15)&(0.05)&(0.05)&(0.09)&(0.14)&& (...)&(0.10)&&\multicolumn{1}{c}{(...)}&\multicolumn{1}{c}{(...)}&\multicolumn{1}{c}{(...)}&\multicolumn{1}{c}{(...)} \\
J1455$+$6107& 21.84& 21.51& 21.33& 21.37& 21.56&& 21.52& 21.70&&\multicolumn{1}{c}{ ... }&\multicolumn{1}{c}{ ... }&\multicolumn{1}{c}{ ... }&\multicolumn{1}{c}{ ... } \\
            &(0.17)&(0.05)&(0.07)&(0.10)&(0.39)&&(0.39)&(0.39)&&\multicolumn{1}{c}{(...)}&\multicolumn{1}{c}{(...)}&\multicolumn{1}{c}{(...)}&\multicolumn{1}{c}{(...)} \\
\hline
\end{tabular}

  \end{table*}

  \begin{table*}
  \caption{Extinction-corrected fluxes and rest-frame equivalent widths of
the emission lines in SDSS spectra
\label{tab3}}
\begin{tabular}{lcrrrrrrrrrr} \hline
 & &\multicolumn{10}{c}{Galaxy}\\
Line &\multicolumn{1}{c}{$\lambda$}&\multicolumn{2}{c}{J0232$-$0426}& \multicolumn{2}{c}{J0919$+$4906}& \multicolumn{2}{c}{J1046$+$5827}& \multicolumn{2}{c}{J1121$+$3806}& \multicolumn{2}{c}{J1127$+$4610} \\
     &&\multicolumn{1}{c}{$I$$^{\rm a}$}&\multicolumn{1}{c}{EW$^{\rm b}$}&\multicolumn{1}{c}{$I$$^{\rm a}$}&\multicolumn{1}{c}{EW$^{\rm b}$}&\multicolumn{1}{c}{$I$$^{\rm a}$}&\multicolumn{1}{c}{EW$^{\rm b}$}&\multicolumn{1}{c}{$I$$^{\rm a}$}&\multicolumn{1}{c}{EW$^{\rm b}$}&\multicolumn{1}{c}{$I$$^{\rm a}$}&\multicolumn{1}{c}{EW$^{\rm b}$} \\
\hline
Mg~{\sc ii}          &2796&  15.8$\pm$4.7&  10& 25.8$\pm$6.7&  19& 29.6$\pm$4.6&  11& 20.7$\pm$4.1&  10& 34.4$\pm$8.0&  26\\
Mg~{\sc ii}          &2803&  14.9$\pm$4.7&   9& 19.2$\pm$6.3&  14& 23.8$\pm$4.3&   9& 15.3$\pm$3.6&  10& 12.4$\pm$6.6&  10\\
$[$O~{\sc ii}$]$     &3727&  44.5$\pm$6.9&  68& 55.0$\pm$4.6& 127&130.6$\pm$8.9& 101& 83.6$\pm$6.8& 130& 89.3$\pm$16.&  72\\
H12                  &3750&   3.8$\pm$3.8&   5&  4.2$\pm$1.5&  17&  5.6$\pm$4.7&   4&\multicolumn{1}{c}{...}& ...&\multicolumn{1}{c}{...}& ...\\
H11                  &3771&   5.0$\pm$4.3&   6&  7.6$\pm$2.0&  21&  7.0$\pm$5.5&   4&\multicolumn{1}{c}{...}& ...&\multicolumn{1}{c}{...}& ...\\
H10                  &3798&   6.9$\pm$3.6&  14&  6.7$\pm$2.1&  20&  7.6$\pm$4.5&   7&\multicolumn{1}{c}{...}& ...&\multicolumn{1}{c}{...}& ...\\
H9                   &3836&  11.1$\pm$4.1&  21&  9.7$\pm$1.9&  28& 13.0$\pm$4.6&  19& 10.3$\pm$3.6&  16& 12.2$\pm$7.4&  13\\
$[$Ne~{\sc iii}$]$   &3869&  37.4$\pm$6.5&  46& 54.2$\pm$4.5& 153& 51.1$\pm$5.3&  40& 49.1$\pm$5.1&  88& 43.2$\pm$11.&  49\\
H8+He~{\sc i}        &3889&  17.8$\pm$5.5&  23& 25.3$\pm$3.3&  55& 21.2$\pm$6.3&  15& 19.4$\pm$4.8&  87& 28.5$\pm$9.0&  40\\
H7+$[$Ne~{\sc iii}$]$&3969&  30.3$\pm$6.7&  40& 33.0$\pm$3.5&  96& 37.3$\pm$6.0&  38& 34.9$\pm$5.1&  60& 33.3$\pm$9.7&  51\\
H$\delta$            &4101&  28.3$\pm$5.7&  87& 22.6$\pm$2.9&  71& 32.3$\pm$5.4&  39& 28.9$\pm$4.8&  51& 29.3$\pm$9.5&  37\\
H$\gamma$            &4340&  40.5$\pm$7.5&  54& 45.4$\pm$4.2& 118& 47.6$\pm$6.0&  66& 47.0$\pm$5.2&  94& 50.8$\pm$12.&  72\\
$[$O~{\sc iii}$]$    &4363&  12.3$\pm$4.3&  17& 15.3$\pm$2.6&  44&  9.7$\pm$2.8&  11& 10.2$\pm$2.7&  19& 10.0$\pm$6.3&  16\\
He~{\sc i}           &4471&   6.4$\pm$2.9&  18&\multicolumn{1}{c}{...}& ...&\multicolumn{1}{c}{...}& ...& 4.8$\pm$2.3&   7&\multicolumn{1}{c}{...}& ...\\
H$\beta$             &4861& 100.0$\pm$11.& 227&100.0$\pm$6.5& 435&100.0$\pm$8.2& 170&100.0$\pm$8.1& 317&100.0$\pm$17.& 158\\
$[$O~{\sc iii}$]$    &4959& 219.6$\pm$16.& 480&209.9$\pm$10.& 992&199.6$\pm$11.& 358&212.6$\pm$12.& 703&169.8$\pm$23.& 201\\
$[$O~{\sc iii}$]$    &5007& 632.4$\pm$35.&1977&635.0$\pm$22.&2547&613.4$\pm$23.&1116&616.2$\pm$23.&2222&531.7$\pm$45.& 685\\
He~{\sc i}           &5876&  10.6$\pm$4.2&  37& 10.4$\pm$2.4&  69& 12.0$\pm$3.0&  67&\multicolumn{1}{c}{...}& ...&\multicolumn{1}{c}{...}& ...\\
H$\alpha$            &6563&255.6$\pm$23.$^{\rm c}$& 472&279.2$\pm$14.&2250&259.1$\pm$16.$^{\rm c}$& 851&274.5$\pm$17.& 901&273.8$\pm$33.&1038\\
$[$N~{\sc ii}$]$     &6583&  \multicolumn{1}{c}{...}& ...&  7.3$\pm$2.0&  76&  5.3$\pm$2.8&  15&  6.1$\pm$2.4&  36&\multicolumn{1}{c}{...}& ...\\
$[$S~{\sc ii}$]$     &6717&  \multicolumn{1}{c}{...}& ...&\multicolumn{1}{c}{...}& ...&\multicolumn{1}{c}{...}& ...&\multicolumn{1}{c}{...}& ...&\multicolumn{1}{c}{...}& ...\\
$[$S~{\sc ii}$]$     &6731&  \multicolumn{1}{c}{...}& ...&\multicolumn{1}{c}{...}& ...&\multicolumn{1}{c}{...}& ...&\multicolumn{1}{c}{...}& ...&\multicolumn{1}{c}{...}& ...\\
$C$(H$\beta$)$_{\rm int}$$^{\rm d}$  &&\multicolumn{2}{c}{0.050$\pm$0.109}&\multicolumn{2}{c}{0.075$\pm$0.061}&\multicolumn{2}{c}{0.050$\pm$0.074}&\multicolumn{2}{c}{0.050$\pm$0.074}&\multicolumn{2}{c}{0.050$\pm$0.156}\\
$C$(H$\beta$)$_{\rm MW}$$^{\rm e}$   &&\multicolumn{2}{c}{0.029}&\multicolumn{2}{c}{0.020}&\multicolumn{2}{c}{0.011}&\multicolumn{2}{c}{0.028}&\multicolumn{2}{c}{0.025}\\
EW(H$\beta$)$^{\rm b}$        &&\multicolumn{2}{c}{227$\pm$26}&\multicolumn{2}{c}{435$\pm$26}&\multicolumn{2}{c}{170$\pm$40}&\multicolumn{2}{c}{317$\pm$34}&\multicolumn{2}{c}{158$\pm$26}\\
$I$(H$\beta$)$^{\rm f}$     &&\multicolumn{2}{c}{4.4$\pm$0.5}&\multicolumn{2}{c}{5.1$\pm$0.4}&\multicolumn{2}{c}{8.8$\pm$0.7}&\multicolumn{2}{c}{7.9$\pm$0.6}&\multicolumn{2}{c}{4.8$\pm$0.8}\\
\hline

 & &\multicolumn{8}{c}{Galaxy}\\
Line &\multicolumn{1}{c}{$\lambda$}&\multicolumn{2}{c}{J1233$+$4959}& \multicolumn{2}{c}{J1349$+$5631}& \multicolumn{2}{c}{J1355$+$1457}& \multicolumn{2}{c}{J1455$+$6107} \\
     &&\multicolumn{1}{c}{$I$$^{\rm a}$}&\multicolumn{1}{c}{EW$^{\rm b}$}&\multicolumn{1}{c}{$I$$^{\rm a}$}&\multicolumn{1}{c}{EW$^{\rm b}$}&\multicolumn{1}{c}{$I$$^{\rm a}$}&\multicolumn{1}{c}{EW$^{\rm b}$}&\multicolumn{1}{c}{$I$$^{\rm a}$}&\multicolumn{1}{c}{EW$^{\rm b}$}\\
\hline
Mg~{\sc ii}          &2796&  17.6$\pm$3.5&  10& 12.7$\pm$4.7&  14& 32.9$\pm$7.2&  28& 32.2$\pm$6.2&  13\\
Mg~{\sc ii}          &2803&   9.6$\pm$3.0&   4& 15.1$\pm$5.2&   9& 34.8$\pm$7.4&  15& 11.6$\pm$4.9&  11\\
$[$O~{\sc ii}$]$     &3727&  60.8$\pm$5.9&  66&144.8$\pm$13.& 158&100.5$\pm$4.8& 162&140.2$\pm$6.5& 212\\
H12                  &3750&\multicolumn{1}{c}{...}& ...&\multicolumn{1}{c}{...}& ...&  4.1$\pm$1.2&   7&\multicolumn{1}{c}{...}& ...\\
H11                  &3771&\multicolumn{1}{c}{...}& ...&\multicolumn{1}{c}{...}& ...&  4.4$\pm$1.4&   8&\multicolumn{1}{c}{...}& ...\\
H10                  &3798&\multicolumn{1}{c}{...}& ...&\multicolumn{1}{c}{...}& ...&  8.1$\pm$1.6&  16&  6.3$\pm$1.5&  11\\
H9                   &3836&\multicolumn{1}{c}{...}& ...&\multicolumn{1}{c}{...}& ...&  8.9$\pm$1.6&  23& 10.5$\pm$1.8&  21\\
$[$Ne~{\sc iii}$]$   &3869&  51.4$\pm$5.4&  46& 47.9$\pm$7.2&  98& 49.5$\pm$3.3&  61& 54.1$\pm$3.8&  78\\
H8+He~{\sc i}        &3889&  20.2$\pm$4.5&  26& 23.6$\pm$6.5&  43& 21.5$\pm$2.2&  36& 28.1$\pm$2.8&  47\\
H7+$[$Ne~{\sc iii}$]$&3969&  31.3$\pm$5.4&  39& 28.9$\pm$8.2&  35& 31.4$\pm$2.7&  51& 28.1$\pm$2.8&  47\\
H$\delta$            &4101&  27.9$\pm$4.6&  49& 22.9$\pm$5.4&  55& 28.5$\pm$2.4&  86& 27.5$\pm$2.9&  67\\
H$\gamma$            &4340&  50.5$\pm$6.0&  80& 44.1$\pm$8.1&  77& 47.0$\pm$3.1& 117& 48.3$\pm$4.1& 157\\
$[$O~{\sc iii}$]$    &4363&   8.5$\pm$2.6&  11& 10.9$\pm$3.9&  13& 15.3$\pm$1.9&  41& 11.3$\pm$1.7&  29\\
He~{\sc i}           &4471&\multicolumn{1}{c}{...}& ...&\multicolumn{1}{c}{...}& ...&\multicolumn{1}{c}{...}& ...& 3.9$\pm$1.3&   7\\
H$\beta$             &4861& 100.0$\pm$7.9& 305&100.0$\pm$12.& 172&100.0$\pm$4.8& 265&100.0$\pm$5.3& 278\\
$[$O~{\sc iii}$]$    &4959& 223.7$\pm$12.& 542&190.2$\pm$16.& 450&202.7$\pm$7.4& 597&193.1$\pm$7.8& 747\\
$[$O~{\sc iii}$]$    &5007& 667.2$\pm$25.&1424&579.4$\pm$32.&1649&614.3$\pm$16.&1714&578.4$\pm$17.&2049\\
He~{\sc i}           &5876&\multicolumn{1}{c}{...}& ...& 16.1$\pm$4.9&  47& 11.3$\pm$1.8&  59&  9.8$\pm$2.0&  26\\
H$\alpha$            &6563& 275.1$\pm$17.&1469&282.6$\pm$24.&1326&280.6$\pm$11.&1851&281.9$\pm$12.& 839\\
$[$N~{\sc ii}$]$     &6583&  11.6$\pm$3.1&  59&  7.5$\pm$4.1&  10&\multicolumn{1}{c}{...}& ...&\multicolumn{1}{c}{...}& ...\\
$[$S~{\sc ii}$]$     &6717&  \multicolumn{1}{c}{...}& ...& 17.5$\pm$5.6&  41&  7.5$\pm$1.7&  49&  7.3$\pm$2.0&  18\\
$[$S~{\sc ii}$]$     &6731&  \multicolumn{1}{c}{...}& ...& 13.0$\pm$5.0&  32&  5.6$\pm$1.6&  35&  8.5$\pm$2.1&  21\\
$C$(H$\beta$)$_{\rm int}$$^{\rm d}$  &&\multicolumn{2}{c}{0.050$\pm$0.073}&\multicolumn{2}{c}{0.110$\pm$0.100}&\multicolumn{2}{c}{0.135$\pm$0.045}&\multicolumn{2}{c}{0.090$\pm$0.050}\\
$C$(H$\beta$)$_{\rm MW}$$^{\rm e}$   &&\multicolumn{2}{c}{0.023}&\multicolumn{2}{c}{0.013}&\multicolumn{2}{c}{0.028}&\multicolumn{2}{c}{0.018}\\
EW(H$\beta$)$^{\rm b}$        &&\multicolumn{2}{c}{305$\pm$30}&\multicolumn{2}{c}{172$\pm$36}&\multicolumn{2}{c}{265$\pm$12}&\multicolumn{2}{c}{278$\pm$10}\\
$I$(H$\beta$)$^{\rm f}$     &&\multicolumn{2}{c}{8.7$\pm$0.7}&\multicolumn{2}{c}{5.0$\pm$0.5}&\multicolumn{2}{c}{12.8$\pm$0.6}&\multicolumn{2}{c}{8.7$\pm$0.4}\\
\hline
  \end{tabular}

\hbox{$^{\rm a}$$I$=100$\times$$I$($\lambda$)/$I$(H$\beta$), where $I$($\lambda$) 
and $I$(H$\beta$) are fluxes of emission lines, corrected for both the 
Milky Way and internal extinction.}

\hbox{$^{\rm b}$Rest-frame equivalent width in \AA.}

\hbox{$^{\rm c}$Clipped line.}

\hbox{$^{\rm d}$Internal galaxy extinction coefficient.}

\hbox{$^{\rm e}$Milky Way extinction coefficient from the NED.}


\hbox{$^{\rm f}$Extinction-corrected flux but not corrected for $f_{\rm esc}$(LyC),
in 10$^{-16}$ erg s$^{-1}$ cm$^{-2}$.}

  \end{table*}

  \begin{table*}
  \caption{Electron temperatures, electron number densities and 
element abundances in H~{\sc ii} regions \label{tab4}}
  \begin{tabular}{lccccc} \hline
Galaxy &J0232$-$0426 &J0919$+$4906 &J1046$+$5827 &J1121$+$3806 &J1127$+$4610  \\ \hline
$T_{\rm e}$ ($[$O {\sc iii}$]$), K      & 14980$\pm$2380       & 16660$\pm$1440       & 13780$\pm$1680     & 14010$\pm$1580& 14890$\pm$2300        \\
$T_{\rm e}$ ($[$O {\sc ii}$]$), K       & 13990$\pm$2070       & 14810$\pm$1190       & 13230$\pm$1510     & 13390$\pm$1410 & 13940$\pm$1750        \\
$N_{\rm e}$ ($[$S {\sc ii}$]$), cm$^{-3}$&   10$\pm$10       &   10$\pm$10  &        10$\pm$10   &    10$\pm$10  &      10$\pm$10       \\ \\
O$^+$/H$^+$$\times$10$^{5}$             &0.50$\pm$0.12        &0.52$\pm$0.06        &1.78$\pm$0.31       &1.10$\pm$0.18 &1.02$\pm$0.38          \\
O$^{2+}$/H$^+$$\times$10$^{5}$          &7.05$\pm$0.48        &5.40$\pm$0.17        &8.36$\pm$0.57      &8.11$\pm$0.49  &5.89$\pm$0.71           \\
O/H$\times$10$^{5}$                    &7.55$\pm$0.50        &5.92$\pm$0.18        &10.14$\pm$0.65      &9.20$\pm$0.52  &6.92$\pm$0.81          \\
12+log O/H                             &7.88$\pm$0.03        &7.77$\pm$0.01        &8.01$\pm$0.03      &7.96$\pm$0.02  &7.84$\pm$0.05          \\ \\
N$^+$/H$^+$$\times$10$^{6}$             &         ...        &0.55$\pm$0.16        &0.51$\pm$0.29      &0.56$\pm$0.24  &       ...          \\
ICF(N)$^{\rm a}$                        &         ...        &10.22                 &5.41               &    7.59      &        ...           \\
N/H$\times$10$^{6}$                     &        ...        &5.62$\pm$1.86        &2.74$\pm$1.63      &4.28$\pm$1.99 &         ...          \\
log N/O                                &        ...        &~$-$1.02$\pm$0.14~~\, &~$-$1.57$\pm$0.26~~\,&~$-$1.33$\pm$0.20~~\, &  ...      \\ \\
Ne$^{2+}$/H$^+$$\times$10$^{5}$          &1.00$\pm$0.24        &1.08$\pm$0.11        &1.77$\pm$0.32       &1.61$\pm$0.27 &1.18$\pm$0.46          \\
ICF(Ne)$^{\rm a}$                       &1.01                 &1.03                 &1.09                &1.04           &1.06                   \\
Ne/H$\times$10$^{5}$                    &1.01$\pm$0.24        &1.11$\pm$0.12        &1.92$\pm$0.35       &1.67$\pm$0.28 &1.25$\pm$0.48          \\
log Ne/O                               &~$-$0.87$\pm$0.11~~\, &~$-$0.73$\pm$0.05~~\, &~$-$0.72$\pm$0.08~~\,&~$-$0.74$\pm$0.08~~\, &~$-$0.74$\pm$0.17~~\,    \\ \\
Mg$^{+}$/H$^+$$\times$10$^{6}$          &0.27$\pm$0.09         &0.32$\pm$0.10       &0.60$\pm$0.15      &0.38$\pm$0.09 &0.42$\pm$0.19          \\
ICF(Mg)$^{\rm a}$                        &     22.89           &    19.23           &    10.47          &14.24        &12.13                  \\
Mg/H$\times$10$^{6}$                    &6.23$\pm$1.96        &6.17$\pm$1.93       &6.31$\pm$1.57      &5.41$\pm$1.32 &5.11$\pm$2.27         \\
log Mg/O                               &~$-$1.06$\pm$0.14~~\, &~$-$1.10$\pm$0.14~~\, &~$-$1.19$\pm$0.11~~\, &~$-$1.21$\pm$0.11~~\, &~$-$1.11$\pm$0.20~~\,   \\
\hline
Galaxy &J1233$+$4959 &J1349$+$5631 &J1355$+$1457 &J1455$+$6107 \\ \hline
$T_{\rm e}$ ($[$O {\sc iii}$]$), K      & 12600$\pm$1490       & 14860$\pm$2420       & 16940$\pm$1090     & 15030$\pm$1060 \\
$T_{\rm e}$ ($[$O {\sc ii}$]$), K       & 12340$\pm$1380       & 13920$\pm$2120       & 14920$\pm$900     & 14020$\pm$920 \\
$N_{\rm e}$ ($[$S {\sc ii}$]$), cm$^{-3}$&   10$\pm$10       &   77$\pm$130  &        76$\pm$59   &      1151$\pm$305        \\ \\
O$^+$/H$^+$$\times$10$^{5}$             &1.06$\pm$0.22        &1.68$\pm$0.35        &0.93$\pm$0.07       &1.78$\pm$0.14 \\
O$^{2+}$/H$^+$$\times$10$^{5}$          &11.75$\pm$1.04        &6.50$\pm$0.47        &5.02$\pm$0.11       &6.33$\pm$0.21  \\
O/H$\times$10$^{5}$                    &12.81$\pm$1.06        &8.18$\pm$0.58        &5.95$\pm$0.13      &8.11$\pm$0.25  \\
12+log O/H                             &8.11$\pm$0.04        &7.91$\pm$0.03        &7.77$\pm$0.01      &7.91$\pm$0.01  \\ \\
N$^+$/H$^+$$\times$10$^{6}$             &1.29$\pm$0.37        &0.64$\pm$0.37        &        ...        &     ...        \\
ICF(N)$^{\rm a}$                        &10.21               &4.73                   &        ...        &    ...         \\
N/H$\times$10$^{6}$                     &1.31$\pm$0.44       &3.05$\pm$1.83        &         ...        &    ...         \\
log N/O                                &~$-$0.99$\pm$0.15~~\, &~$-$1.43$\pm$0.26~~\, &      ...        &     ...       \\ \\
Ne$^{2+}$/H$^+$$\times$10$^{5}$          &2.38$\pm$0.49        &1.32$\pm$0.30        &0.94$\pm$0.08       &1.44$\pm$0.14 \\
ICF(Ne)$^{\rm a}$                       &0.99                 &1.11                 &1.07                &1.12           \\
Ne/H$\times$10$^{5}$                    &2.37$\pm$0.49        &1.46$\pm$0.33        &1.01$\pm$0.08       &1.61$\pm$0.16 \\
log Ne/O                               &~$-$0.73$\pm$0.10~~\, &~$-$0.75$\pm$0.10~~\, &~$-$0.77$\pm$0.04~~\,&~$-$0.70$\pm$0.05~~\, \\ \\
Mg$^{+}$/H$^+$$\times$10$^{6}$          &0.41$\pm$0.12         &0.26$\pm$0.10       &0.47$\pm$0.13      &0.40$\pm$0.12  \\
ICF(Mg)$^{\rm a}$                        &     19.25           &    9.55            &    12.08          &  8.84          \\
Mg/H$\times$10$^{6}$                    &7.95$\pm$2.24        &2.51$\pm$0.94       &5.62$\pm$1.58      &3.52$\pm$1.07  \\
log Mg/O                               &~$-$1.19$\pm$0.13~~\, &~$-$1.51$\pm$0.17~~\, &~$-$1.03$\pm$0.12~~\, &~$-$1.36$\pm$0.13~~\, \\
\hline
\end{tabular}

\hbox{$^{\rm a}$Ionisation correction factor.}
  \end{table*}

Direct observations of high-redshift galaxies are
difficult for several reasons, including their faintness, the increase of
IGM opacity, and contamination by 
lower-redshift interlopers \citep[e.g. ][]{V10,V12,Inoue14,Gr15}.
Therefore, it is important to identify
and study local proxies of this galaxy population.  
It has been argued that low-mass compact galaxies
at low redshifts $z$~$<$~1 with very active star formation
may be promising candidates for escaping ionising radiation
\citep{JO13,NO14}. Low-redshift galaxies, because of their proximity,
 can be studied in more detail than high-redshift ones and their 
$f_{\rm esc}$(LyC) can be derived with a higher reliability. In particular, the
intrinsic LyC flux produced by hot stars in a low-redshift galaxy can 
directly be derived from the observable H$\beta$ flux, and together with the
observed LyC flux, allows the direct determination of $f_{\rm esc}$(LyC). 
This technique, which is not possible for most of high-redshift 
galaxies because of the unobservable H$\beta$ emission, 
has been applied in particular by \citet{I16a,I16b,I18a,I18b}.

The general
characteristic of compact SFGs is the presence of strong emission lines in the 
optical spectra of their H~{\sc ii} regions, powered by numerous O-stars,
which produce plenty of ionising radiation. Different subsets of compact
SFGs, depending on redshift, photometric
characteristics and luminosities, have been variously called
``blue compact dwarf'' (BCD) galaxies \citep[e.g. ][]{TM81,ITL94,I18c}, 
``blueberry'' galaxies \citep{Ya17}, ``green pea'' (GP) galaxies \citep{Ca09}
or ``luminous compact'' star-forming galaxies (LCGs) \citep{I11}.

Stellar masses, star formation rates (SFR) and metallicities of compact SFGs
vary in wide ranges and they are similar to
those of high-redshift star-forming galaxies including Lyman-alpha emitting 
galaxies \citep{I15}. Many
low-redshift compact SFGs are characterised by high line ratios O$_{32}$ =
[O~{\sc iii}]$\lambda$5007/[O~{\sc ii}]$\lambda$3727 $\ga$ 4 -- 5, reaching 
values of up to 60 in some galaxies \citep{S15}. These high values are not yet 
observed in high-$z$ galaxies, However, their {\sl Spitzer} colours suggest 
strong [O~{\sc iii}] + H$\beta$ equivalent 
widths \citep{La13,Sm14,B19,E20}. Such high values
may indicate that the ISM is predominantly ionised,
allowing the escape of Lyman continuum photons \citep{JO13,NO14}. Indeed,
\citet{I16a,I16b,I18a,I18b} obtained {\sl HST}/COS observations
of eleven compact SFGs at redshifts $z\sim$ 0.3 -- 0.4 and with O$_{32}$ = 5
-- 28 and $M_\star$ $\sim$ 6$\times$10$^7$ -- 6$\times$10$^9$ M$_\odot$, and 
found all these galaxies to be leaking LyC radiation,
with an escape fraction in the range of 2~--~72~ per cent. We note that recent
X-shooter observations of one of the strong high-$z$ leakers, {\em Ion2},
reveal the emission-line characteristics in the optical range to be similar to those
of $\sim$ 0.3 -- 0.4 LyC leakers, in particular, a high O$_{32}$ $\sim$ 9 
\citep{Va20}.
Furthermore, the Ly$\alpha$ profile of ten galaxies shows a  double peak 
with one having a triple peak. All 
these profiles show a small velocity separation of the peaks ($<$ 450 km s$^{-1}$), as 
predicted by \citet{V15} for low H~{\sc i} column densities.

Finally, an analysis of UV absorption lines, including hydrogen lines of the 
Lyman series and heavy element lines such as Si~{\sc ii}\,$\lambda1260$ can 
provide a consistent and accurate measure of the Lyman continuum escape 
fraction \citep[e.g. ][]{Ga18,Ch18,Ga20}. Mg~{\sc ii} $\lambda$2796, 2803 
emission provides also a strong constraint of the LyC escape and the doublet ratio 
can be used to infer the neutral gas column density \citep{Hen18,Ch20}. 

\begin{figure*}
\hbox{
\includegraphics[angle=-90,width=0.325\linewidth]{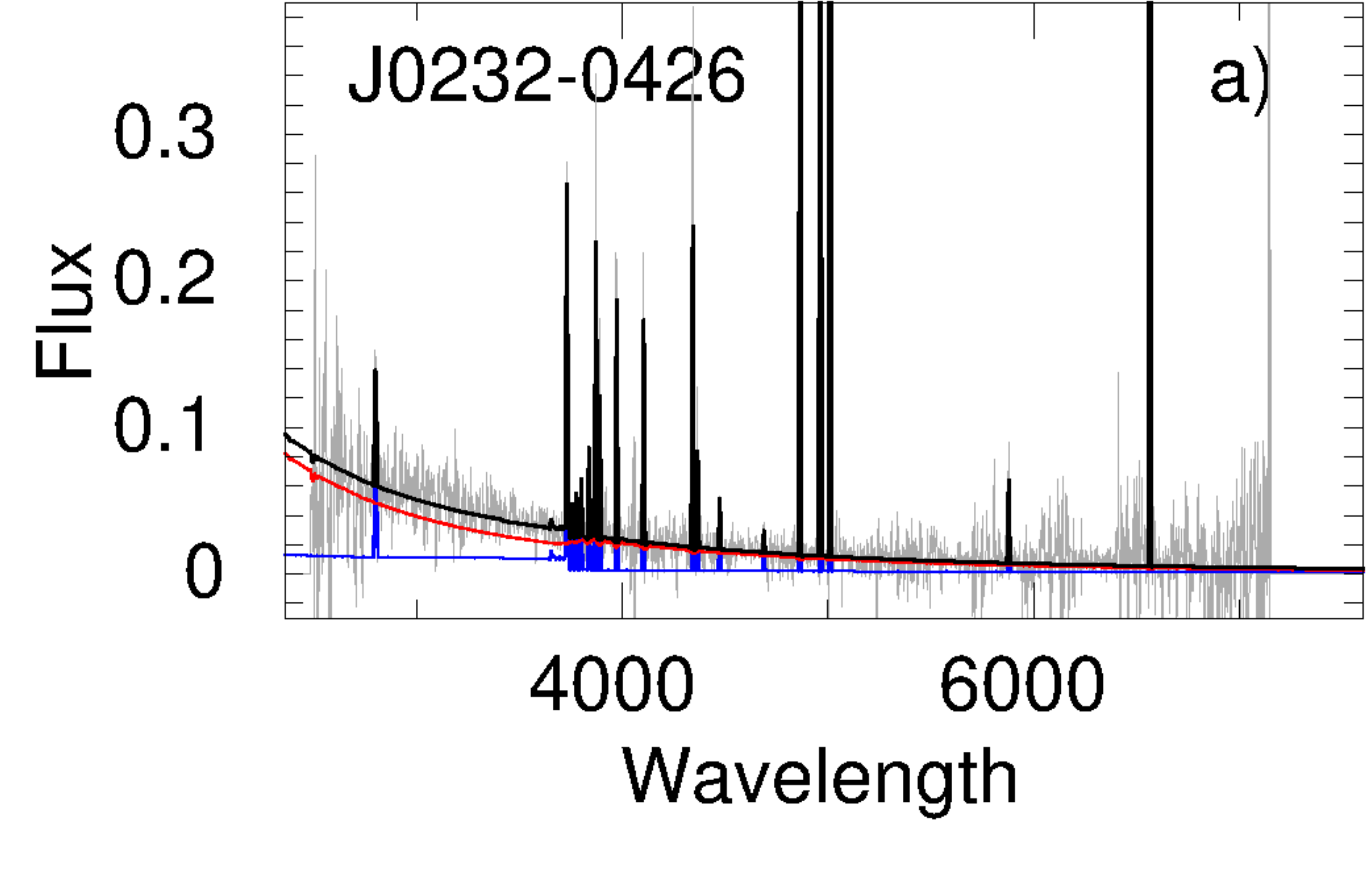}
\includegraphics[angle=-90,width=0.325\linewidth]{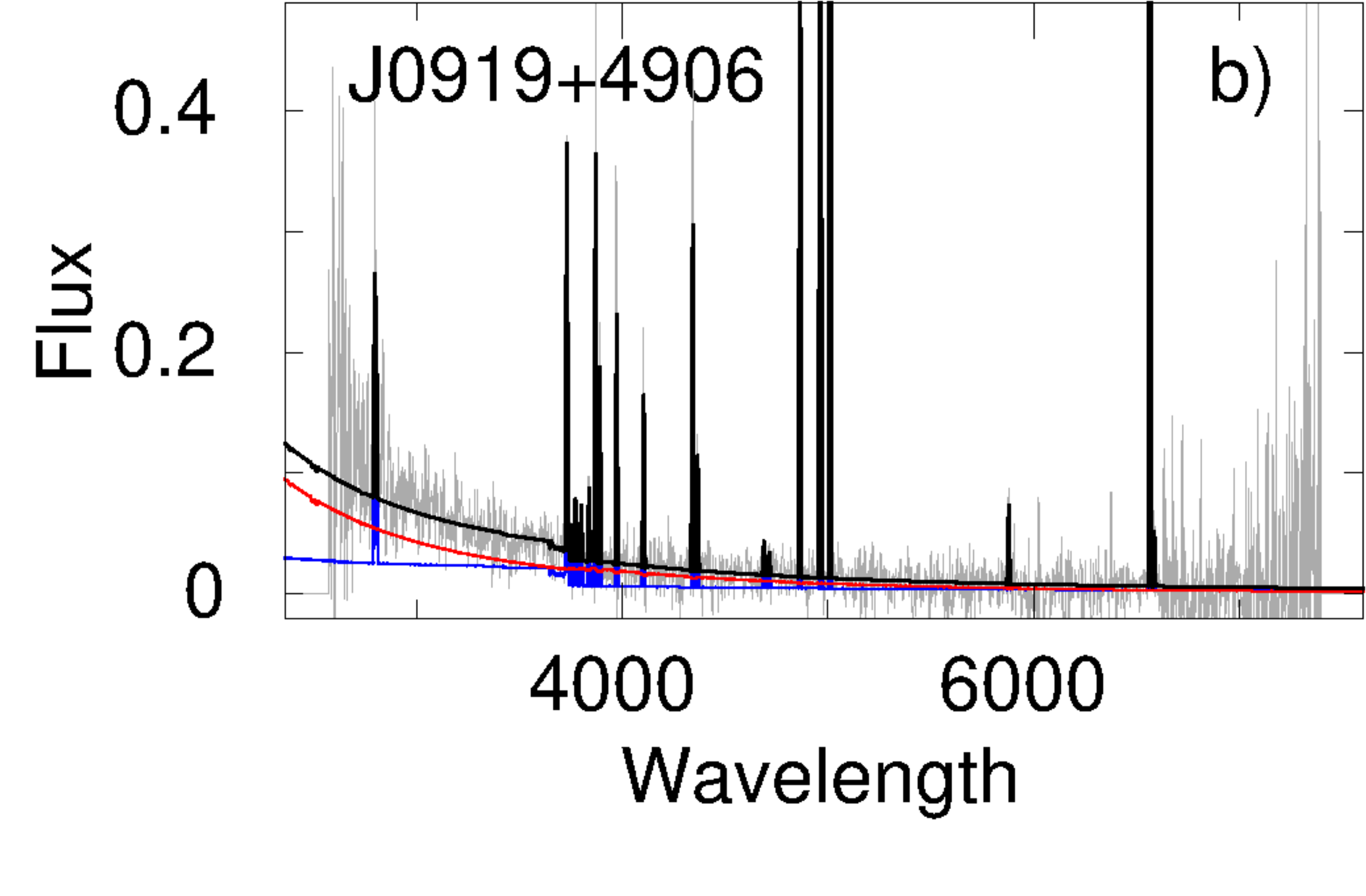}
\includegraphics[angle=-90,width=0.325\linewidth]{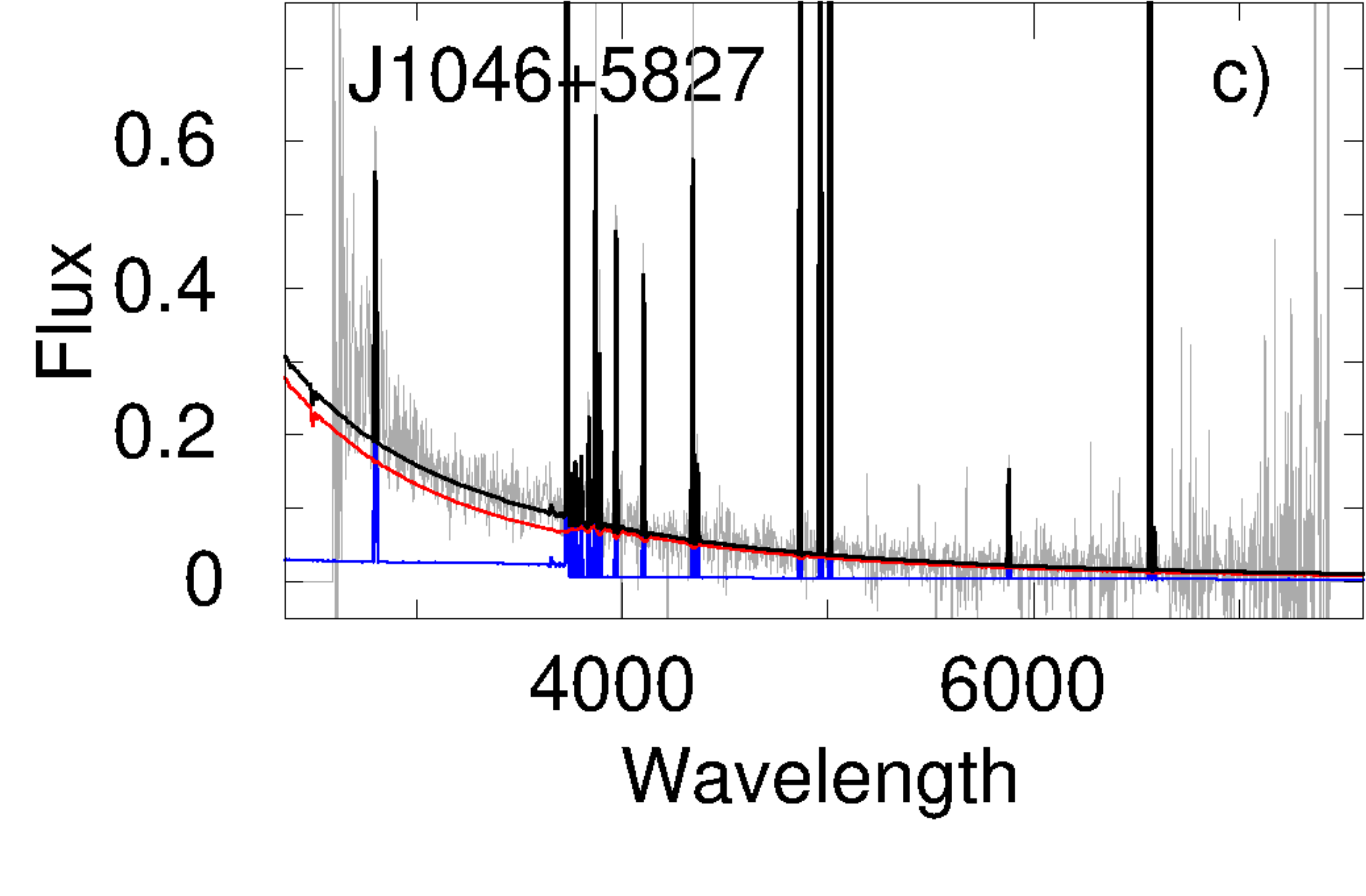}
 }
 \hbox{
\includegraphics[angle=-90,width=0.325\linewidth]{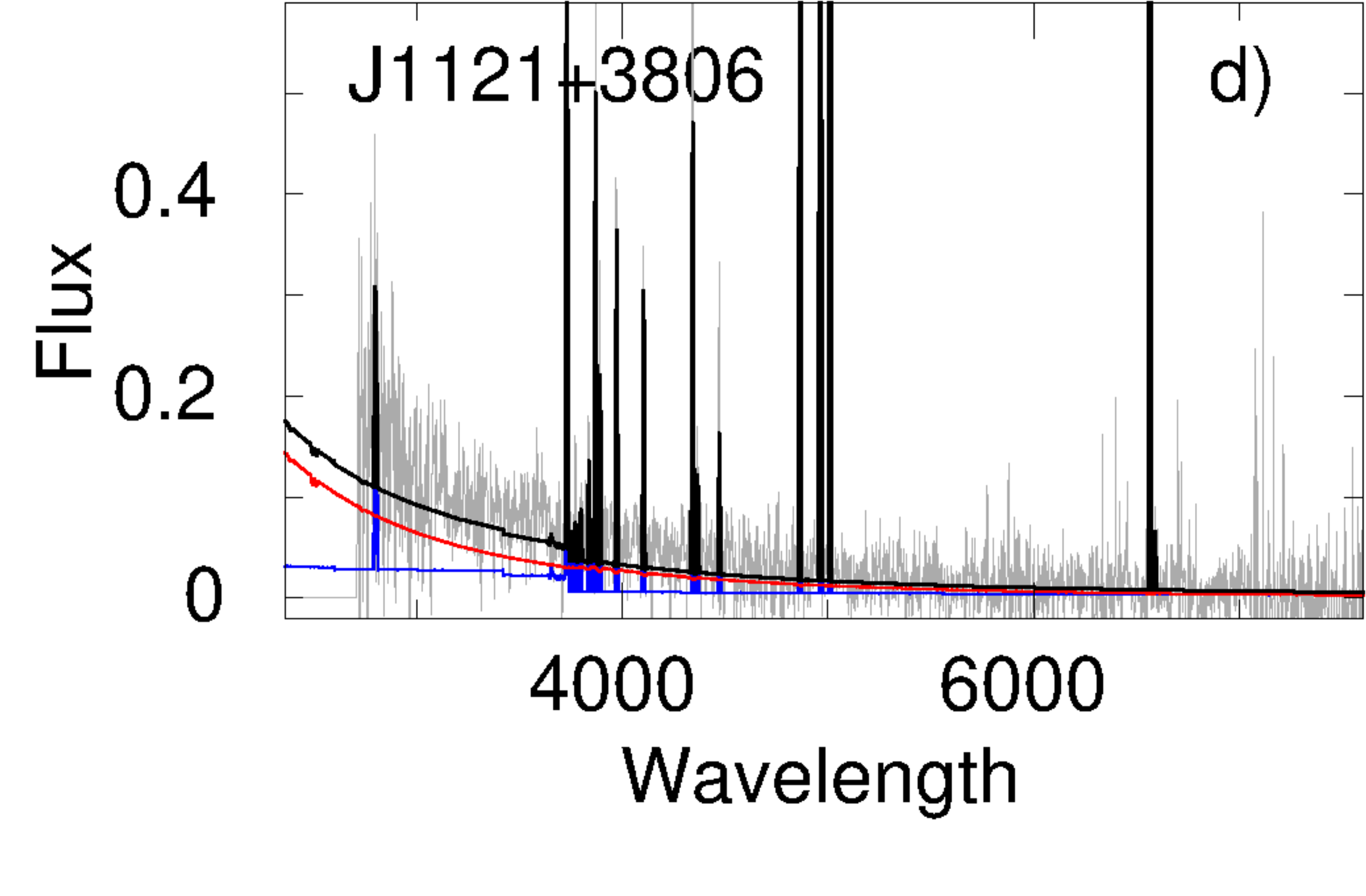}
\includegraphics[angle=-90,width=0.325\linewidth]{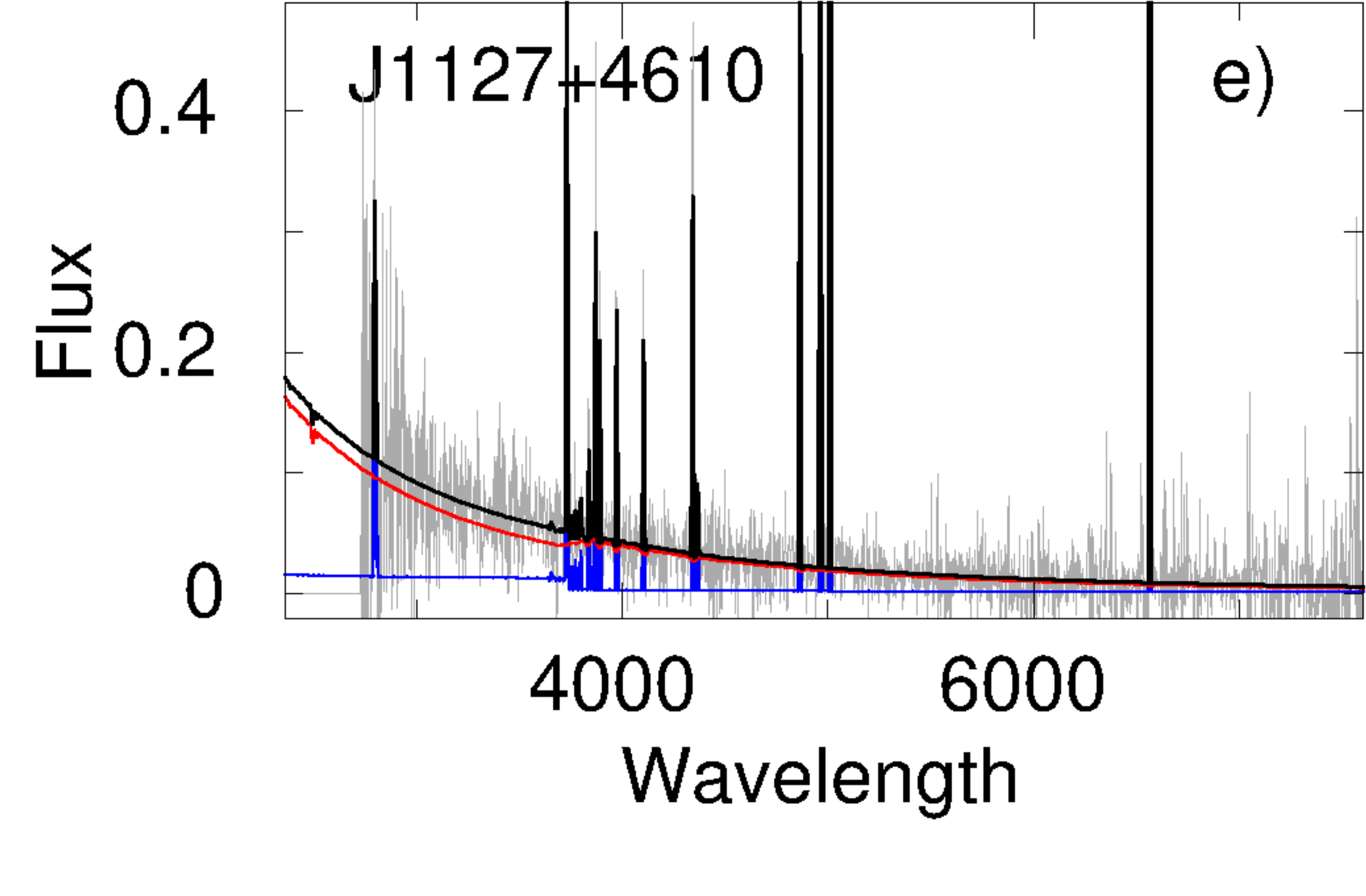}
\includegraphics[angle=-90,width=0.325\linewidth]{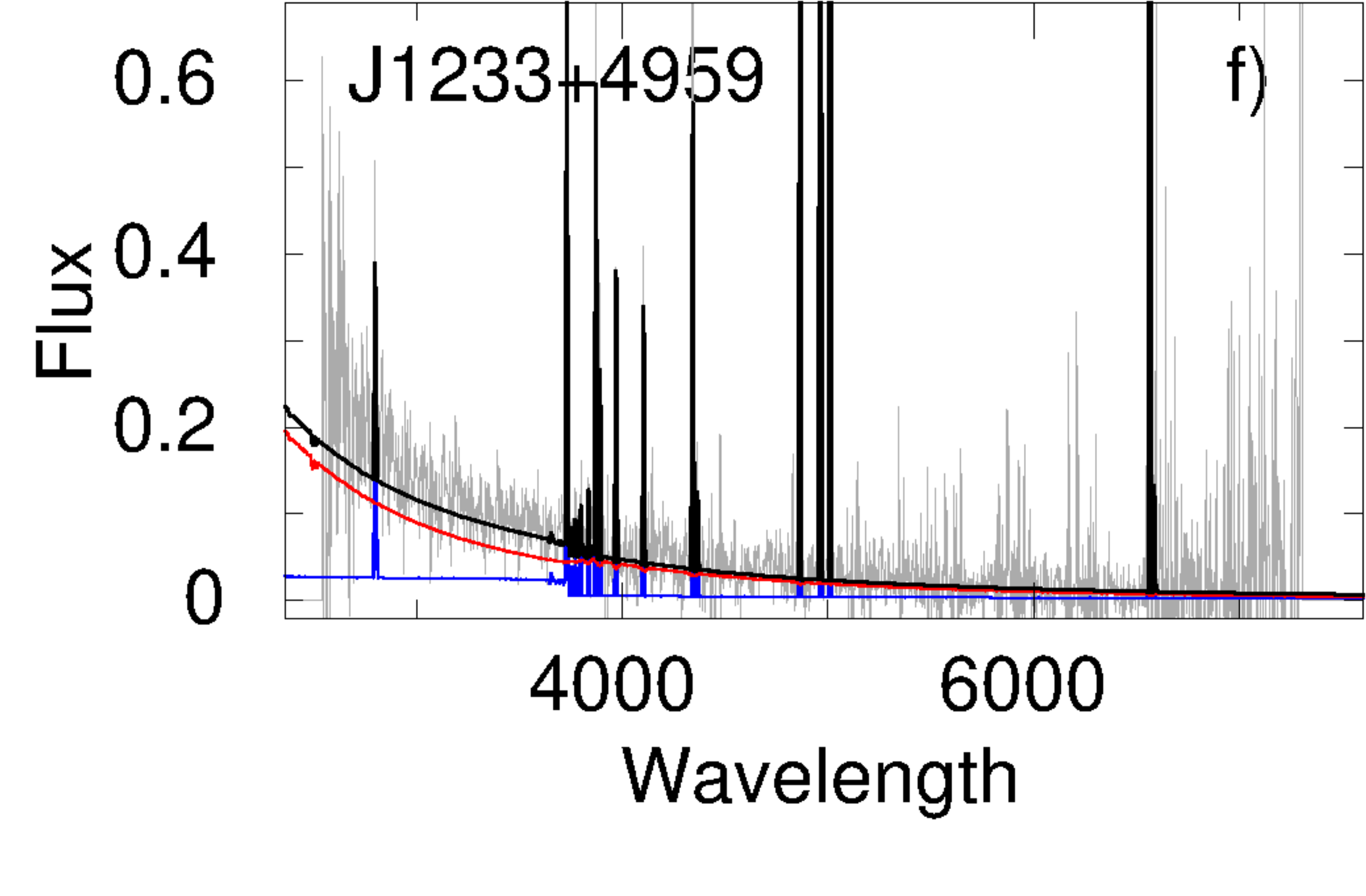}
}
 \hbox{
\includegraphics[angle=-90,width=0.325\linewidth]{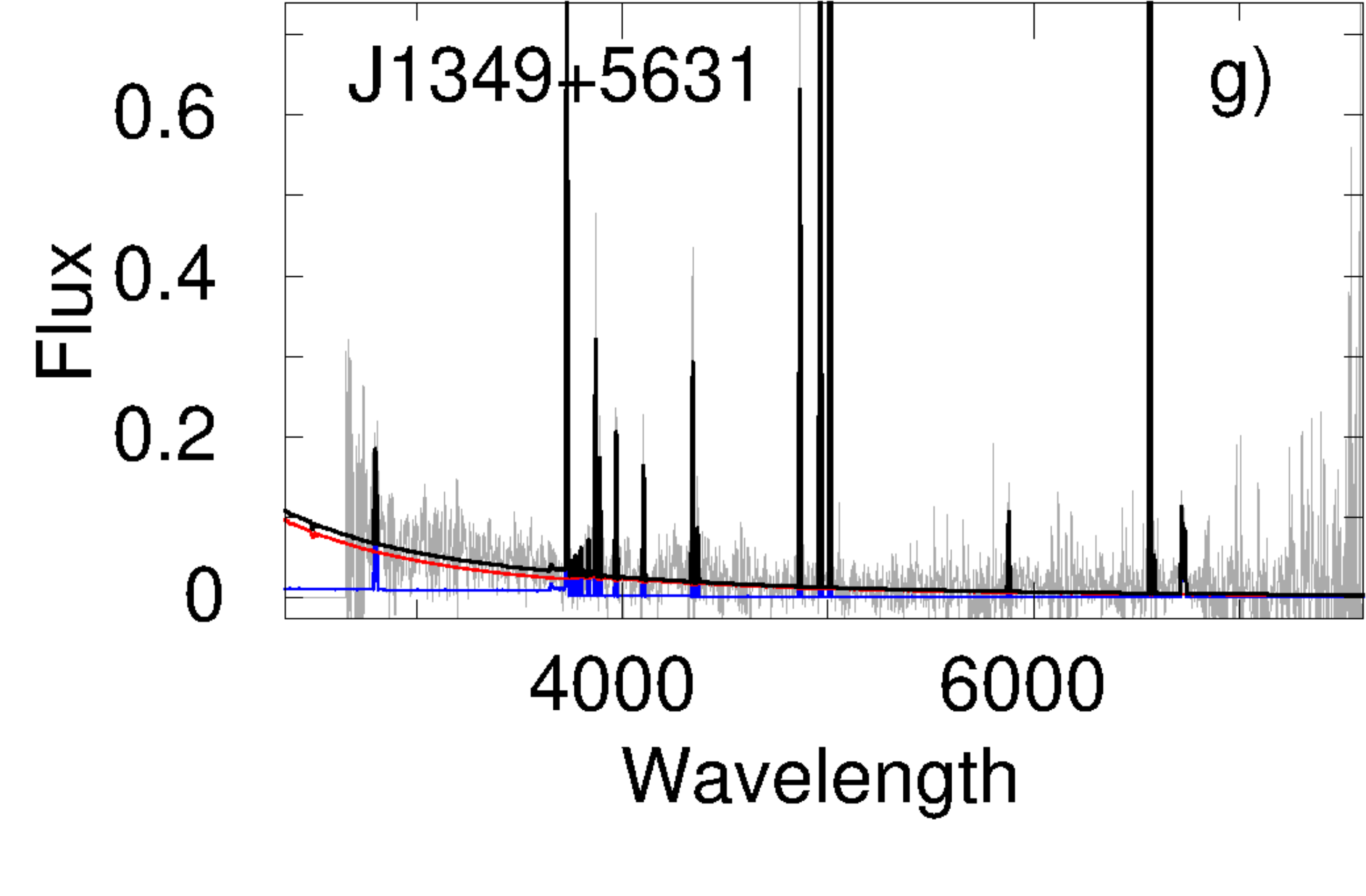}
\includegraphics[angle=-90,width=0.325\linewidth]{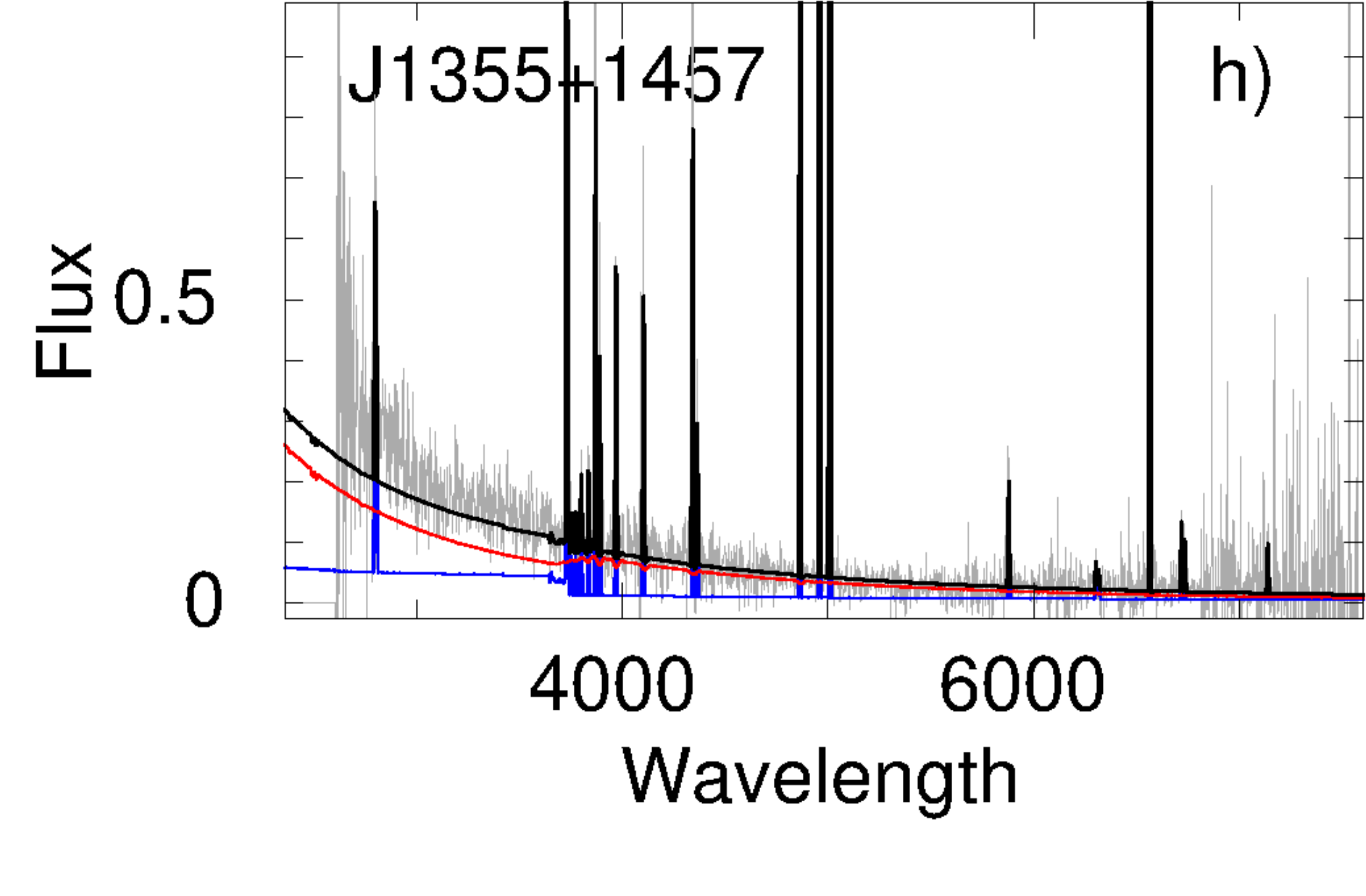}
\includegraphics[angle=-90,width=0.325\linewidth]{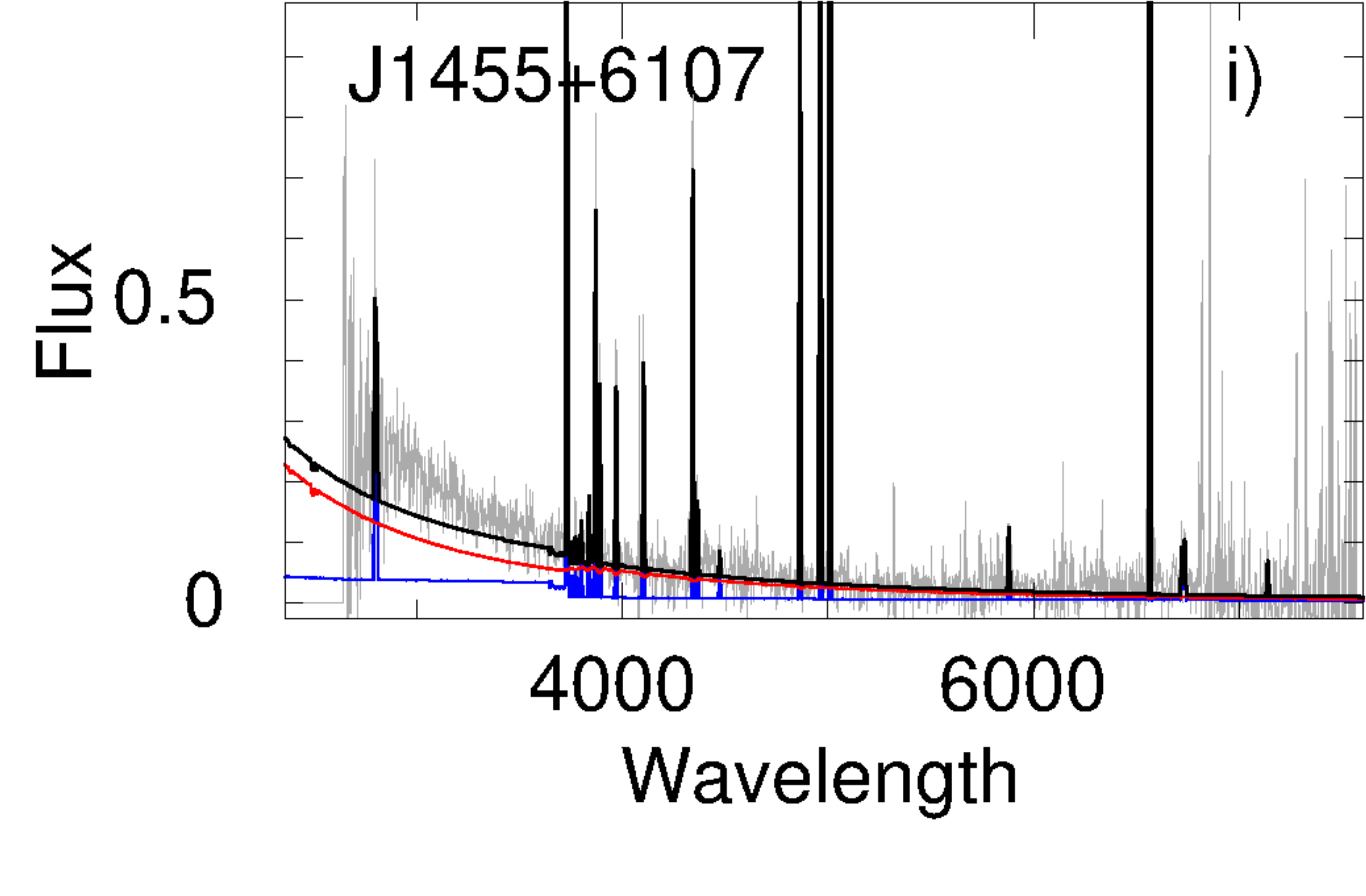}
}
\caption{SED fitting of the galaxy SDSS spectra. The restframe spectra, 
corrected for Milky Way and internal extinction, are shown by grey
lines. The total, nebular and stellar modelled intrinsic SEDs are shown 
by black, blue and red lines, respectively. Fluxes and wavelengths are
expressed in 10$^{-16}$ erg s$^{-1}$ cm$^{-2}$ \AA$^{-1}$ and \AA, respectively.
\label{fig1}}
\end{figure*}

The goal of this paper is to determine $f_{\rm esc}$(LyC) for nine
low-mass galaxies with $M_\star$ $<$ 10$^8$ M$_\odot$. We wish to extend the range
of stellar masses down to 10$^7$ M$_\odot$ as high-redshift galaxies, at the epoch of reionisation,
are thought to be low-mass objects.
We also wish to enlarge the known 
sample of low-redshift LyC leakers and to search for reliable
diagnostics for the indirect determination of $f_{\rm esc}$(LyC). 

\citet{I18b} using the sample of known low-redshift LyC leakers
have considered three indicators for diagnostics, the velocity
separation $V_{\rm sep}$ between the Ly$\alpha$ peaks, the O$_{32}$ ratios and
the stellar mass $M_\star$. 
They
have found a tight anti-correlation between $f_{\rm esc}$(LyC) and $V_{\rm sep}$,  
making $V_{\rm sep}$ one of the best indirect indicators for escaping
ionising radiation. 

As mentioned before, a high O$_{32}$ ratio has 
been suggested to be a promising indirect 
indicator of escaping LyC emission in galaxies at any redshift. 
Although \citet{I18b} and \citet{Na20} did find some
trend of increasing $f_{\rm esc}$(LyC) with increasing O$_{32}$, the dependence
is weak, with a large scatter. 

It has also been suggested that $f_{\rm esc}$(LyC) tends
to be higher in low-mass galaxies \citep{W14,T17}. 
Stellar masses, $M_\star$, are available for a
large number of star-forming galaxies because ground-based observations
are sufficient for their determination.
\citet{I18b} found only a relatively weak anti-correlation
between $f_{\rm esc}$(LyC) and $M_\star$. 
However, this conclusion is based mostly on objects with 
$M_\star$ $\geq$ 10$^{8}$ M$_\odot$, with the exception of one galaxy. 
Intriguingly, the unique galaxy in the LyC leaker sample with a stellar mass 
$<$ 10$^{8}$ M$_\odot$, J1243$+$4646, is also the one 
with the highest $f_{\rm esc}$(LyC), equal to 72 per cent. 
The question then arises: 
are lower mass compact galaxies stronger LyC leakers? 

\citet{Fl19} have found more LyC leakers at
$z$ $\sim$ 3.1 with $M_\star$~$<$~10$^{8}$~M$_\odot$. 
However, they derived $M_\star$ from UV rest-frame photometry,
which is dominated by the radiation from hot massive stars, while the stellar 
masses of low-redshift LyC leakers are derived from the rest-frame spectra in 
the optical range which also includes the emission of lower-mass cooler stars.
Furthermore, the stellar masses derived by \citet{Fl19} are up to 1.5 dex
lower than the values derived by \citet{Na18} for the same galaxies.
Therefore, the $M_\star$ of \citet{Fl19} might be underestimated.

  \begin{table*}
  \caption{Integrated characteristics \label{tab5}}
  \begin{tabular}{lcccccccccccc} \hline
Name&$M_{\rm FUV}$$^{\rm a}$&log $M_\star$$^{\rm b}$ &log $M_{\rm y}$$^{\rm b}$ &log $M_{\rm o}$$^{\rm b}$ &log $L$(H$\beta$)$^{\rm c}$&$t_b$$^{\rm d}$&SFR$^{\rm e}$&sSFR$^{\rm f}$&$\alpha$$^{\rm g}$&$r_{50}$$^{\rm h}$&$\Sigma$$_1^{\rm i}$&$\Sigma$$_2^{\rm j}$\\
\hline   
J0232$-$0426&$-$19.17&7.49&7.40&6.76&41.53&4.4& 7.5  &250&0.47&0.13&10.8~\,&141~\,\\
J0919$+$4906&$-$19.15&7.51&7.46&6.57&41.48&2.6& 8.4  &250&0.59&0.12& 6.1&148~\,\\
J1046$+$5827&$-$20.11&7.89&7.84&6.88&41.70&4.8&11.0\,~~&160& ...$^{\rm k}$& ...$^{\rm k}$& ...$^{\rm k}$&...$^{\rm k}$\\
J1121$+$3806&$-$18.85&7.20&7.14&6.32&41.42&2.7&10.0\,~~&630&0.46&0.16& 8.9& 73\\
J1127$+$4610&$-$19.00&7.44&7.36&6.66&41.22&4.7& 4.2  &160&0.52&0.15& 4.4& 52\\
J1233$+$4959&$-$19.97&7.79&7.57&7.38&41.75&3.3&14.4\,~~&250&0.59&0.17&11.5~\,&139~\,\\
J1349$+$5631&$-$18.76&7.36&7.29&6.58&41.36&4.7& 5.1  &200&0.39&0.19&10.7~\,& 45\\
J1355$+$1457&$-$19.96&7.74&7.61&7.13&41.77&3.6&13.1\,~~&250&0.34&0.21&36.0~\,& 95\\
J1455$+$6107&$-$19.77&7.90&7.54&7.65&41.61&3.3& 9.1  &250&0.39&0.15&19.1~\,&129~\,\\
\hline
  \end{tabular}

\hbox{$^{\rm a}$Absolute FUV magnitude derived from the intrinsic rest-frame SED in mag.}


\hbox{$^{\rm b}$$M_{\rm y}$, $M_{\rm o}$, $M_\star$ are the masses of young, old stellar populations, and total 
stellar mass ($M_{\rm y}$ $+$ $M_{\rm o}$).}

\hbox{$^{\rm c}$$L$(H$\beta$) is the H$\beta$ luminosity corrected for the Milky Way and internal extinction
in erg s$^{-1}$.}

\hbox{$^{\rm d}$$t_b$ is the starburst age in Myr.}

\hbox{$^{\rm e}$Star-formation rate corrected for the Milky Way and internal extinction, and escaping LyC radiation in M$_\odot$ yr$^{-1}$.}

\hbox{$^{\rm f}$Specific star-formation rate in Gyr$^{-1}$.}

\hbox{$^{\rm g}$Exponential disc scale length in kpc.}

\hbox{$^{\rm h}$Galaxy radius, at which NUV intensity equal to half of maximal intensity in kpc.}

\hbox{$^{\rm i}$Star-formation rate surface density assuming galaxy radius equal to $\alpha$ 
in M$_\odot$ yr$^{-1}$kpc$^{-2}$.}

\hbox{$^{\rm j}$Star-formation rate surface density assuming galaxy radius equal to $r_{50}$ 
in M$_\odot$ yr$^{-1}$kpc$^{-2}$.}

\hbox{$^{\rm k}$Acquisition image not obtained.}
  \end{table*}

  \begin{table}
  \caption{{\sl HST}/COS observations \label{tab6}}
  \begin{tabular}{lcccc} \hline
\multicolumn{1}{c}{}&\multicolumn{1}{c}{}&\multicolumn{3}{c}{Exposure time (s)} \\ 
\multicolumn{1}{c}{Name}&\multicolumn{1}{c}{Date}&\multicolumn{3}{c}{(Central wavelength (\AA))} \\ 
    &    &MIRRORA&G140L&G160M \\ \hline
J0232$-$0426&2019-09-25&2$\times$800     &  8211& 3410\\
            &          &         &(800)&(1623)\\
J0919$+$4906&2020-04-30&2$\times$900     &  8641& 3499\\
            &          &         &(800)&(1611)\\
J1046$+$5827&2019-10-08&2$\times$0$^{\rm a}$       &  7407$^{\rm b}$& 3910\\
            &          &         &(800)&(1600)\\
J1121$+$3806&2020-04-16&2$\times$900     &  5551$^{\rm b}$& 3326\\
            &          &         &(800)&(1533)\\ 
J1127$+$4610&2020-01-10&2$\times$900     &  8641& 3497\\
            &          &         &(800)&(1533)\\ 
J1233$+$4959&2020-01-17&2$\times$900     &  8641& 3497\\
            &          &         &(800)&(1623)\\ 
J1349$+$5631&2020-03-12&2$\times$800     &  8953& 3916\\
            &          &         &(800)&(1577)\\ 
J1355$+$1457&2020-05-28&2$\times$800     &  8232& 3436\\
            &          &         &(800)&(1577)\\ 
J1455$+$6107&2020-03-09&2$\times$900     &  9019& 3760\\
            &          &         &(800)&(1577)\\ 
\hline
\end{tabular}

\hbox{$^{\rm a}$Failed exposure.}

\hbox{$^{\rm b}$Partially executed observation.}

  \end{table}

\begin{figure*}
\hbox{
\includegraphics[angle=-90,width=0.32\linewidth]{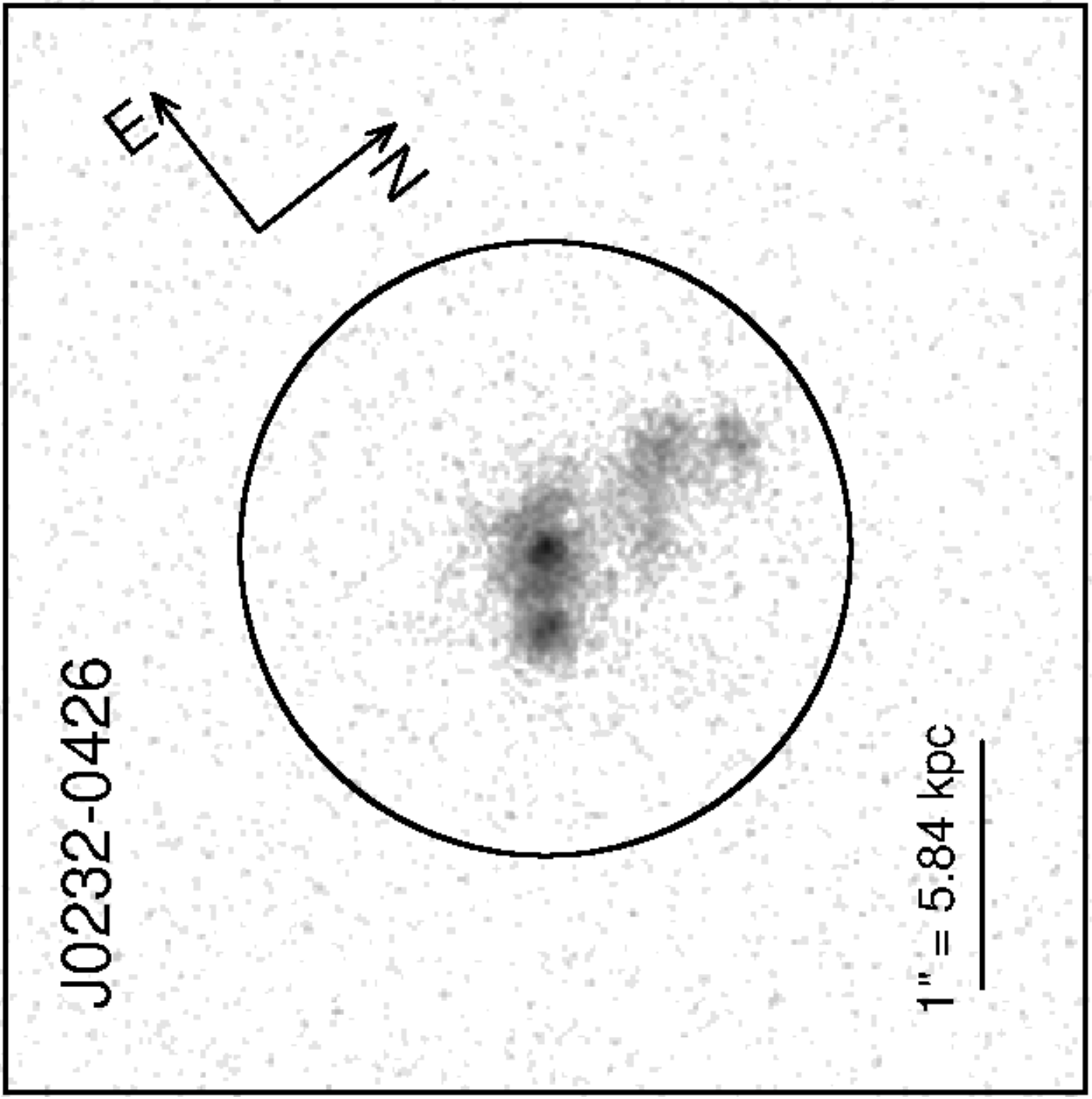}
\includegraphics[angle=-90,width=0.32\linewidth]{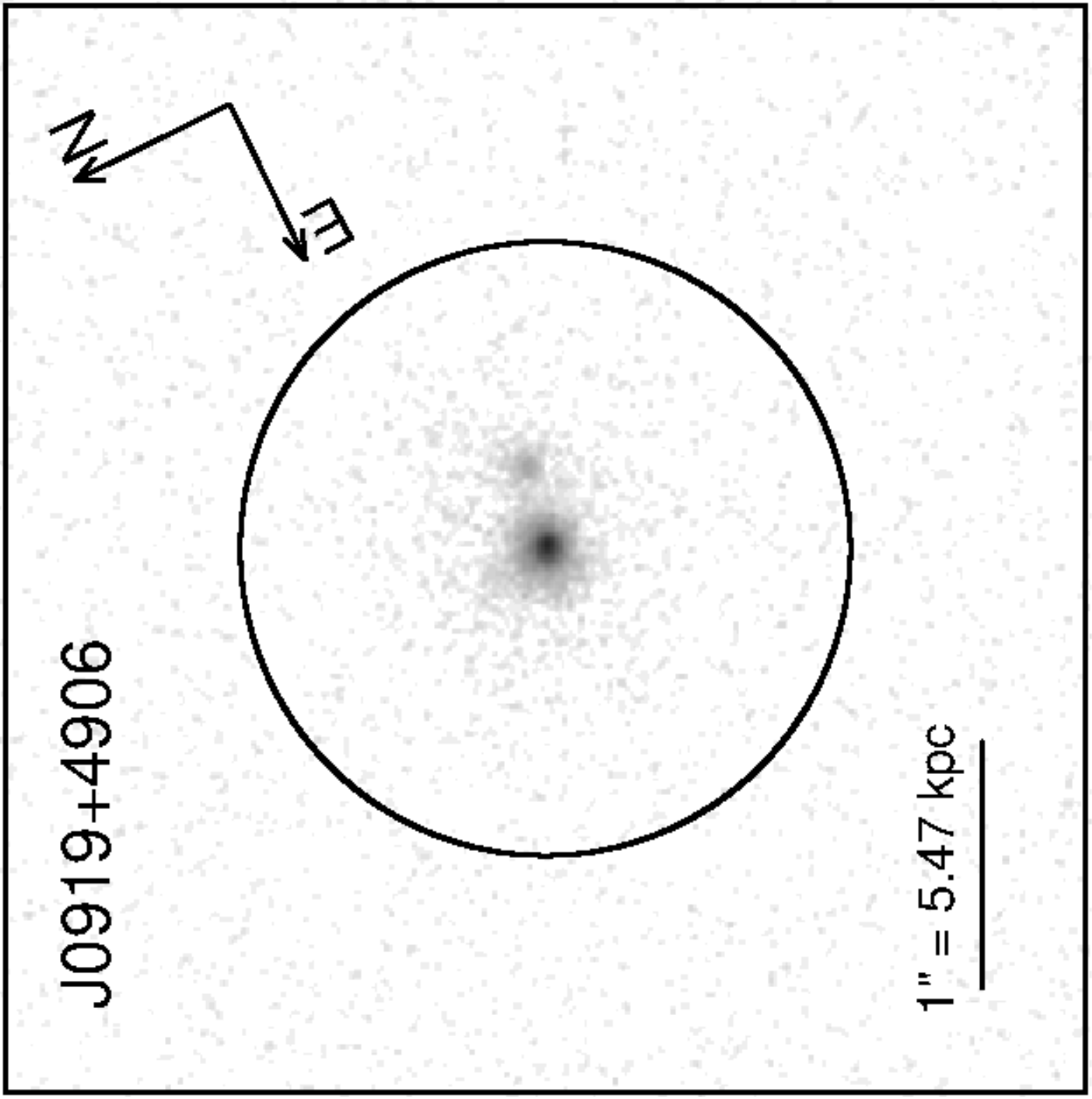}
}
\hbox{
\includegraphics[angle=-90,width=0.32\linewidth]{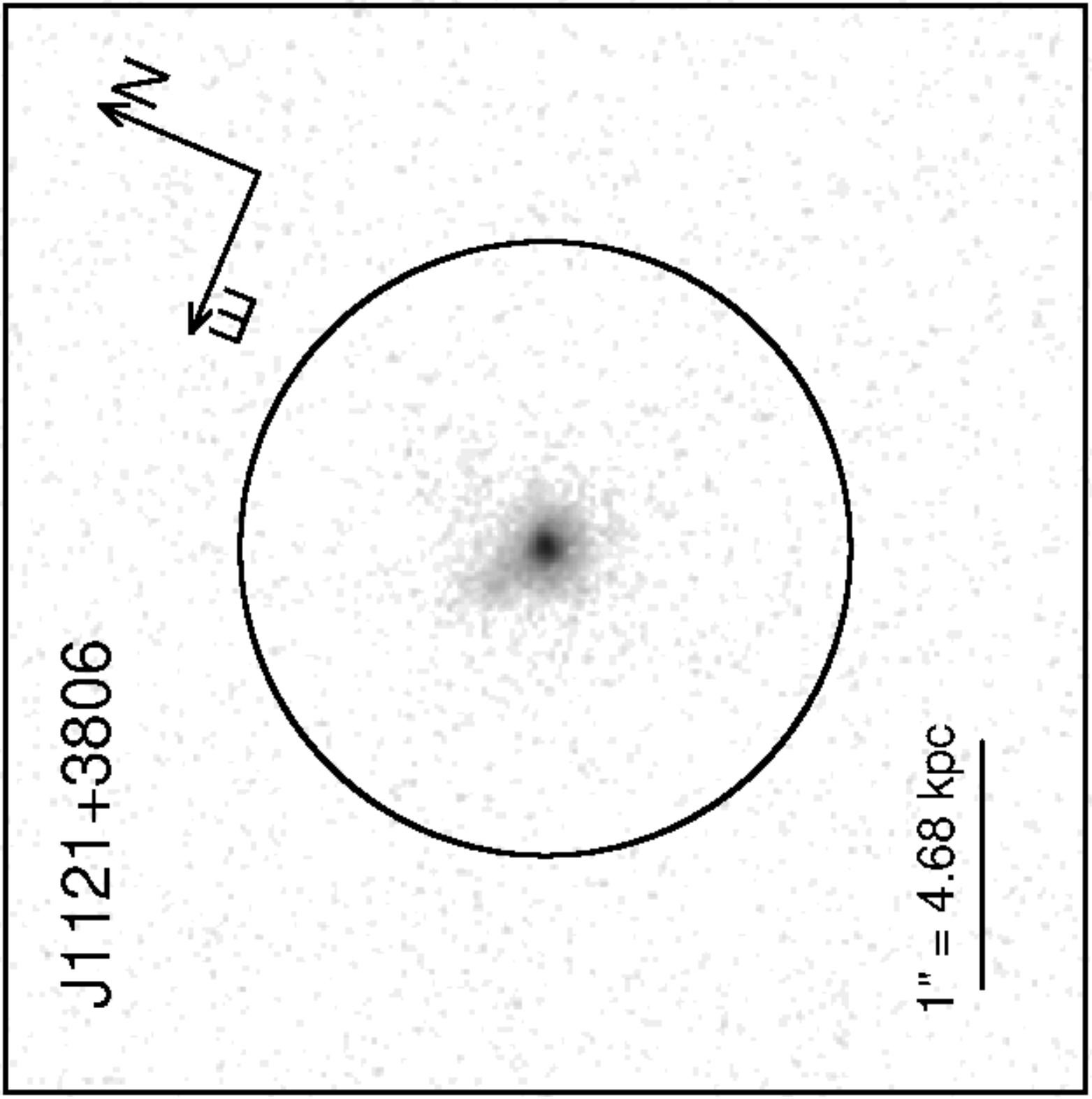}
\includegraphics[angle=-90,width=0.32\linewidth]{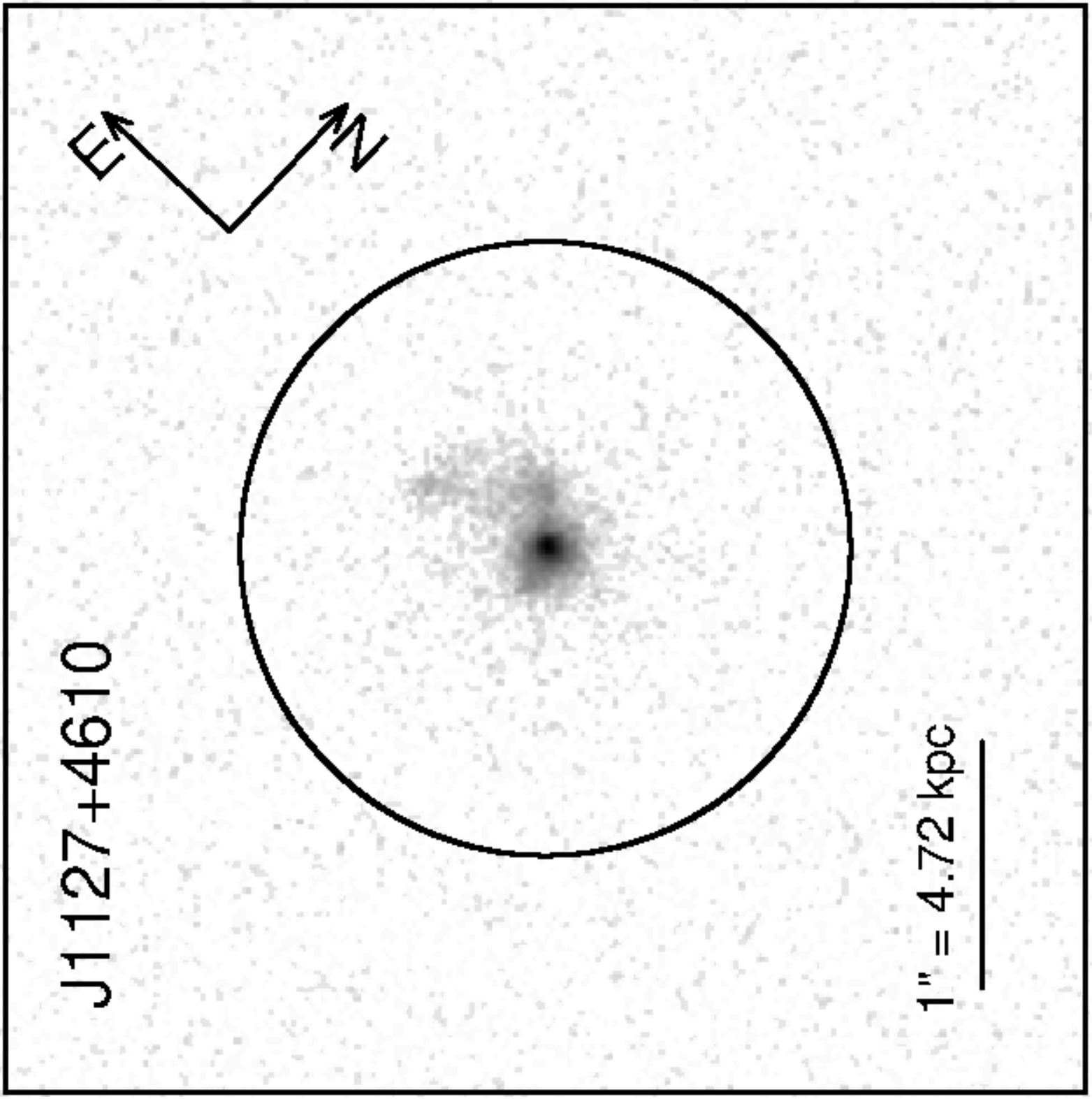}
\includegraphics[angle=-90,width=0.32\linewidth]{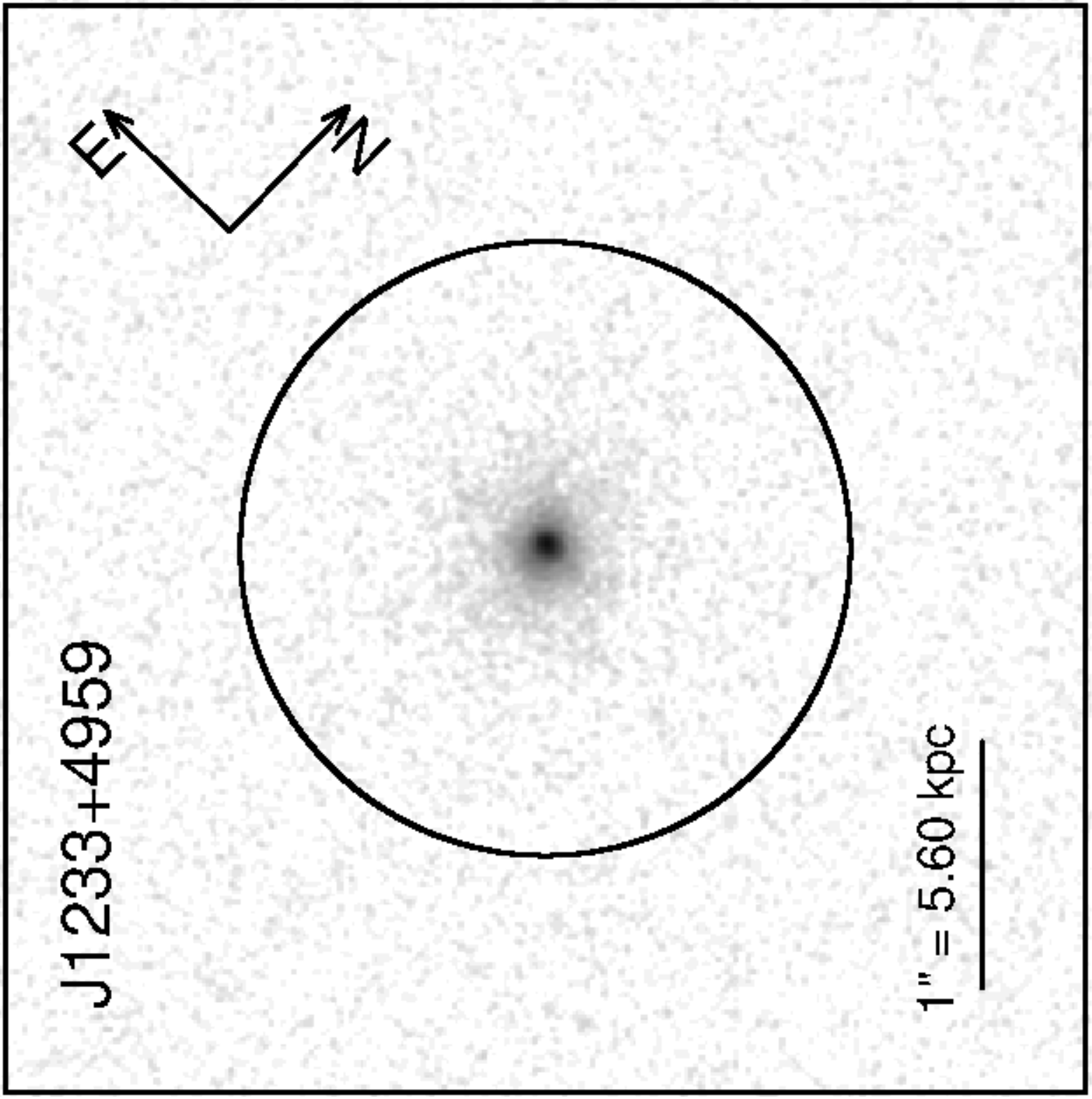}
}
\hbox{
\includegraphics[angle=-90,width=0.32\linewidth]{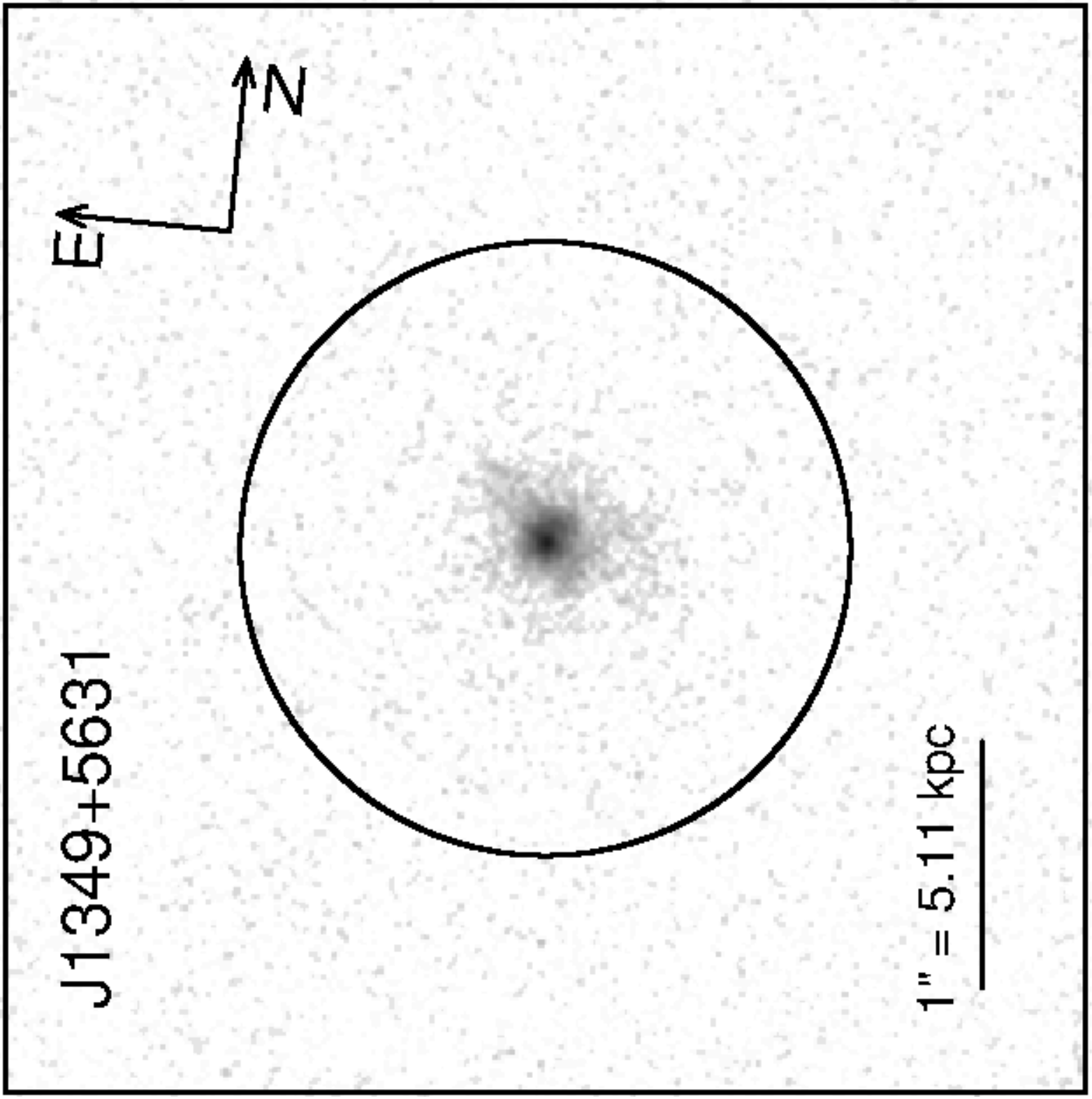}
\includegraphics[angle=-90,width=0.32\linewidth]{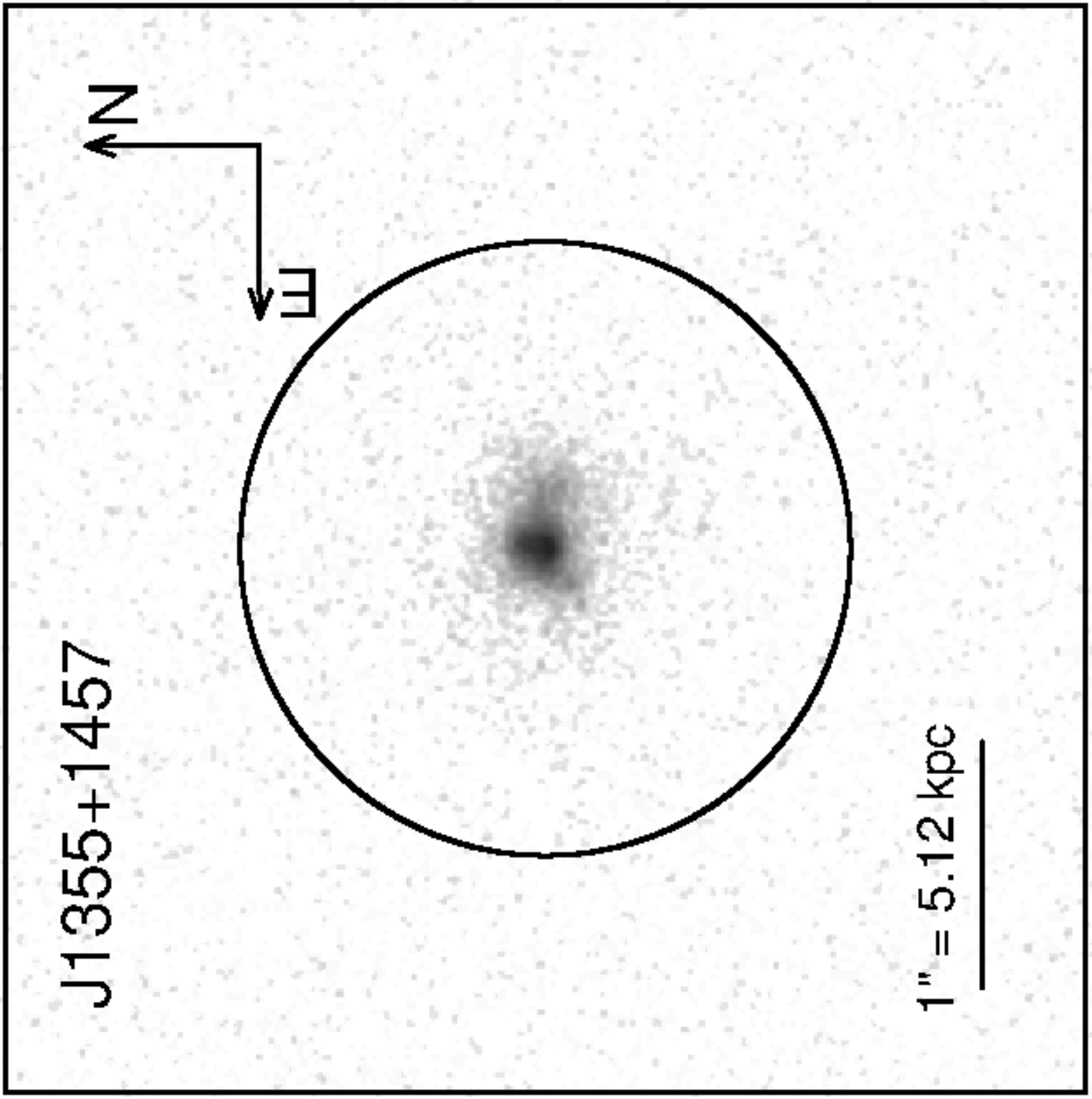}
\includegraphics[angle=-90,width=0.32\linewidth]{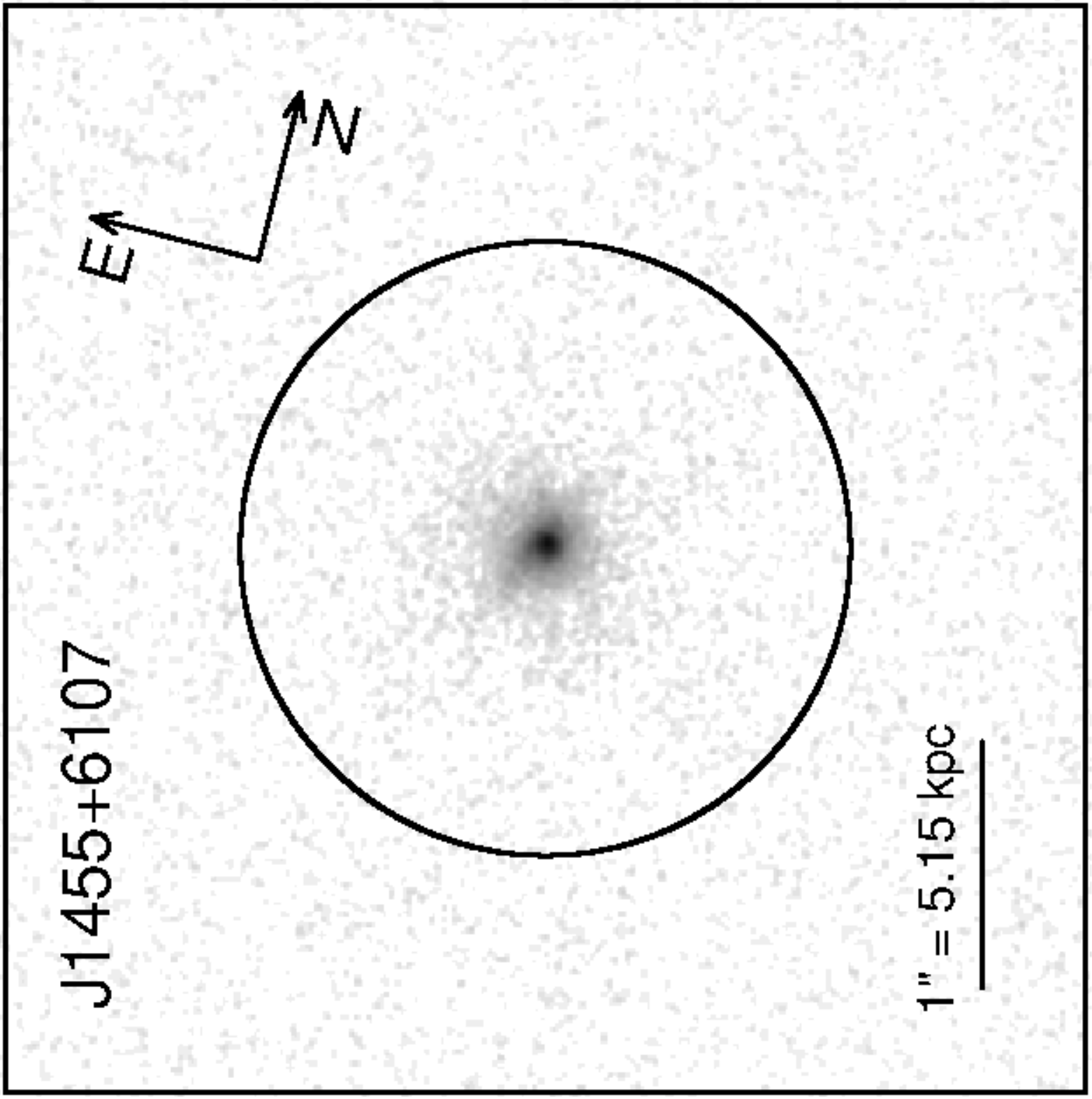}
}
\caption{The {\sl HST} NUV acquisition images of the candidate LyC leaking galaxies in a
log surface brightness scale. The COS spectroscopic aperture with a diameter of 
2.5 arcsec is shown in all panels by a circle.
The linear scale in each panel is derived adopting an angular size distance.
\label{fig2}}
\end{figure*}

\begin{figure*}
\hbox{
\includegraphics[angle=-90,width=0.328\linewidth]{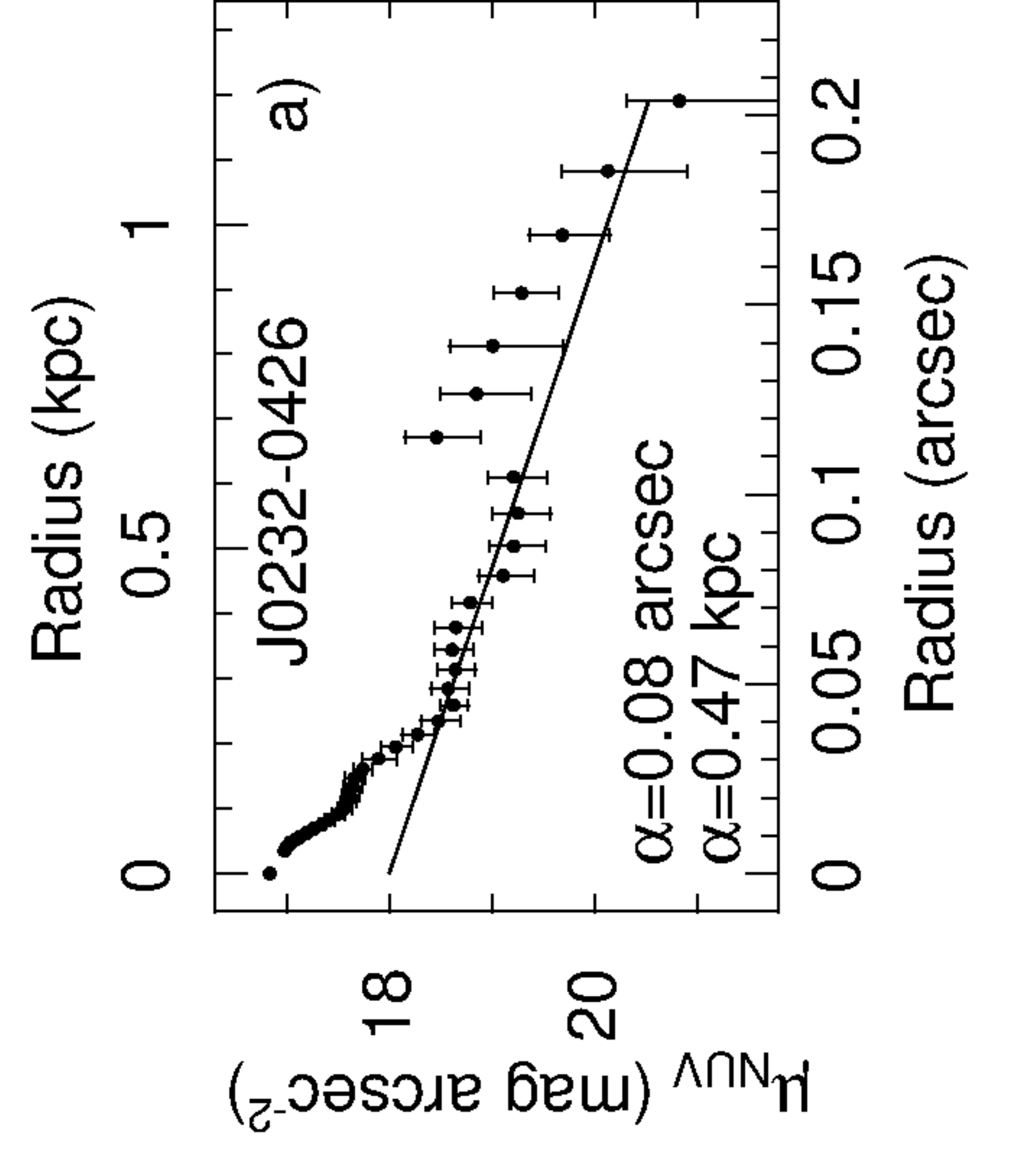}
\includegraphics[angle=-90,width=0.328\linewidth]{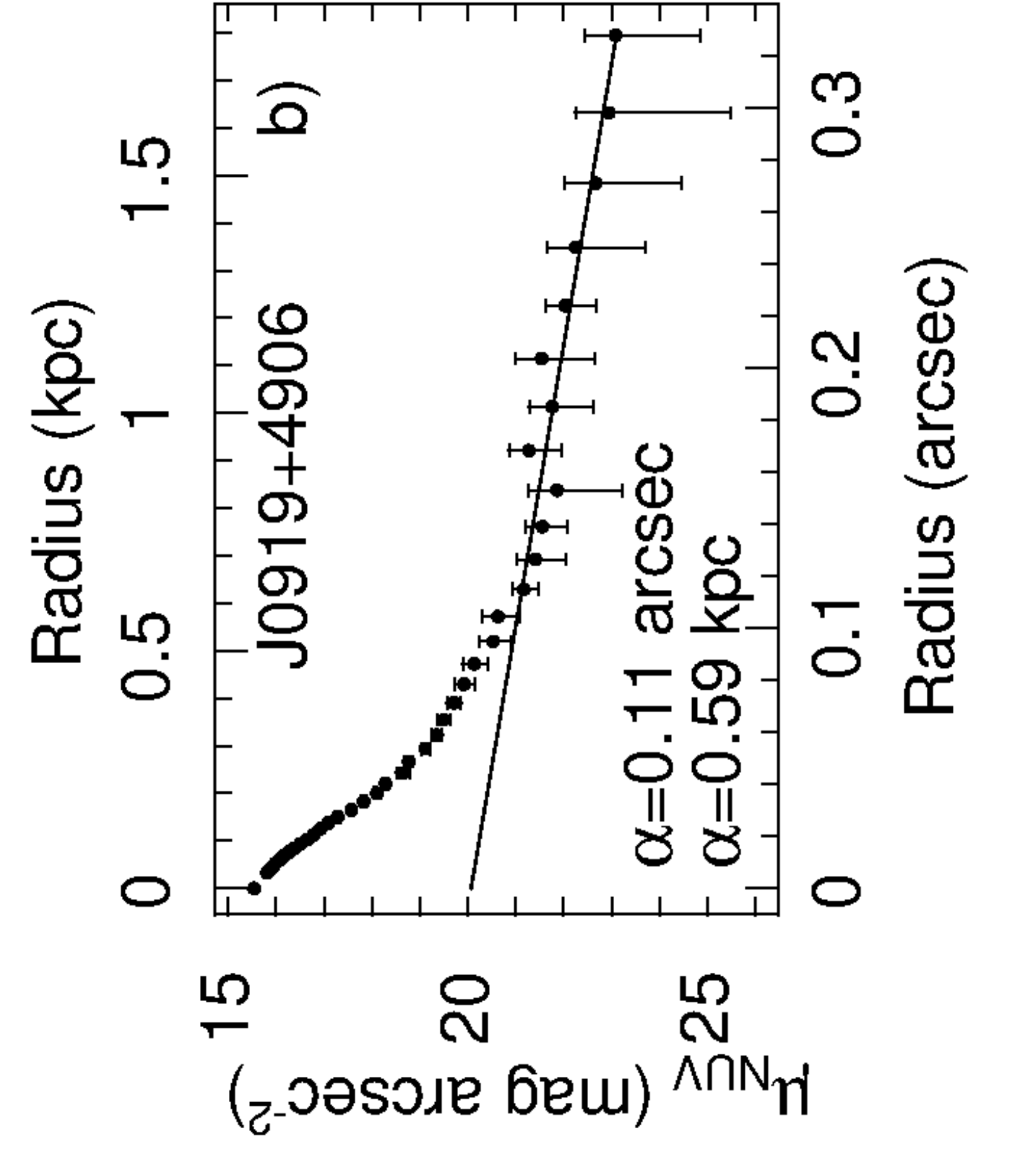}
 }
 \hbox{
\includegraphics[angle=-90,width=0.328\linewidth]{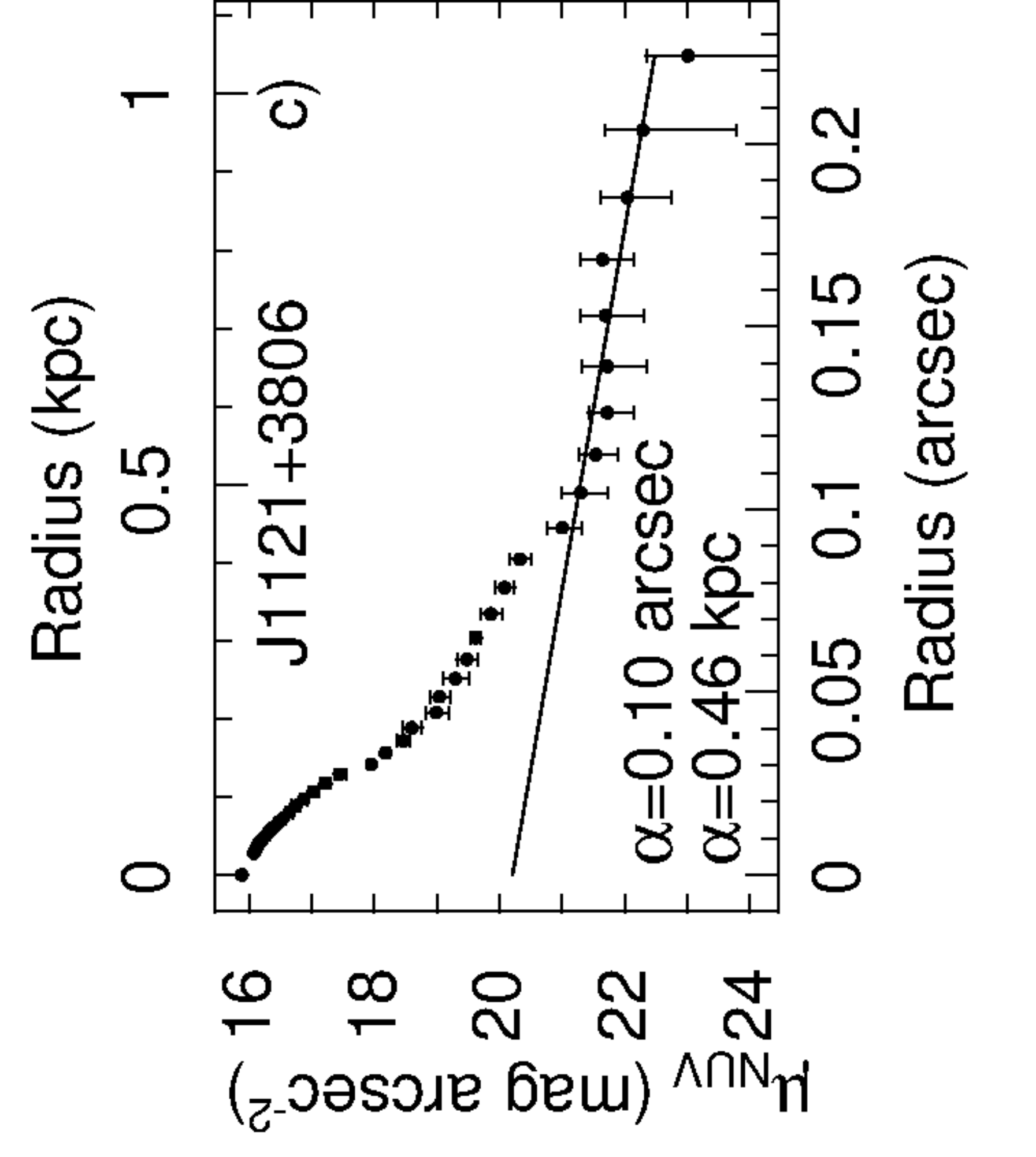}
\includegraphics[angle=-90,width=0.328\linewidth]{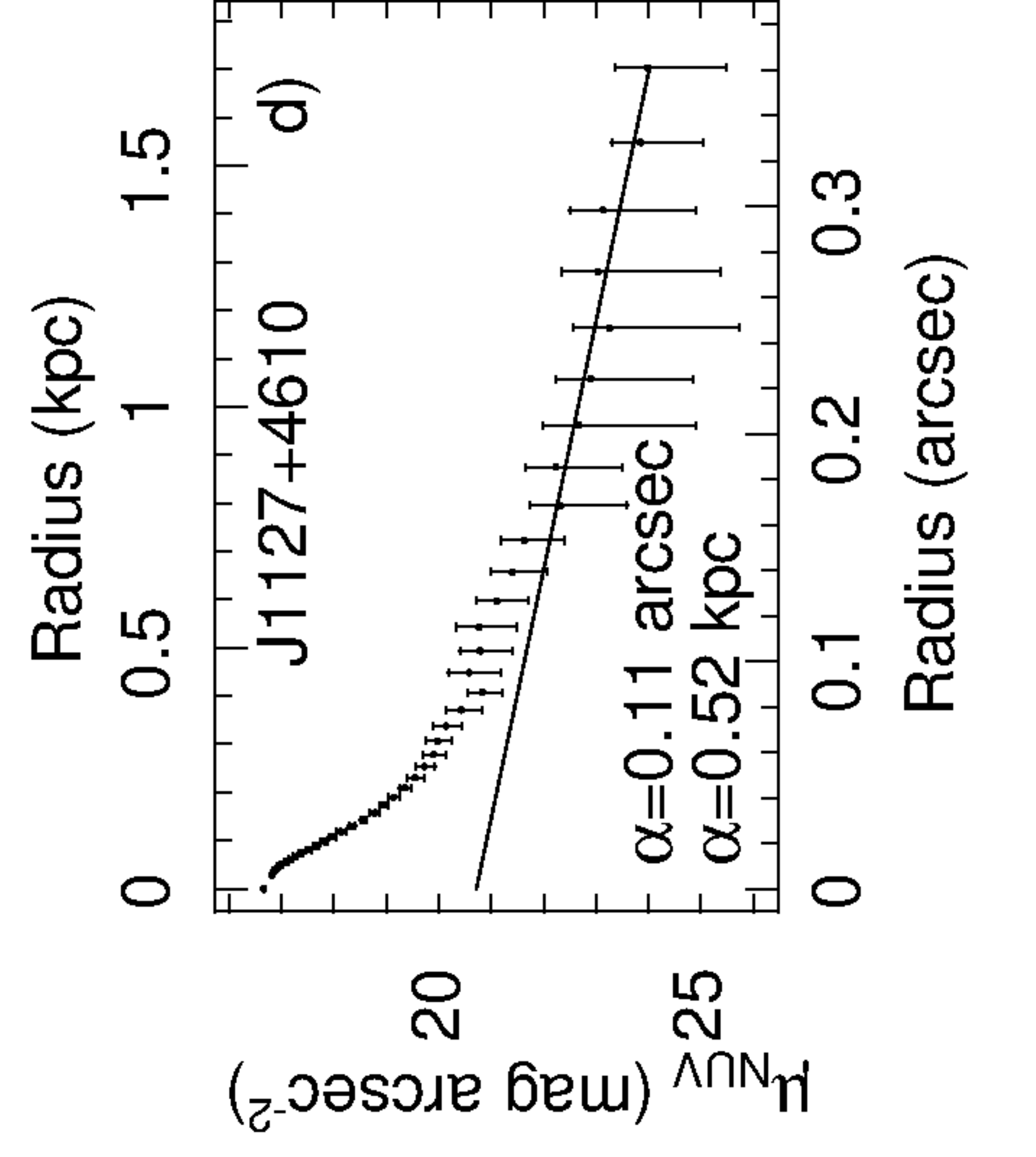}
\includegraphics[angle=-90,width=0.328\linewidth]{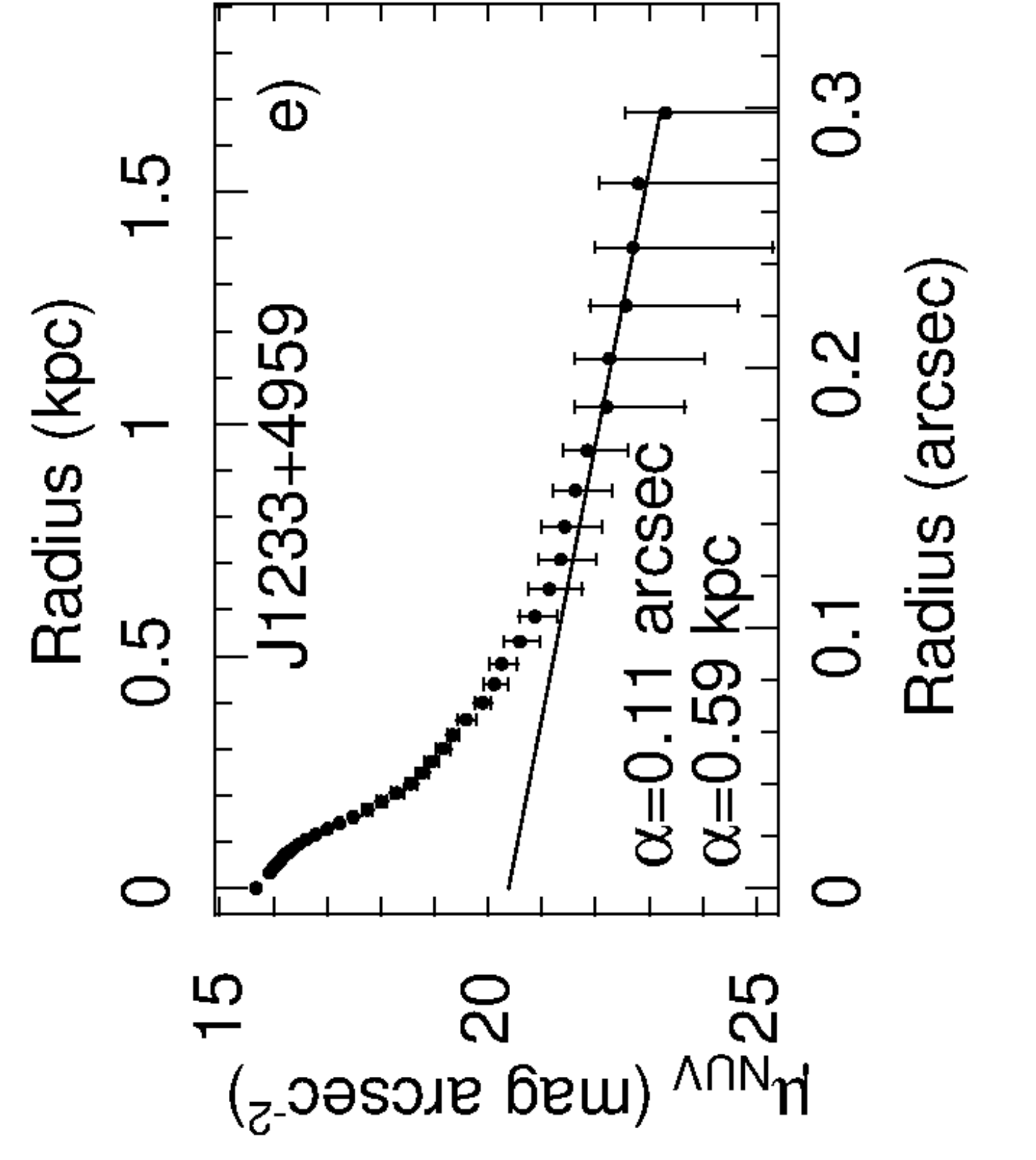}
}
 \hbox{
\includegraphics[angle=-90,width=0.328\linewidth]{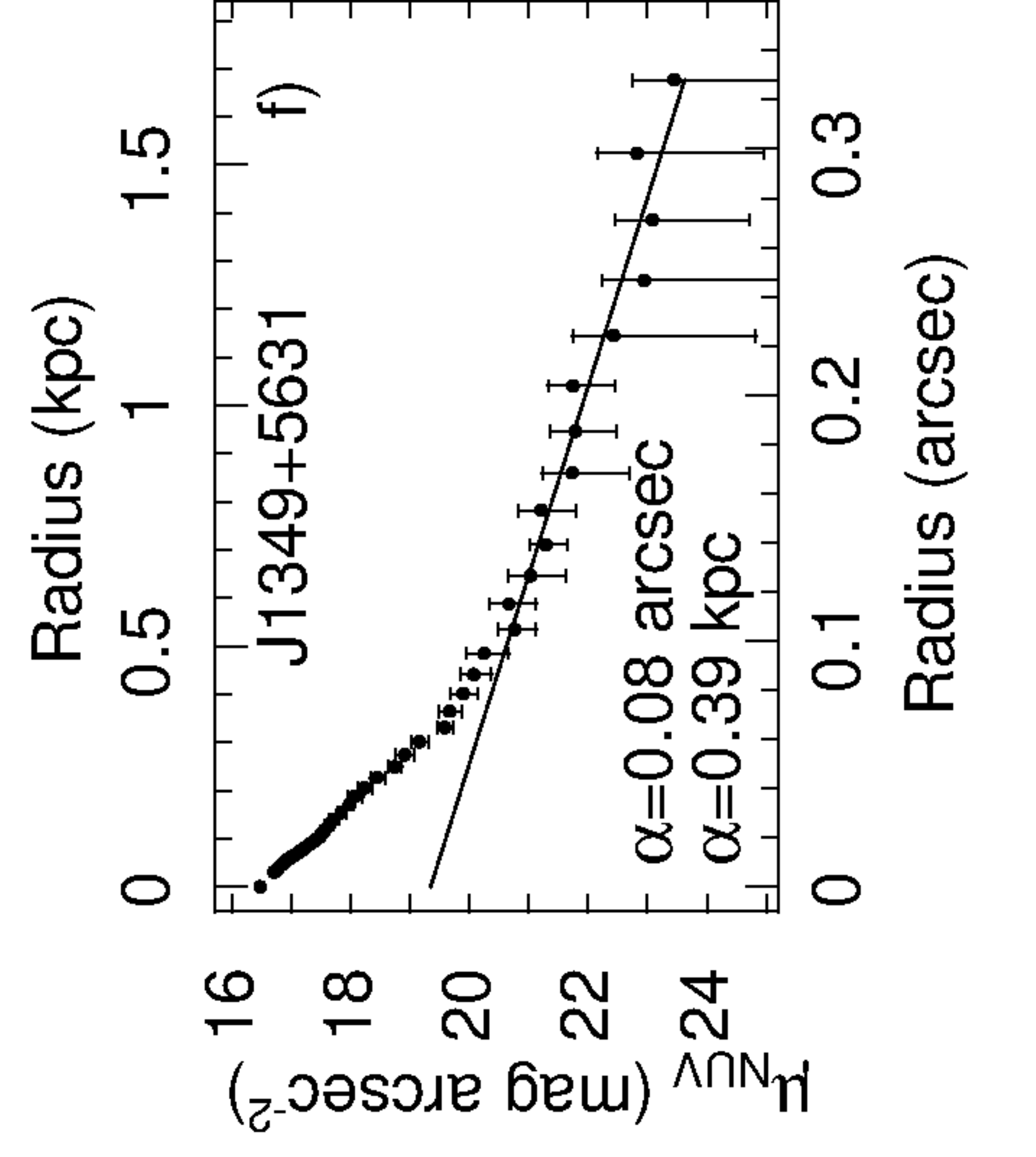}
\includegraphics[angle=-90,width=0.328\linewidth]{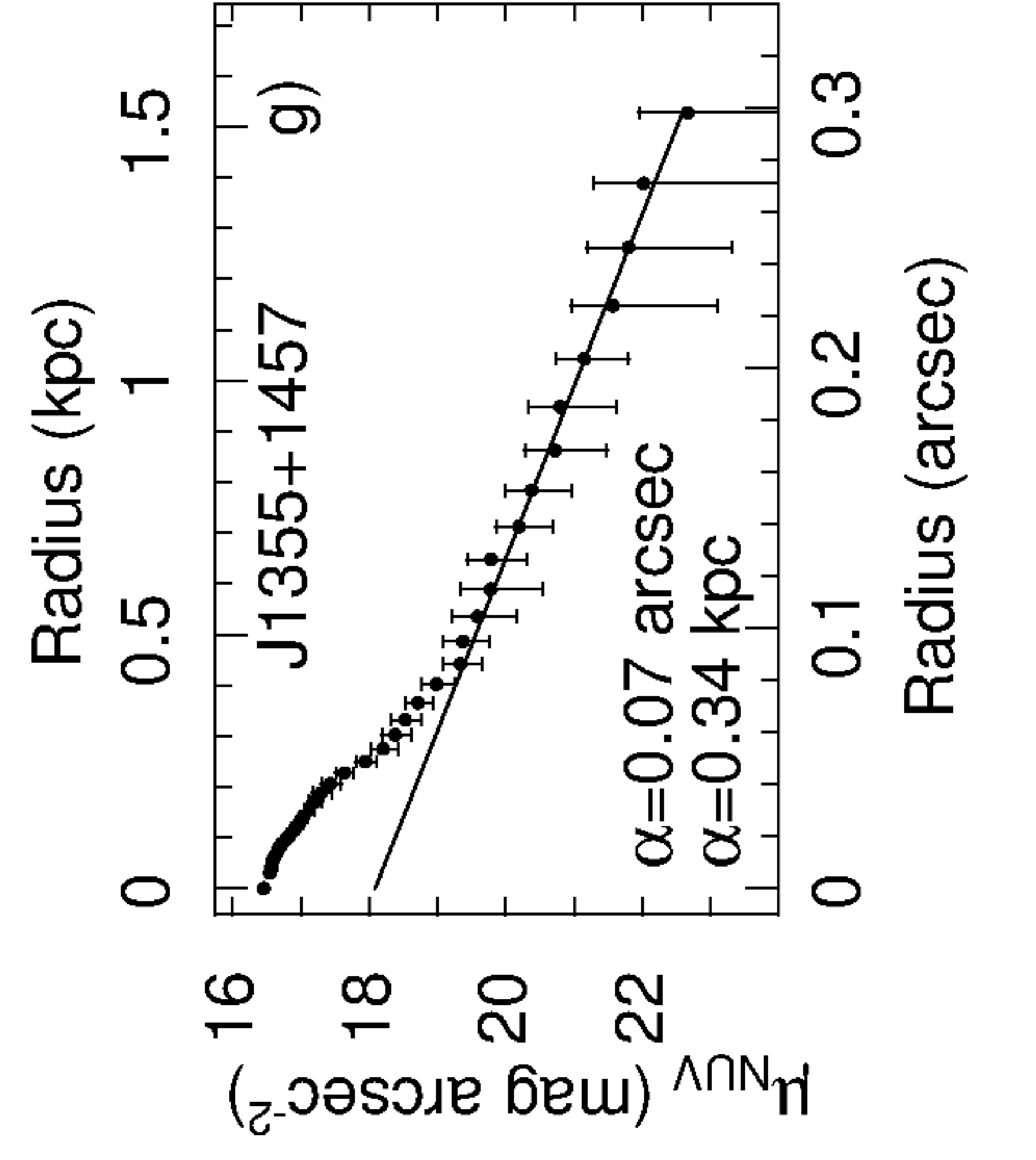}
\includegraphics[angle=-90,width=0.328\linewidth]{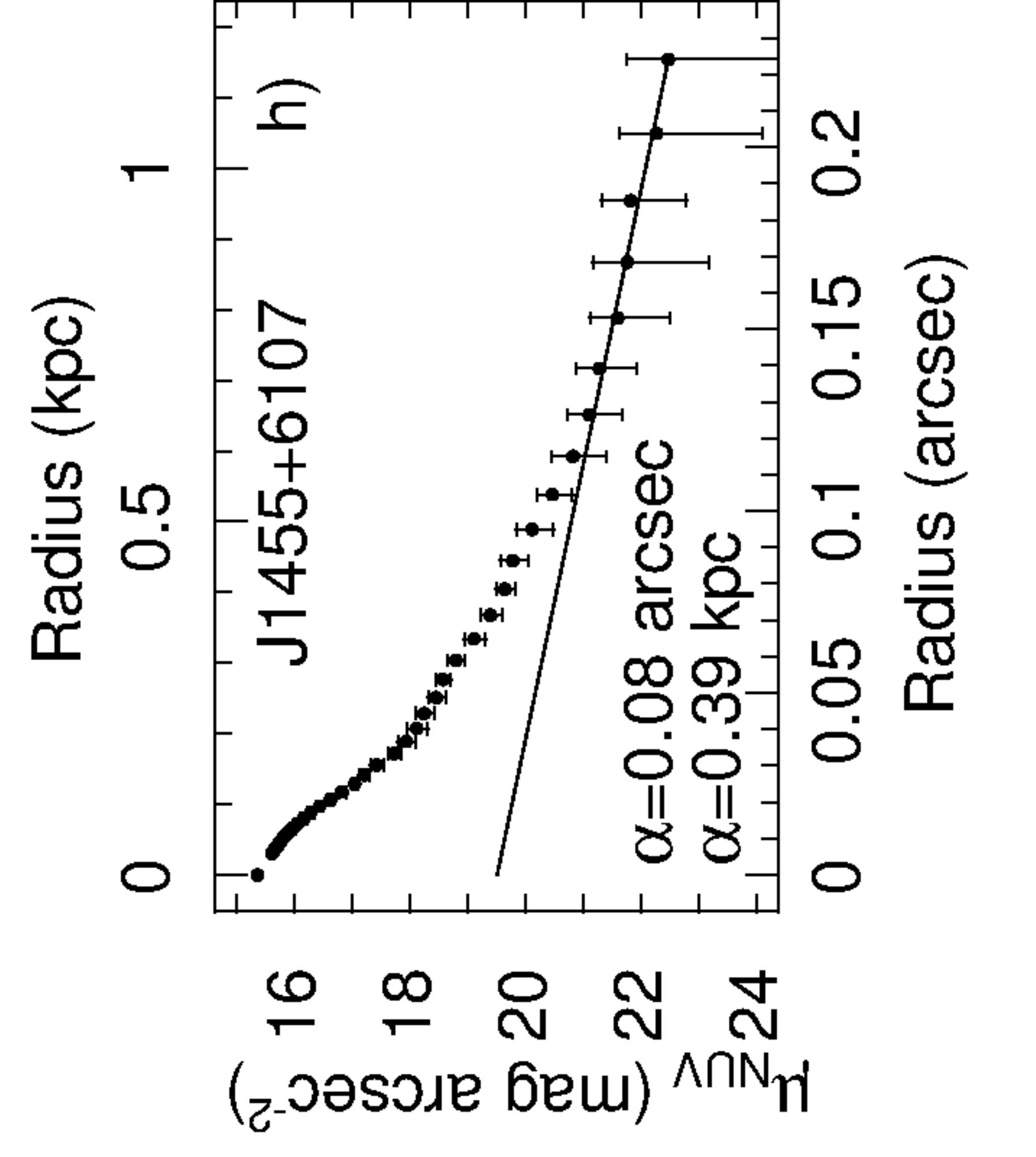}
}
\caption{NUV surface brightness profiles of galaxies. 
\label{fig3}}
\end{figure*}

In this paper, we present new {\sl HST}/COS observations of the LyC 
in 9 compact SFGs, with the lowest $M_\star$~$<$~10$^8$~M$_\odot$
ever observed at $z$ $\sim$ 0.3 -- 0.4. We wish to detect their ionising radiation, 
and examine its behaviour over a wide range of stellar masses. 
The properties of the selected SFGs derived from observations in the optical range
are presented in Section~\ref{sec:select}. The {\sl HST} observations 
and data reduction are described
in Section~\ref{sec:obs}. The surface brightness profiles 
in the UV range are discussed in Section~\ref{sec:sbp}. 
In Section~\ref{sec:global}, we compare the {\sl HST}/COS spectra
with the extrapolation of the modelled SEDs to the UV range.
Ly$\alpha$ emission is considered in 
Section~\ref{sec:lya}. The escaping Lyman continuum emission is discussed in 
Section~\ref{sec:lyc} together with the corresponding
escape fractions. The indirect indicators of escaping LyC emission are 
considered in Section~\ref{Ind}.
We summarize our findings in Section~\ref{summary}.

\section{Properties of selected galaxies derived from
observations in the optical range}\label{sec:select}

We selected a sample of local compact low-mass SFGs 
($M_\star$~$<$~10$^8$~M$_\odot$) to observe their Ly$\alpha$ and LyC emission with
{\sl HST}/COS. The sources are selected from SDSS in the redshift range 
$z = 0.32-0.45$, which allows efficient observations of the LyC with COS.  
They are also chosen to be the brightest and have the highest O$_{32}$ ratios and 
the highest equivalent widths EW(H$\beta$) of the H$\beta$ emission line so that 
a galaxy can be acquired and observed with low- and medium-resolution 
gratings in one visit, consisting of 5 orbits.
These selection criteria yield a total sample of 9 galaxies. 
They are listed in Table \ref{tab1}.
The important features of the objects in this sample compared to the previous 
ones are 1) the lower stellar masses, extending the mass range down to
$\sim$ 2$\times$10$^7$~M$_\odot$, which is more typical of high-$z$
star-forming galaxies; 2) the large range of 
O$_{32}$ $\sim$ 4 -- 14, similar to that of 
the LyC leakers of \citet{I16a,I16b,I18a,I18b};  and 3) the 
fainter UV absolute magnitudes $M_{\rm FUV}$ $\sim$ --19 - --20 mag, whereas
the galaxies in \citet{I16a,I16b,I18a,I18b} are generally brighter, with
$M_{\rm FUV}$ $\sim$ --20 - --21 mag. This is fainter than the characteristic 
$M_{\rm FUV}$ $\sim$ --21 mag for galaxies at $z$ $\sim$ 6. Thus, our galaxies
represent the faint end of the luminosity function.
It is also worth noting that
all objects from the selected sample have low metallicities (the range of 
oxygen abundances is $\sim$ 7.8 -- 8.1) and high equivalent 
widths of the H$\beta$ emission line (158 - 435~\AA). 
All galaxies are nearly unresolved by the SDSS $g$-band images, which have 
FWHMs of 1.0 -- 1.6 arcsec, so that all the galaxy's light 
falls within the 2.5 arcsec diameter COS aperture, ensuring that 
global quantities can be derived. 

The SDSS, {\sl GALEX} and {\sl WISE} apparent magnitudes of the selected
galaxies are shown in Table \ref{tab2}, indicating that these SFGs are
among the faintest low-redshift LyC leaker candidates selected 
for {\sl HST} observations. In fact, only four out of nine galaxies have been 
detected in 
the FUV range and only two galaxies are present in the AllWISE catalogue.

To derive absolute magnitudes and other integrated parameters we adopted
luminosity and angular size distances \citep[NASA Extragalactic 
Database (NED)\footnote{NASA/IPAC Extragalactic Database (NED) is operated by 
the Jet Propulsion Laboratory, California Institute of Technology, under 
contract with the National Aeronautics and Space Administration.},][]{W06} with the cosmological 
parameters $H_0$=67.1 km s$^{-1}$Mpc$^{-1}$, $\Omega_\Lambda$=0.682, 
$\Omega_m$=0.318 \citep{P14}. These distances are presented in Table~\ref{tab1}.

\subsection{Interstellar extinction and element abundances}\label{sec:ext}

Internal interstellar extinction has been derived from the observed decrement
of all hydrogen emission lines, which are measurable in the SDSS spectra 
\citep*{ITL94}. First, the emission-line fluxes in the observed SDSS spectra, 
uncorrected for redshift, were corrected for 
the Milky Way extinction with $A(V)_{\rm MW}$ from the NED, adopting the
\citet*{C89} reddening law and $R(V)_{\rm MW}$~=~3.1.
Second, the fluxes of emission lines at the rest-frame 
wavelengths were corrected for the internal 
extinction of galaxies with $R(V)_{\rm int}~=~3.1$ and 
$A(V)_{\rm int}$~=~3.1$\times$$E(B-V)_{\rm int}$, where 
$E(B-V)_{\rm int}$~=~$C$(H$\beta$)$_{\rm int}$/1.47 \citep{A84}.
Finally, the extinction-corrected emission lines are
used to derive ionic and total element abundances following the methods 
described in \citet{I06} and \citet{G13}. 

The emission-line fluxes $I$($\lambda$) relative to the H$\beta$ flux corrected 
for both the Milky Way and internal extinctions, the rest-frame equivalent 
widths, the Milky Way ($C$(H$\beta$)$_{\rm MW}$) and internal 
($C$(H$\beta$)$_{\rm int}$) extinction coefficients,
and extinction-corrected H$\beta$ fluxes are shown in Table \ref{tab3}.
We note that H$\alpha$ emission lines in SDSS spectra of two galaxies,
J0232$-$0426 and J1046$+$5827 are clipped and therefore their fluxes are lower
than the theoretical recombination values. Consequently, we excluded these 
lines from the determination of internal extinctions in J0232$-$0426 and 
J1046$+$5827. The fluxes and the direct $T_{\rm e}$ method are used to derive 
the physical conditions (electron temperature and electron number density) and 
the element abundances in the H~{\sc ii} regions.
These quantities are shown in Table \ref{tab4}. The oxygen abundances are 
comparable to those in known low-redshift LyC leakers by
\citet{I16a,I16b,I18a,I18b}. The ratios of the $\alpha$-element 
(neon and magnesium)
abundances to oxygen abundance are similar to those in
dwarf emission-line galaxies \citep[e.g. ][]{I06,G13}. On the other hand, the 
nitrogen-to-oxygen abundance ratios in some galaxies are somewhat elevated, 
similar to those in other LyC leakers at $z$~$\ga$~0.3.

\begin{figure*}
\hbox{
\includegraphics[angle=-90,width=0.325\linewidth]{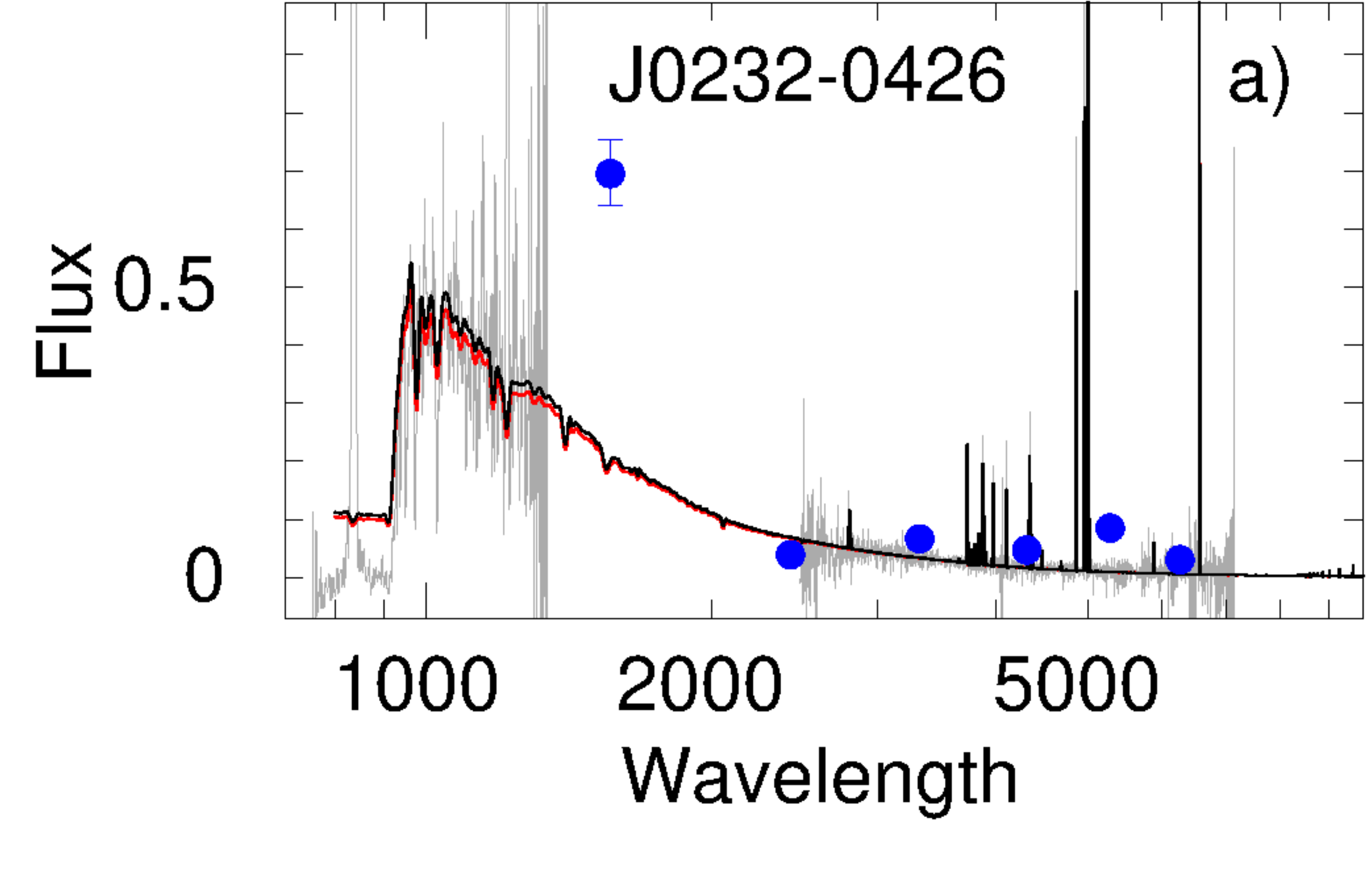}
\includegraphics[angle=-90,width=0.325\linewidth]{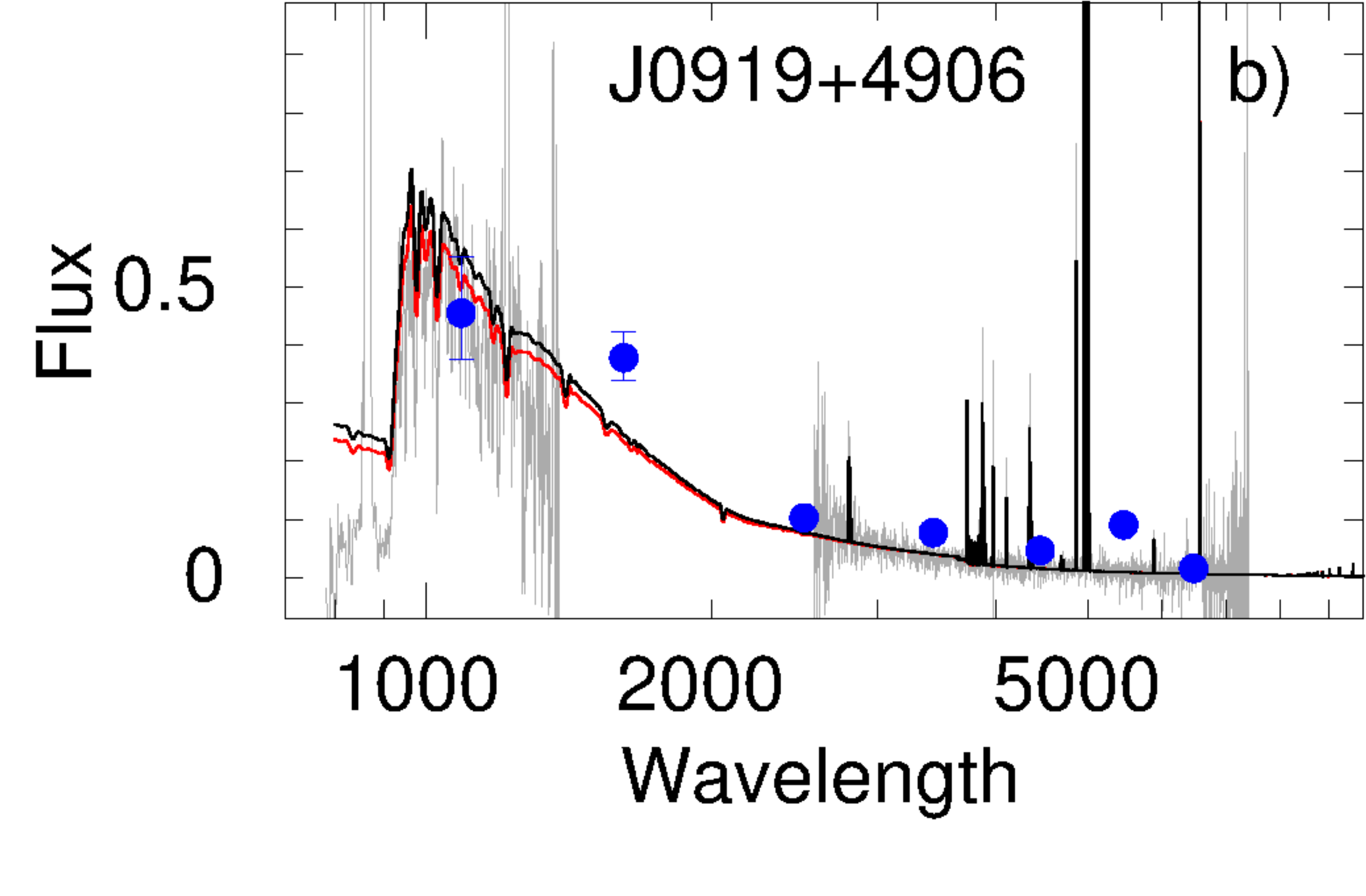}
\includegraphics[angle=-90,width=0.325\linewidth]{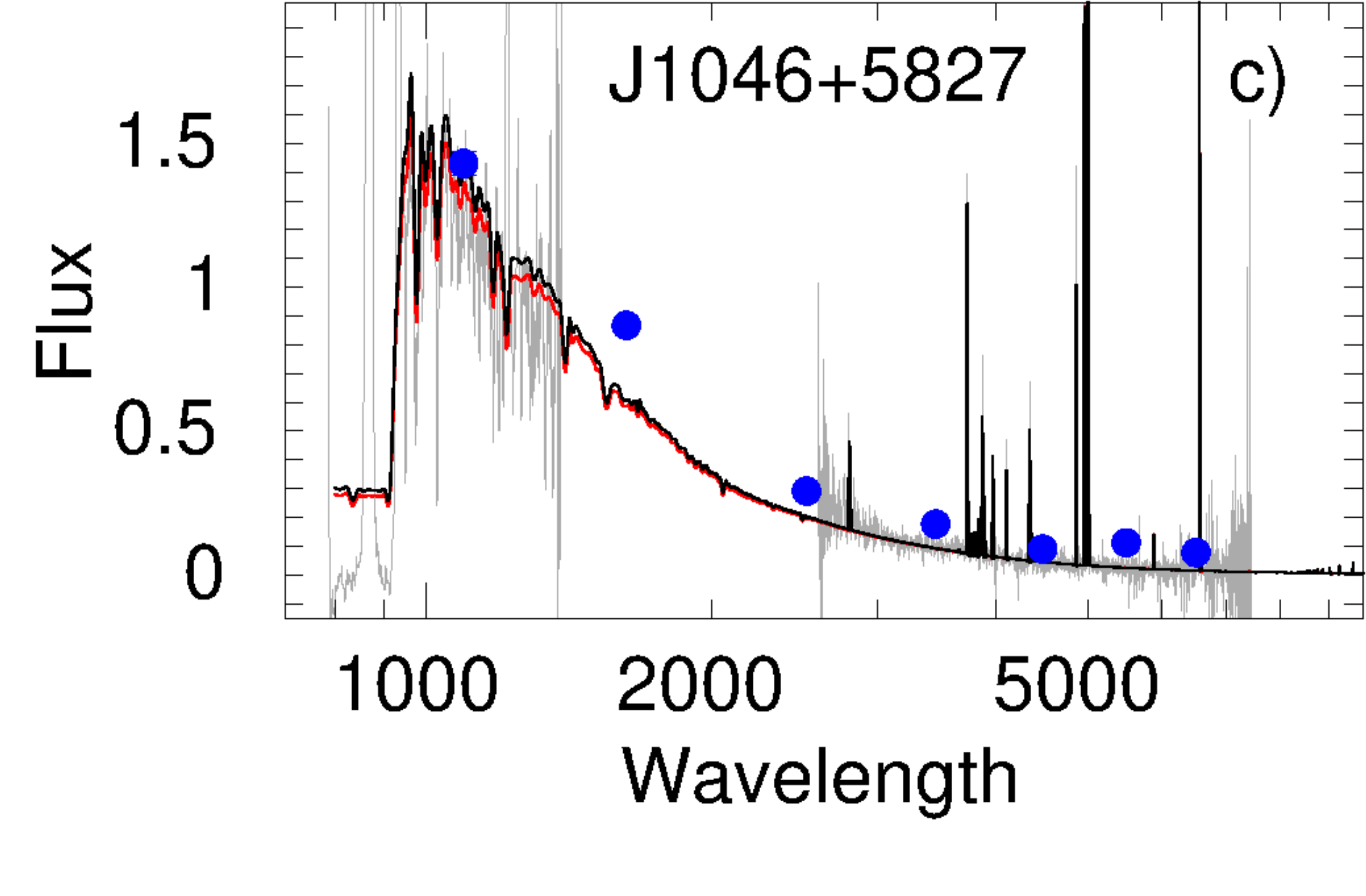}
 }
 \hbox{
\includegraphics[angle=-90,width=0.325\linewidth]{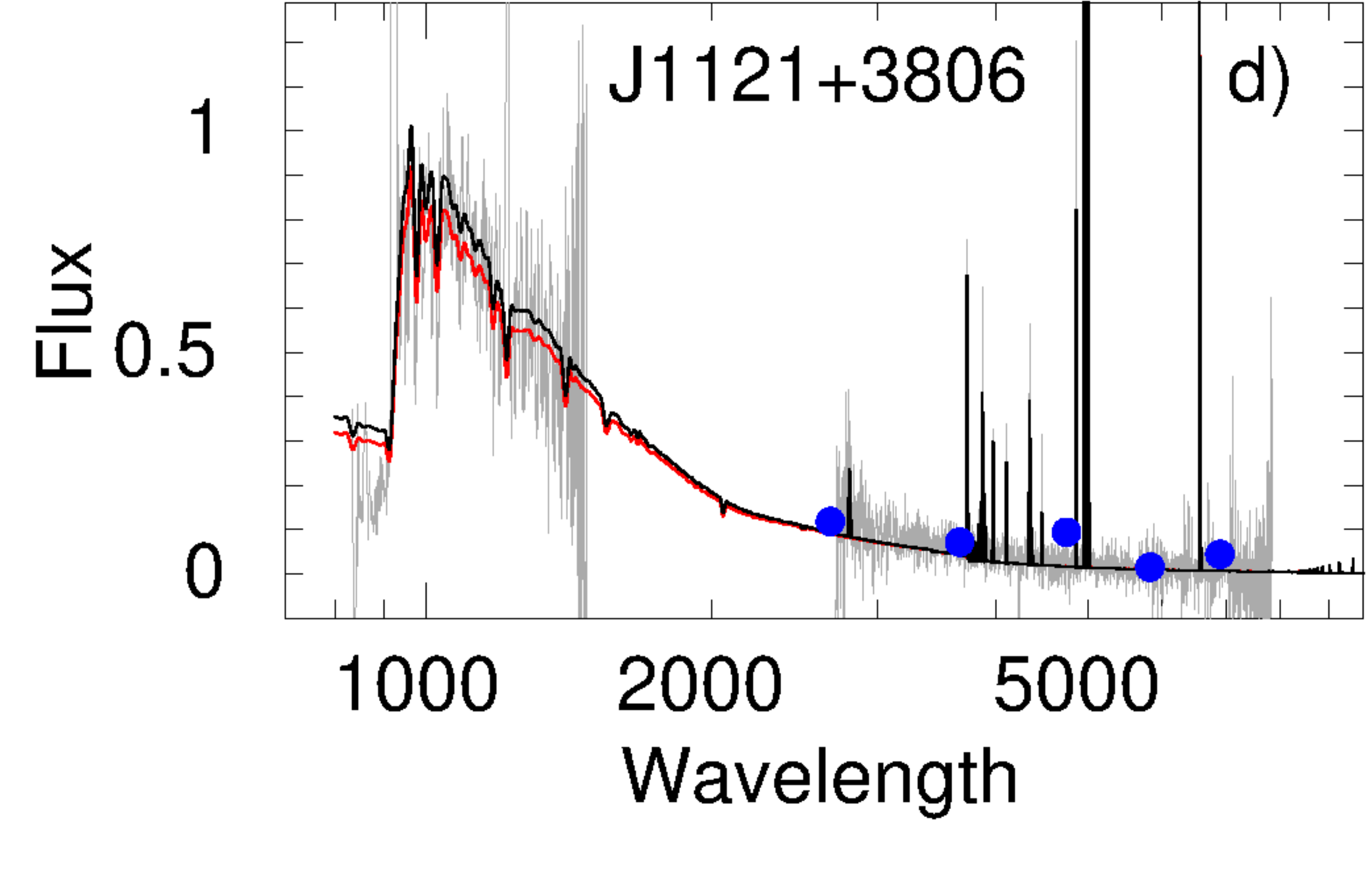}
\includegraphics[angle=-90,width=0.325\linewidth]{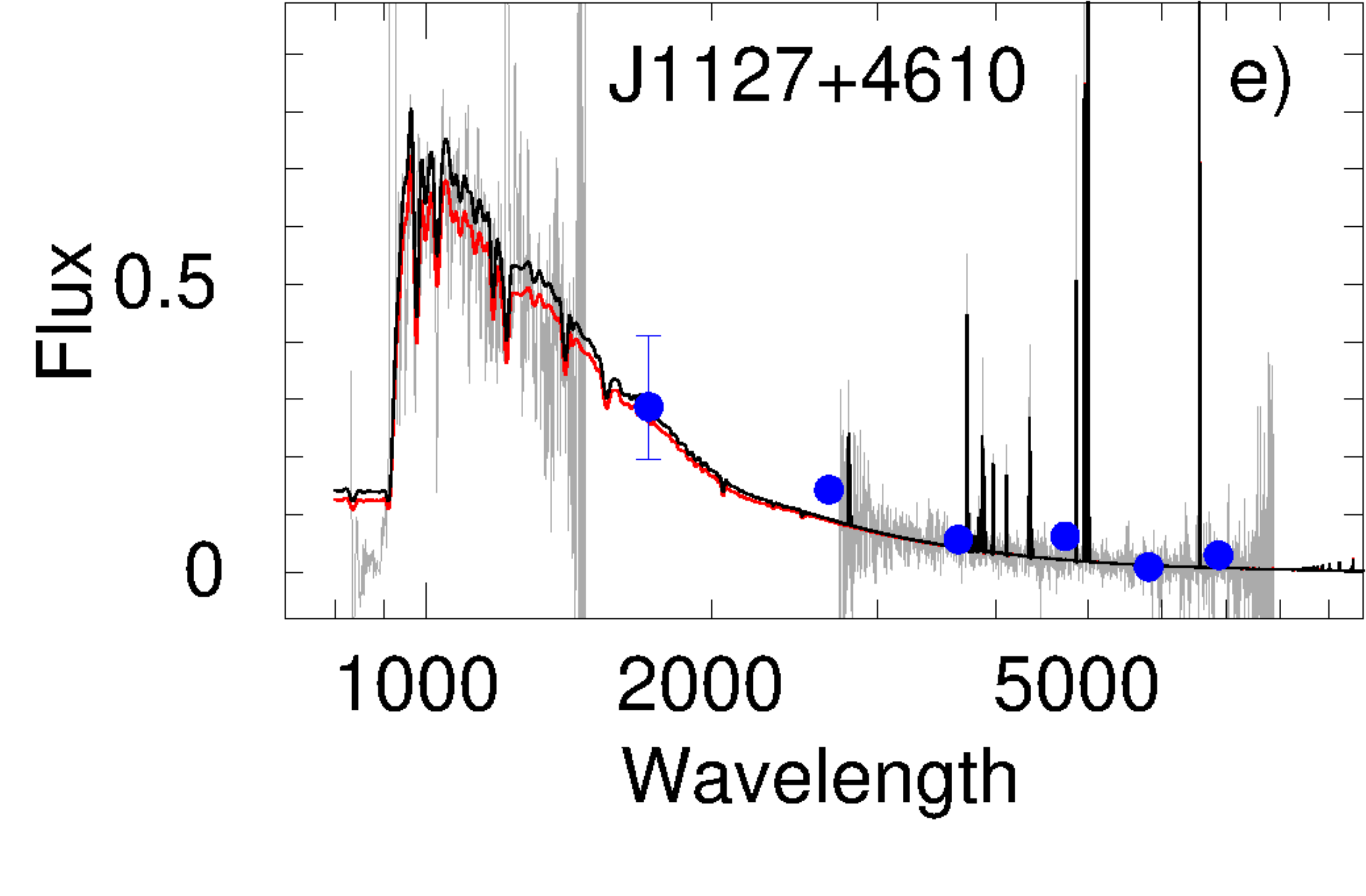}
\includegraphics[angle=-90,width=0.325\linewidth]{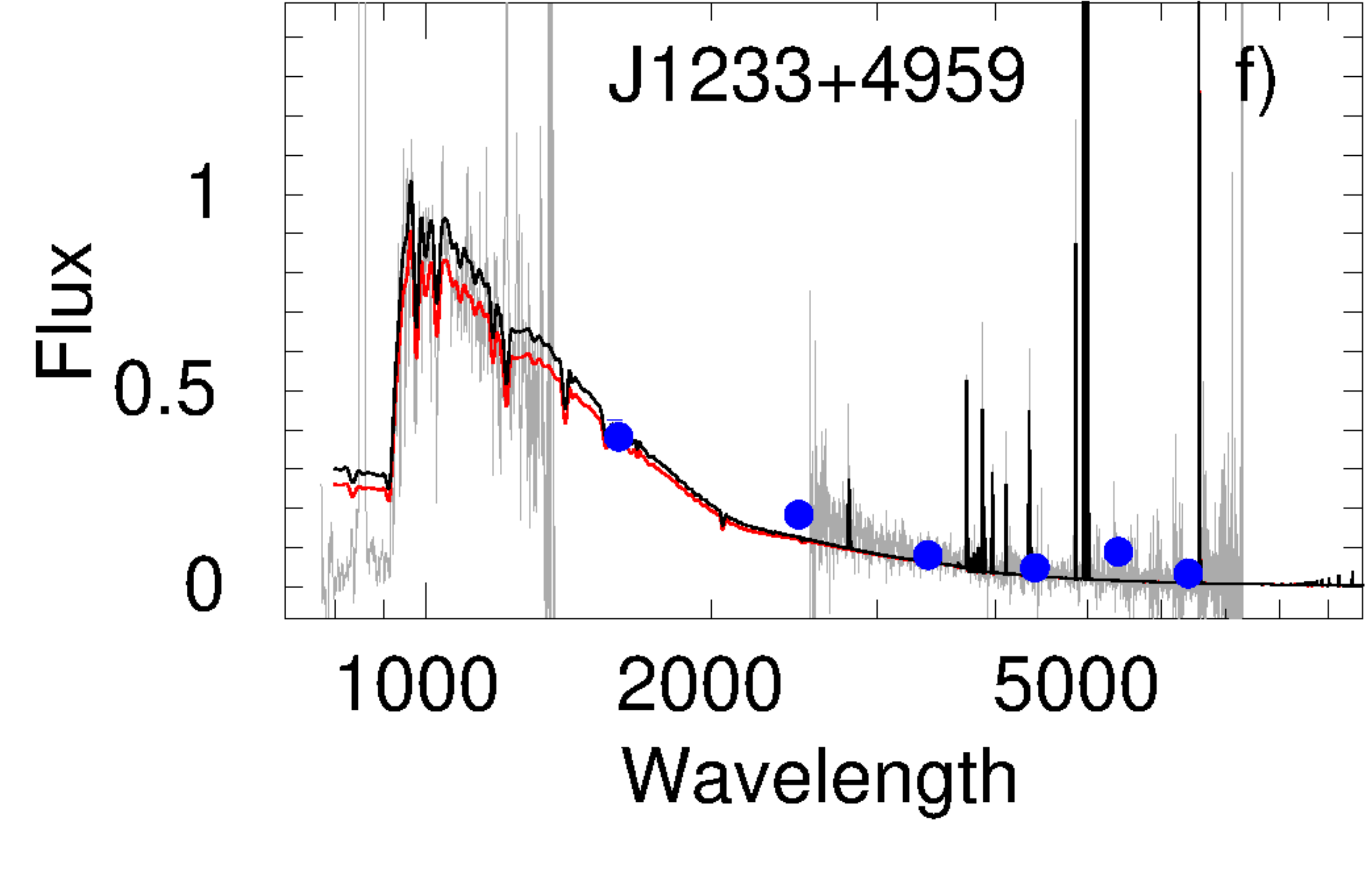}
}
 \hbox{
\includegraphics[angle=-90,width=0.325\linewidth]{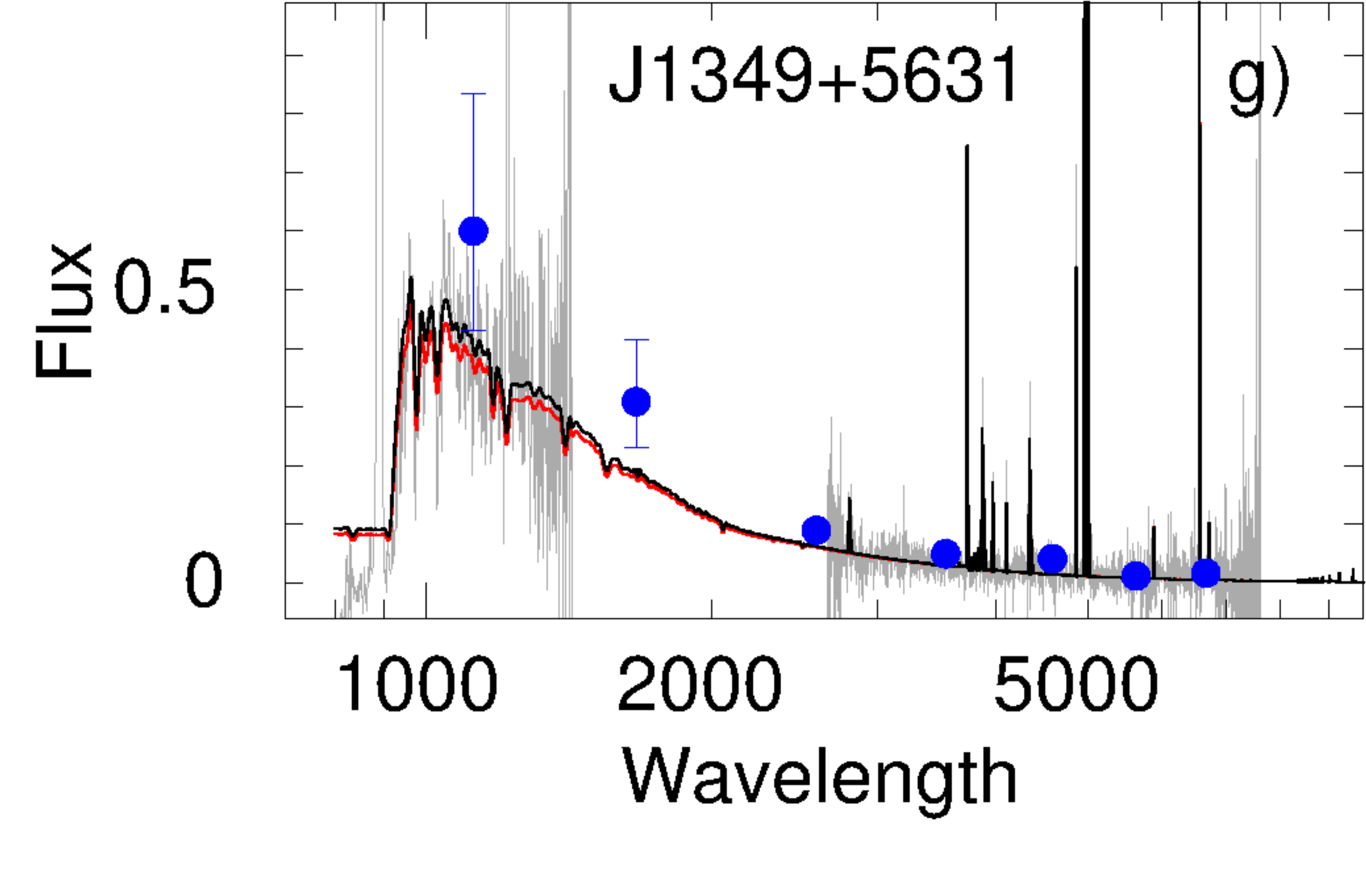}
\includegraphics[angle=-90,width=0.325\linewidth]{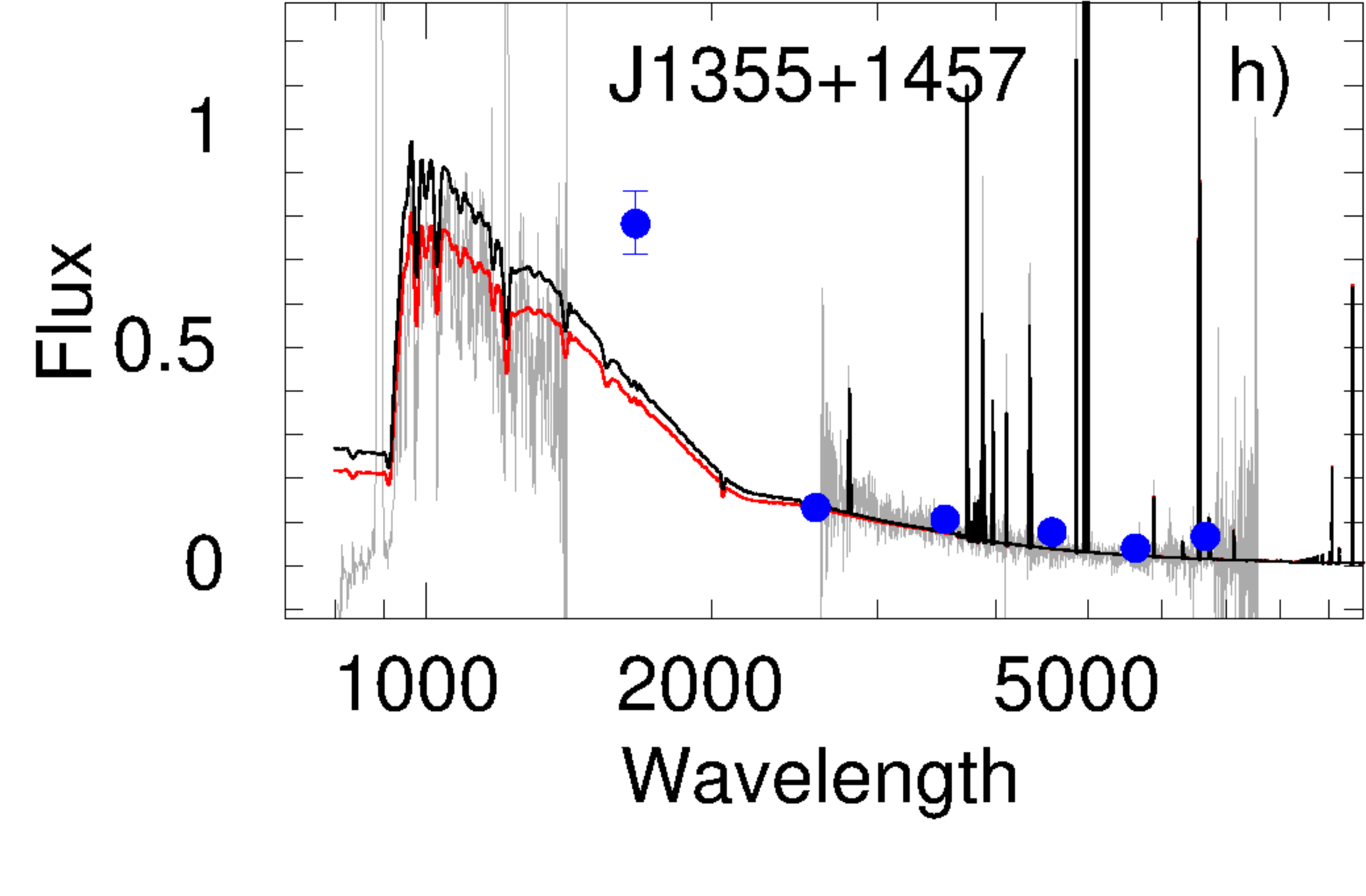}
\includegraphics[angle=-90,width=0.325\linewidth]{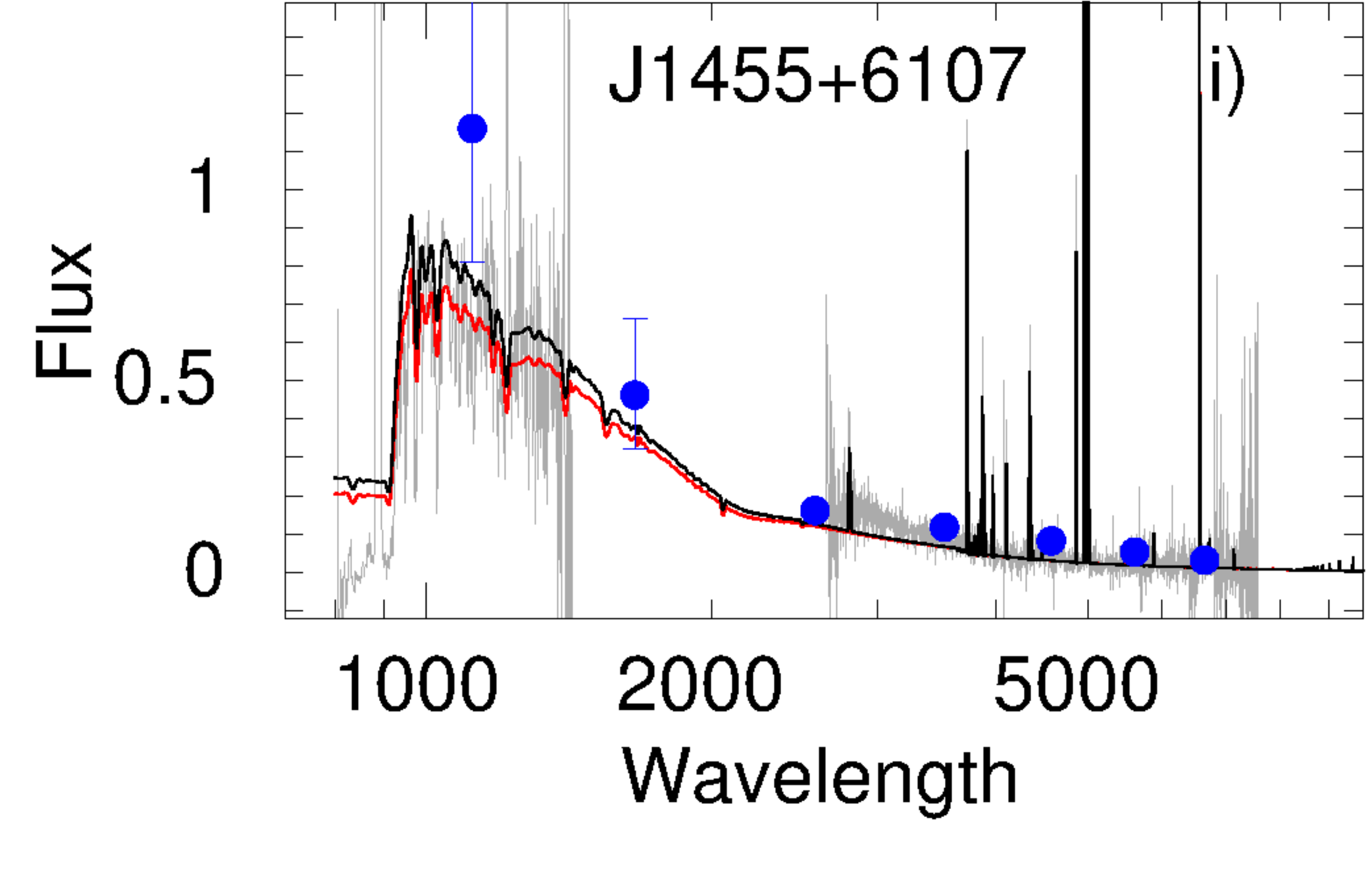}
}
\caption{A comparison of the COS G140L and SDSS spectra (grey lines), and
photometric data (blue filled circles) with the modelled SEDs.
{\sl GALEX} FUV and NUV fluxes with 1$\sigma$ error bars and SDSS fluxes 
in $u,g,r,i,z$ bands are shown by blue-filled circles.
Modelled intrinsic SEDs, which are reddened by the Milky Way extinction 
with $R(V)_{\rm MW}$ = 3.1 and
internal extinction with $R(V)_{\rm int}$ = 3.1 and 2.7, are shown by black and 
red solid lines, respectively. Fluxes are in 
10$^{-16}$ erg s$^{-1}$ cm$^{-2}$\AA$^{-1}$, wavelengths are in \AA. \label{fig4}}
\end{figure*}

  \begin{table*}
  \caption{Parameters for the Ly$\alpha$ emission line \label{tab7}}
  \begin{tabular}{rcrcrrcc} \hline
Name&$A$(Ly$\alpha$)$_{\rm MW}$$^{\rm a}$&\multicolumn{1}{c}{$I$$^{\rm b}$}&log $L$$^{\rm c}$&\multicolumn{1}{c}{EW$^{\rm d}$}&\multicolumn{1}{c}{$V_{\rm sep}$$^{\rm e}$}&\multicolumn{1}{c}{blue/red$^{\rm f}$}&\multicolumn{1}{c}{$f_{\rm esc}$(Ly$\alpha$)$^{\rm g}$} \\ 
\hline
J0232$-$0426&0.106& 43.5$\pm$6.4&42.53& 90.6$\pm$1.8& 446.0$\pm$51.8&\,~9.8&42.5$\pm$5.3 \\
J0919$+$4906&0.110& 81.7$\pm$9.5&42.68&219.5$\pm$3.7& 369.2$\pm$44.9&11.6&68.7$\pm$8.9 \\
J1046$+$5827&0.059& 65.3$\pm$8.2&42.57& 65.3$\pm$2.2& 512.6$\pm$58.7&19.0&31.8$\pm$4.3 \\
J1121$+$3806&0.154& 79.5$\pm$8.6&42.43&142.3$\pm$2.9& 229.3$\pm$79.5&23.1&43.2$\pm$5.2 \\
J1127$+$4610&0.123& 44.4$\pm$5.6&42.18& 94.3$\pm$2.2& 259.4$\pm$51.8&28.8&39.7$\pm$8.5 \\
J1233$+$4959&0.124& 83.6$\pm$9.7&42.74&152.0$\pm$6.7& 271.9$\pm$27.6&16.9&41.2$\pm$3.9 \\
J1349$+$5631&0.070& 47.0$\pm$5.6&42.33& 95.1$\pm$3.2& 386.5$\pm$56.3&\,~9.1&40.3$\pm$4.4 \\
J1355$+$1457&0.152& 68.9$\pm$7.3&42.50&151.6$\pm$2.9& 430.4$\pm$69.1&63.3&23.1$\pm$2.8 \\
J1455$+$6107&0.098& 74.0$\pm$8.4&42.54&144.0$\pm$2.7& 388.9$\pm$41.5&41.5&36.5$\pm$4.5 \\
\hline
  \end{tabular}

\hbox{$^{\rm a}$Milky Way extinction at the observed wavelength of the Ly$\alpha$
emission line in mags}

\hbox{\, adopting \citet{C89} reddening law with $R(V)$=3.1.}

\hbox{$^{\rm b}$Flux in 10$^{-16}$ erg s$^{-1}$ cm$^{-2}$ measured in
the COS spectrum and corrected for the Milky Way extinction.}

\hbox{$^{\rm c}$$L$ is Ly$\alpha$ luminosity in erg s$^{-1}$ corrected for the
Milky Way extinction.}

\hbox{$^{\rm d}$Rest-frame equivalent width in \AA.}

\hbox{$^{\rm e}$Ly$\alpha$ peak separation in km s$^{-1}$.}

\hbox{$^{\rm f}$Flux ratio of blue-to-red peaks in per cent.}

\hbox{$^{\rm g}$Ly$\alpha$ escape fraction in per cent.}

  \end{table*}

\begin{figure*}
\hbox{
\includegraphics[angle=-90,width=0.325\linewidth]{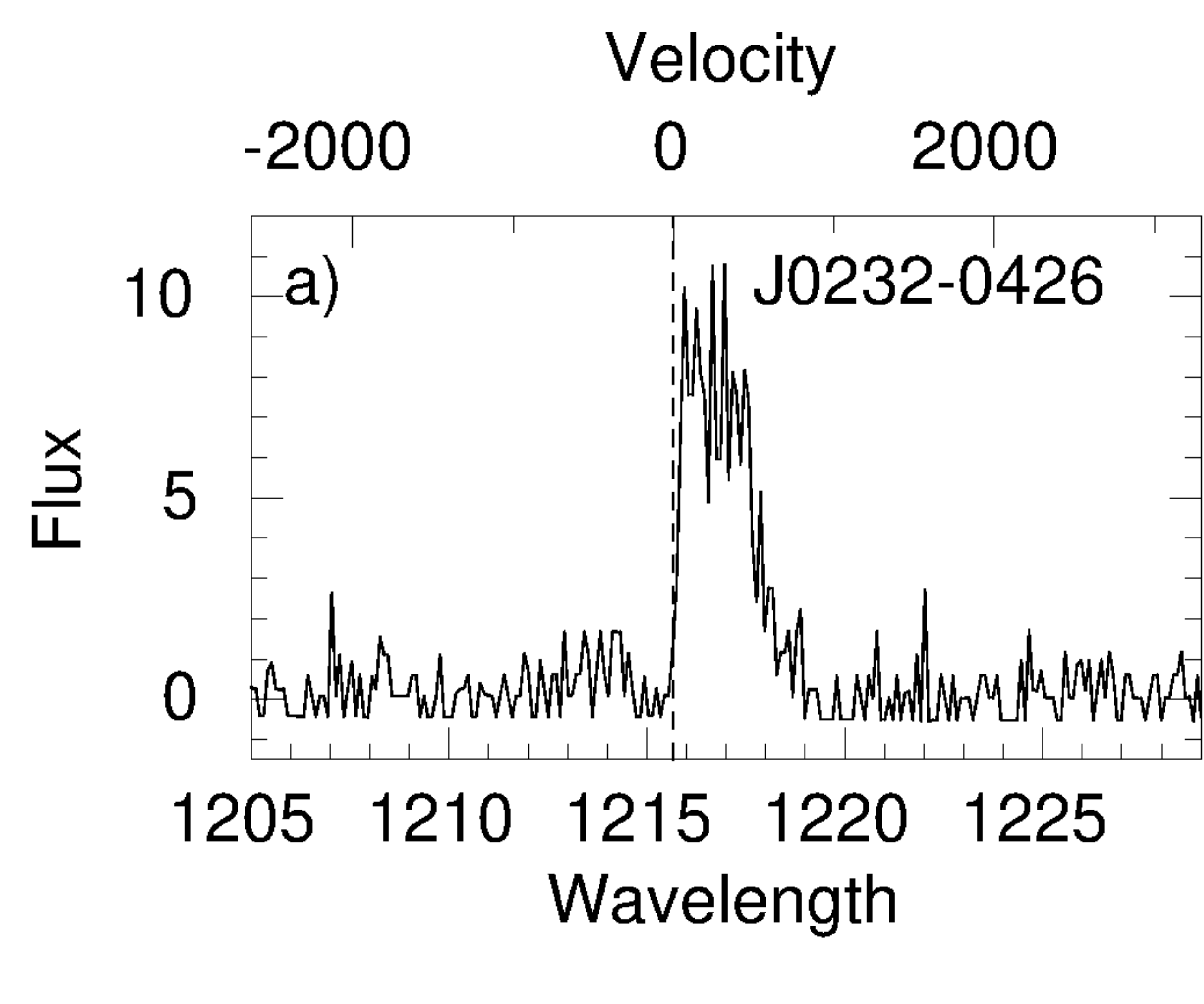}
\includegraphics[angle=-90,width=0.325\linewidth]{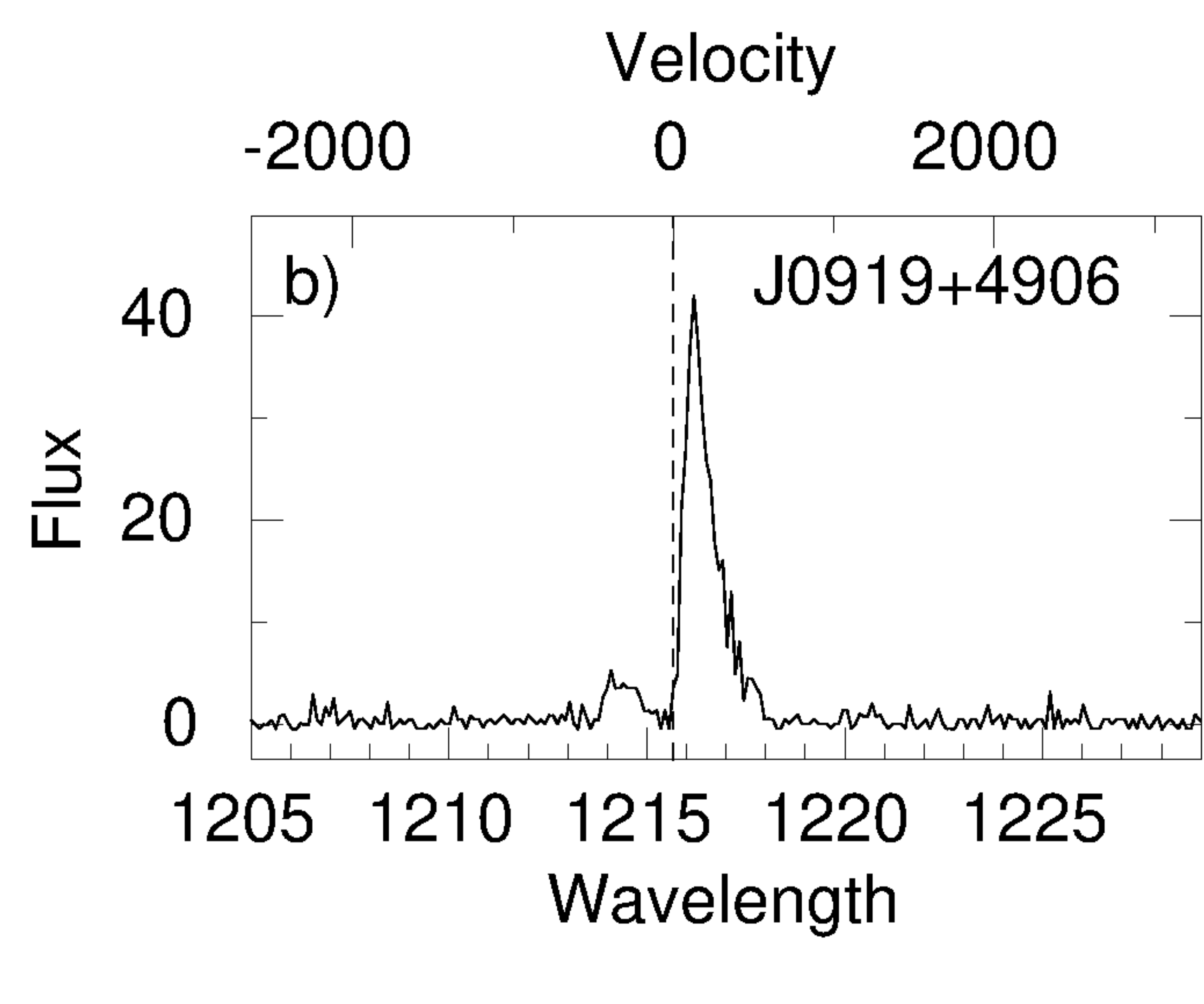}
\includegraphics[angle=-90,width=0.325\linewidth]{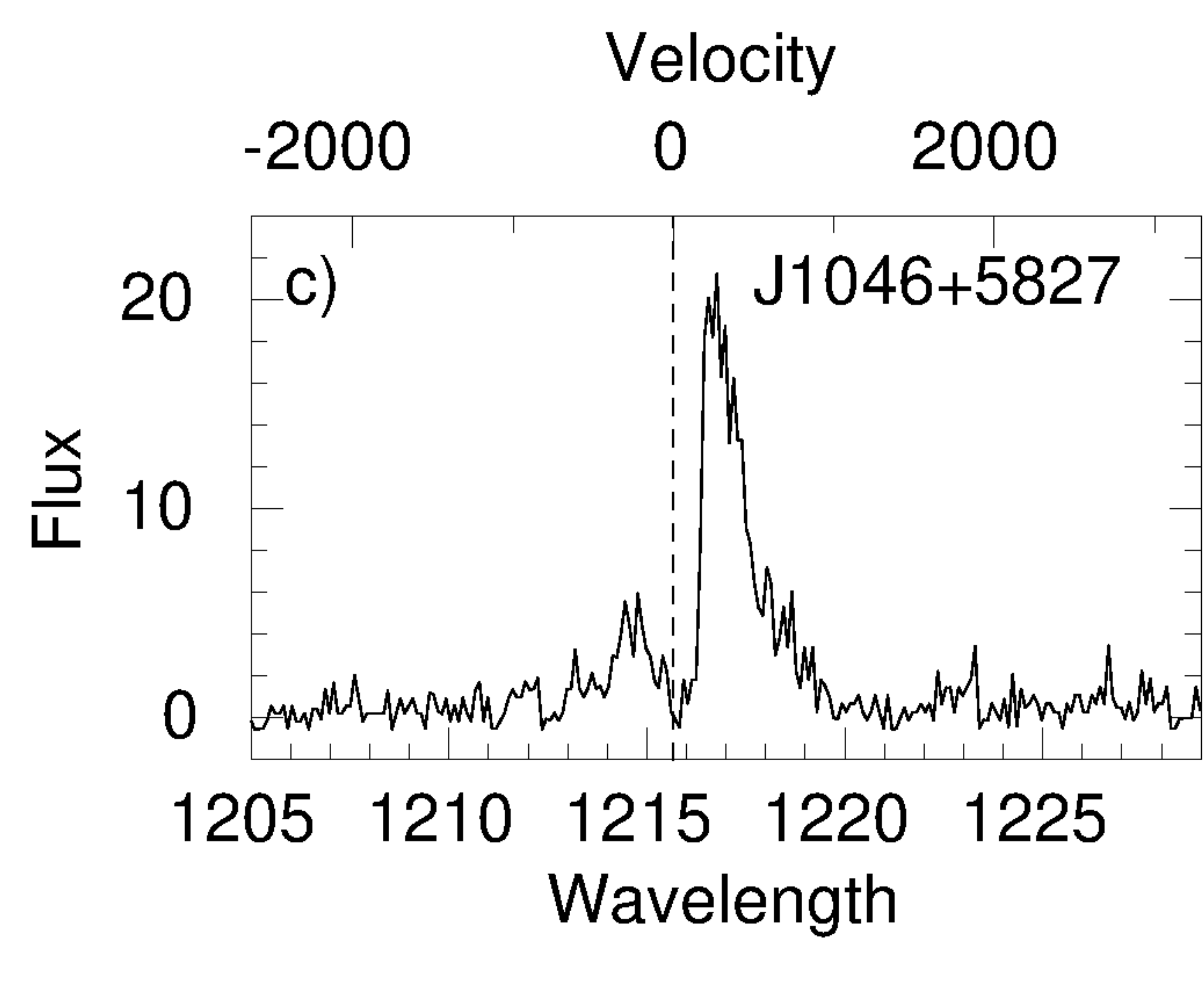}
}
\vspace{-0.3cm}
\hbox{
\includegraphics[angle=-90,width=0.325\linewidth]{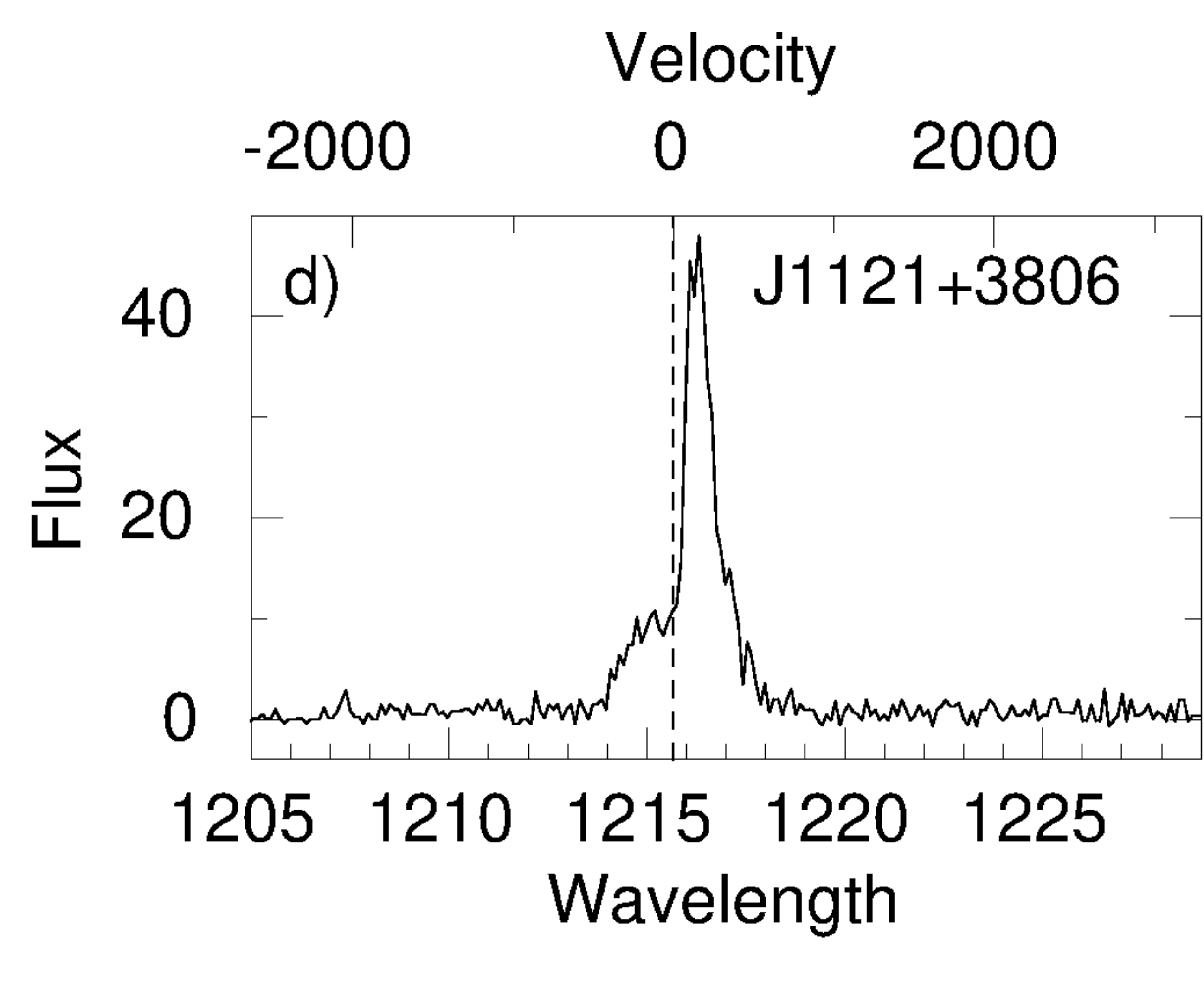}
\includegraphics[angle=-90,width=0.325\linewidth]{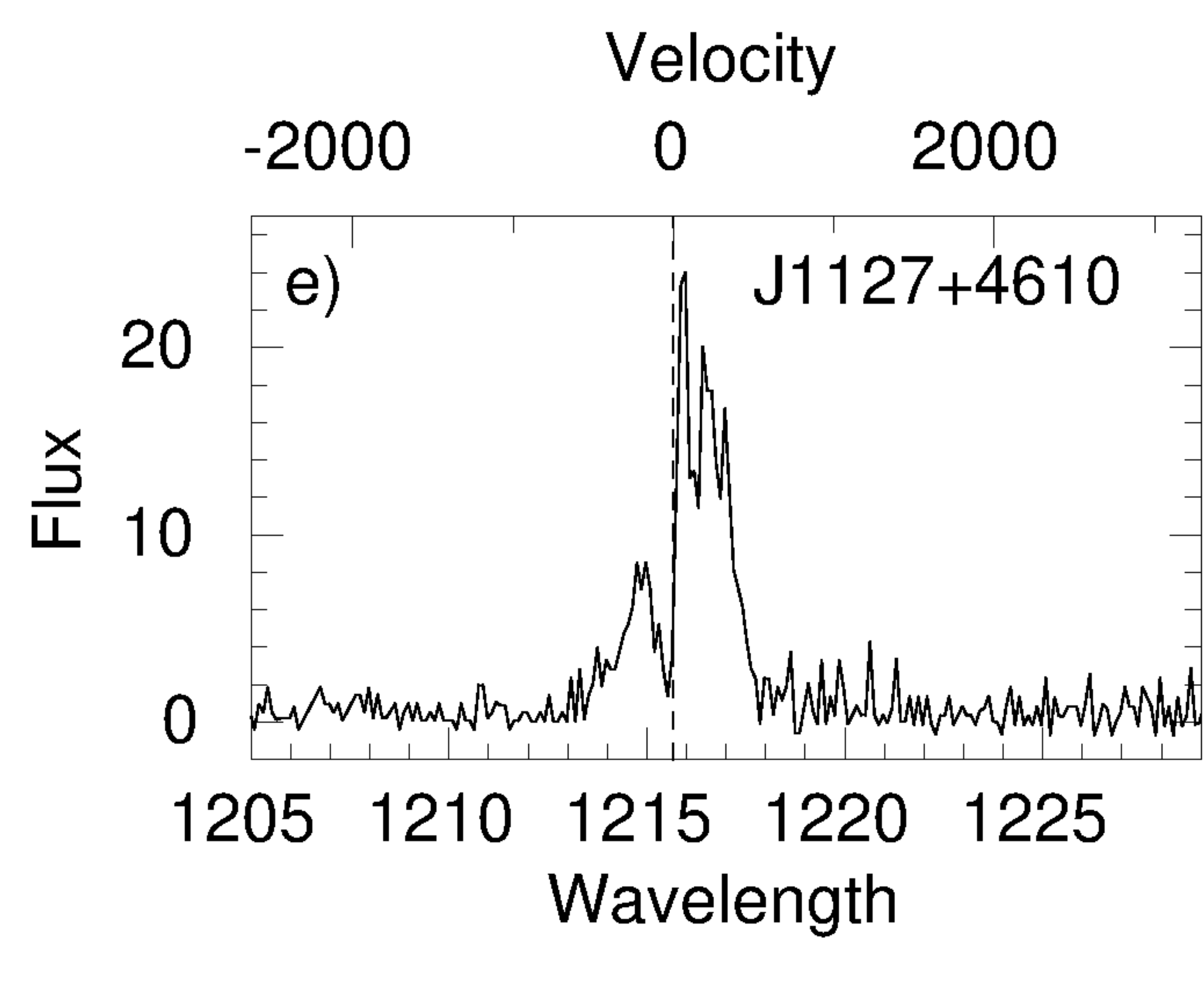}
\includegraphics[angle=-90,width=0.325\linewidth]{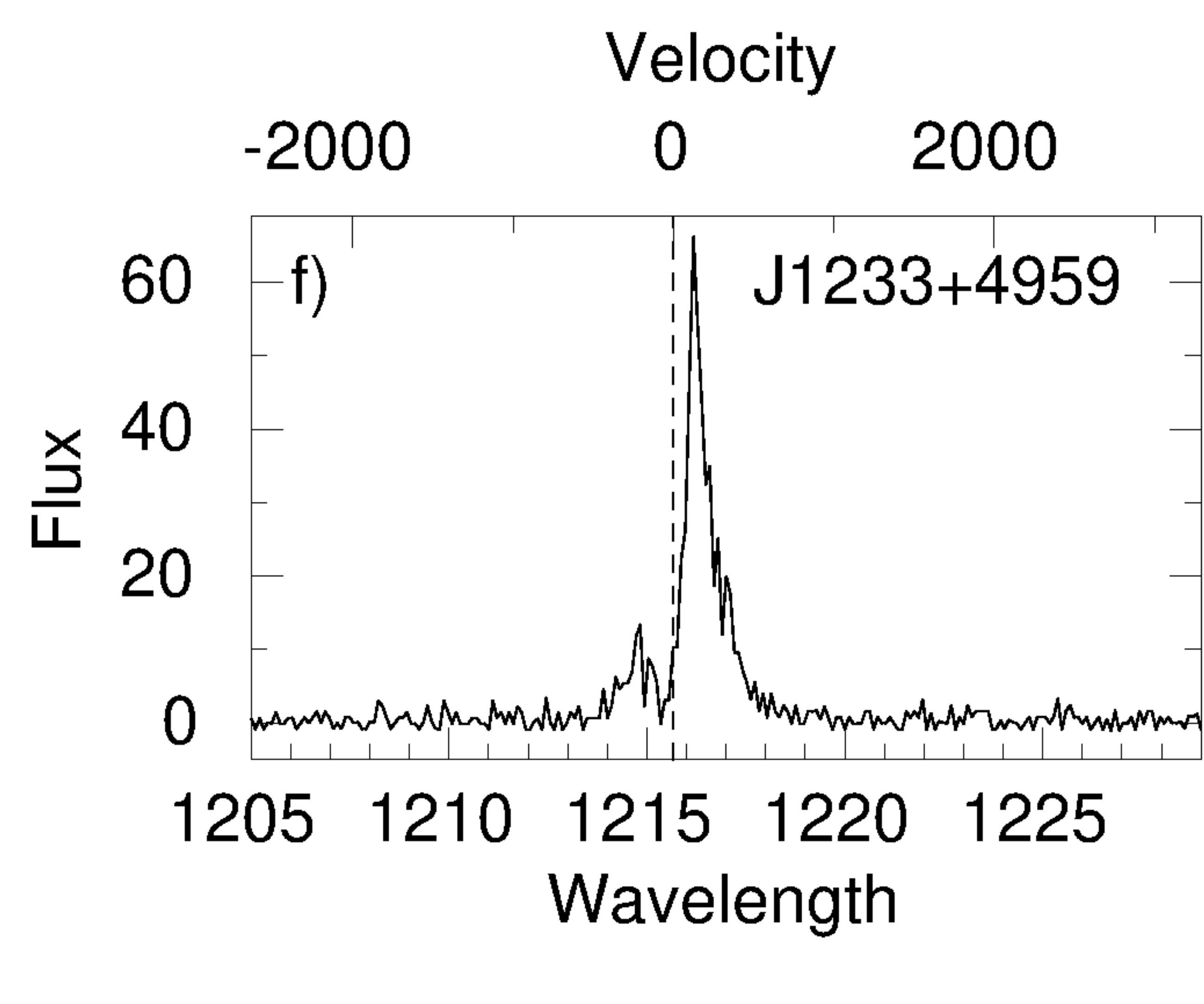}
}
\vspace{-0.3cm}
\hbox{
\includegraphics[angle=-90,width=0.325\linewidth]{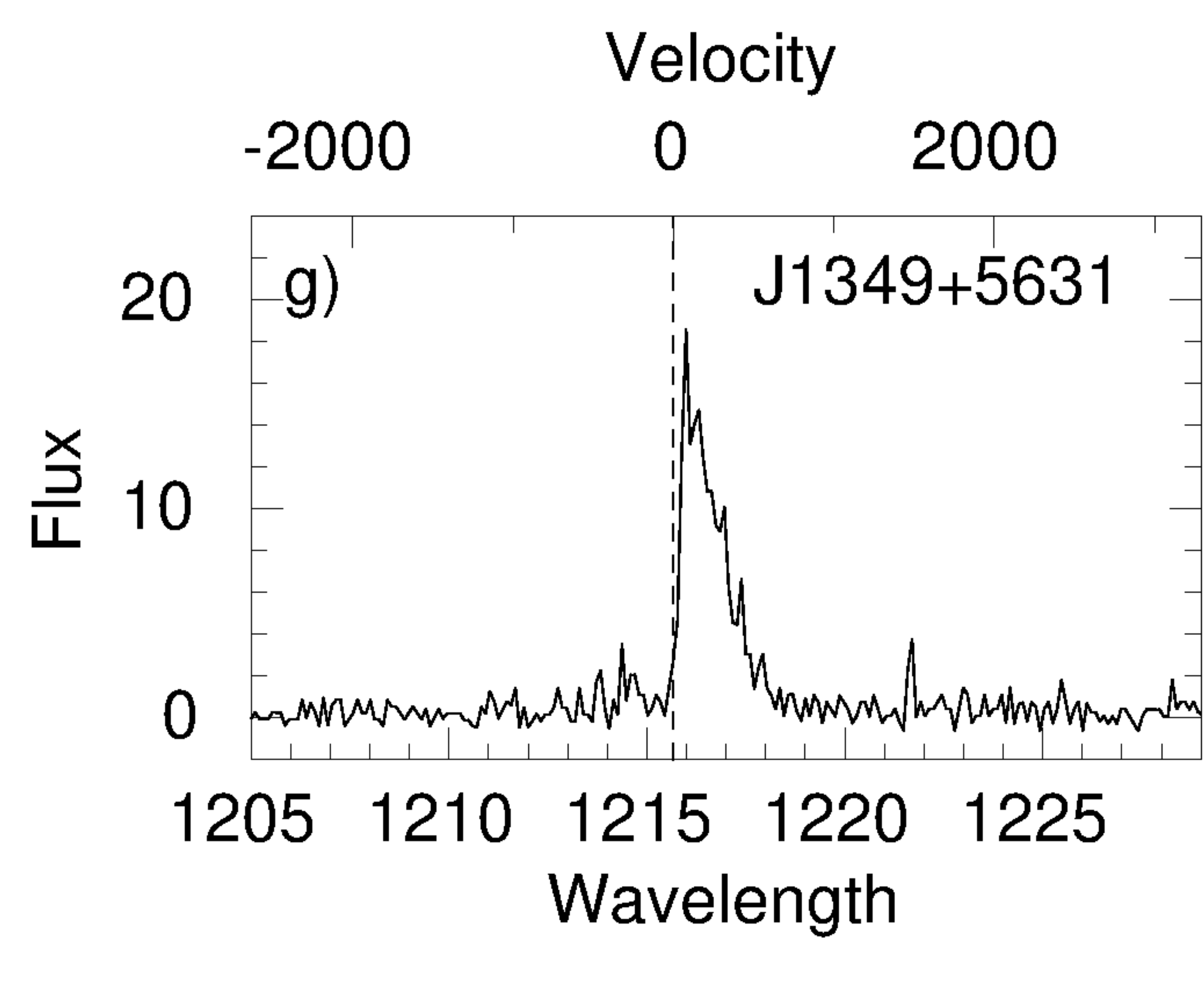}
\includegraphics[angle=-90,width=0.325\linewidth]{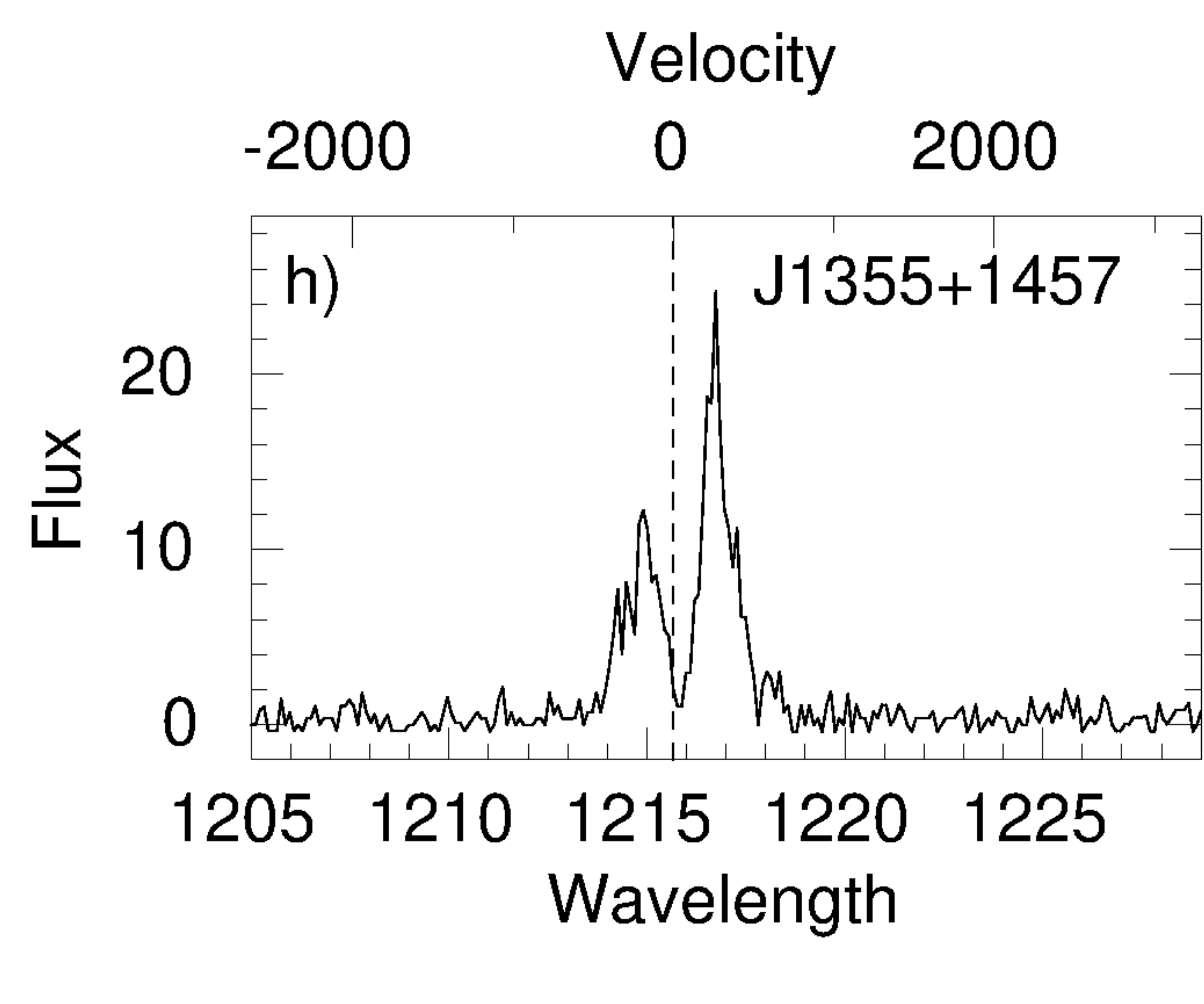}
\includegraphics[angle=-90,width=0.325\linewidth]{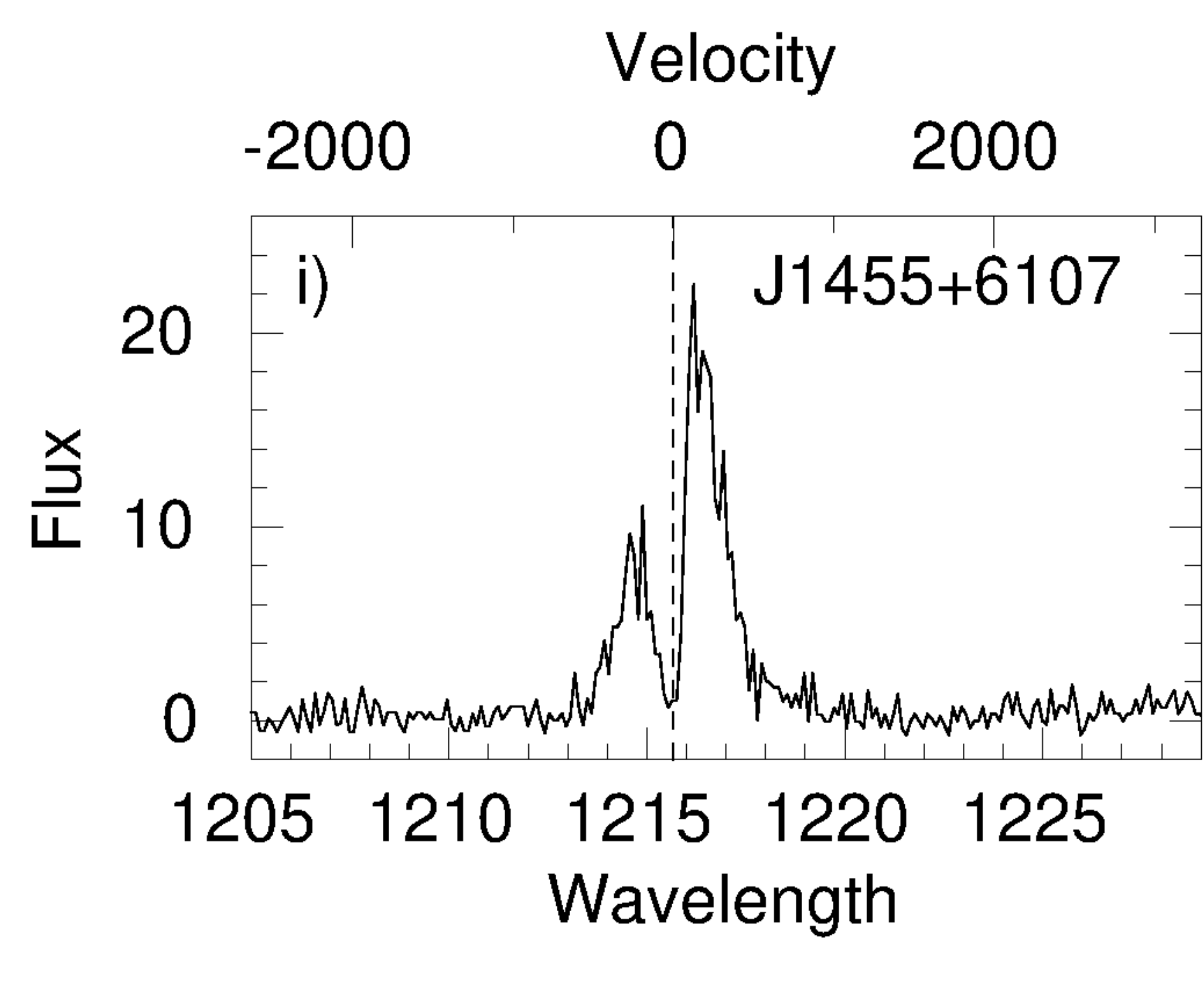}
}
\caption{Ly$\alpha$ profiles. Vertical dashed lines indicate the restframe 
wavelength of 1215.67\AA\ for Ly$\alpha$. 
Fluxes are in 10$^{-16}$ erg s$^{-1}$ cm$^{-2}$\AA$^{-1}$, restframe 
wavelengths are in \AA\ and velocities are in km s$^{-1}$. \label{fig5}}
\end{figure*}

\subsection{Absolute magnitudes, H$\beta$ luminosities and stellar masses}
\label{sec:absmag}

For absolute FUV magnitudes, we do not use the observed {\sl GALEX} FUV apparent
magnitudes for two reasons: 1) only four galaxies are detected by {\sl GALEX} in
the FUV range; 2) for the remaining galaxies, FUV magnitudes are given for
emission at somewhat different wavelengths. Therefore, for homogeneity, we
derive absolute magnitudes from the fluxes of the extinction-corrected spectral
distribution (SED) at the rest-frame wavelength $\lambda$ = 1500~\AA. We 
designate these absolute magnitudes as $M_{\rm FUV}$. They are, on average, 
$\sim$~0.7 mag fainter than $M_{\rm FUV}$s for other $z$~$\sim$~0.3~--~0.4
LyC leakers.

The H$\beta$ luminosity $L$(H$\beta$) and the corresponding
star-formation rates SFR were obtained from the extinction-corrected H$\beta$ 
fluxes, using the relation of \citet{K98} for the SFR. 
SFRs are increased by a factor 1/[1 $-$ $f_{\rm esc}$(LyC)] to take into account 
the escaping ionising radiation which is discussed later. The SFRs corrected for
escaping LyC radiation are shown in Table~\ref{tab5}. 
They are somewhat below the range of 14 -- 80 M$_\odot$ yr$^{-1}$
for the other LyC leakers studied by \citet{I16a,I16b,I18a,I18b}, presumably because of
their smaller stellar masses. On the other hand, their specific star 
formation rates, sSFR = SFR/$M_\star$, are $>$ 150 Gyr$^{-1}$, with the highest 
value of $\sim 630$~Gyr$^{-1}$ for J1121+3806, which has also the lowest stellar
mass (Table~\ref{tab5}). These sSFR are among the largest sSFR for dwarf SFGs at
any redshift and they are significantly higher than the sSFRs observed in 
other LyC leakers \citep{I16a,I16b,I18a,I18b}. 

We use the SDSS spectra of our LyC leakers to fit the SED and derive their 
stellar masses. The fitting method, using a two-component model, is described 
in \citet{I18a,I18b}. Spectral energy distributions of instantaneous 
bursts in the range between 0 and 10 Gyr with evolutionary tracks of 
non-rotating stars by 
\citet{G00} and a combination of stellar atmosphere models \citep*{L97,S92} 
are used to produce the integrated SED for 
each galaxy. The adoption of other sets of evolutionary tracks of rotating and
non-rotating stars and stellar atmosphere models would, in particular, change
the level of Lyman continuum by $\sim$~10 per cent at most \citep{I16a}.
The star formation history is approximated by a young 
burst with a randomly varying age $t_b$ in the range $<$ 10 Myr,  and  
a continuous star formation for older 
ages between times $t_1$ and $t_2$, randomly varying  
in the range 10 Myr - 10 Gyr, and adopting a constant SFR. 
The contribution of the two components is determined by randomly 
varying the ratio of their stellar masses, $b$ = $M_{\rm o}/M_{\rm y}$, in the 
range 0.1 -- 1000, where $M_{\rm o}$ and $M_{\rm y}$ are the masses of the old and 
young stellar populations.

The nebular continuum emission, including free-free and free-bound hydrogen
and helium emission, and two-photon emission, is also taken into account 
using the observed H$\beta$ flux, the ISM temperature and density.
The fraction of nebular emission in the observed continuum near H$\beta$ 
is determined by the ratio of the observed H$\beta$ equivalent width 
EW(H$\beta$)$_{\rm obs}$, reduced to the rest frame, to the equivalent
EW(H$\beta$)$_{\rm rec}$ for pure nebular emission. EW(H$\beta$)$_{\rm rec}$ varies
from $\sim$ 900\AA\ to $\sim$ 1100\AA, for electron temperatures in the range
$T_{\rm e}$ = 10000 -- 20000K. For example, the fraction of nebular continuum
near H$\beta$ in J0919$+$4906 with EW(H$\beta$)$_{\rm obs}$ = 435\AA\ 
(Table~\ref{tab3}) is $\sim$ 40 per cent and the remaining 60 per cent of 
emission in the continuum is stellar.
We note that non-negligible nebular emission, including both
the continuum and emission lines, is produced only by the burst.

The Salpeter initial mass
function (IMF) is adopted, with a slope of --2.35, upper and lower mass limits
$M_{\rm up}$ and $M_{\rm low}$ of 100~M$_\odot$ and 0.1~M$_\odot$, respectively.
A $\chi ^2$ minimisation technique was used 1) to fit the continuum 
in such parts of the wavelength range 3600 -- 6500\AA, where the SDSS spectrum 
is least noisy and free of nebular emission lines, and 2) to reproduce 
the observed H$\beta$ and H$\alpha$ equivalent widths. 

To illustrate the quality of our SED fitting, Fig. \ref{fig1} shows
the modelled stellar, nebular and total SEDs superposed upon the rest-frame
extinction-corrected SDSS spectra. For all galaxies we find good 
agreement. The masses of young ($M_{\rm y}$) and old  ($M_{\rm o}$) stellar 
populations, and total stellar masses ($M_\star$ = $M_{\rm y}$ + $M_{\rm o}$)
of our LyC leakers derived from SED fitting are 
presented in Table~\ref{tab5}. They are derived in exactly the same way as the 
stellar masses for the other LyC leakers studied by our team 
\citep{I16a,I16b,I18a,I18b}, permitting a direct comparison. They are 
considerably lower, being all less than 10$^{8}$ M$_\odot$. 

We find that the 
masses of the old stellar population are lower or similar to those of the young stellar
population. However, the determination of the mass $M_{\rm o}$ of the old stellar 
population is very uncertain because of the strong dominance of the young stellar
population in the SDSS optical spectrum. For example, the luminosity of a 
young stellar population with an age of 3 Myr at the wavelength 5000\AA\ 
is $\sim$ 250 times higher than the luminosity of an old population with an age of 
1 Gyr. Therefore, to have a non-negligible contribution of $\sim$ 5 per cent
to emission at 5000\AA, the mass of the old stellar population should be 
$\sim$ 10 times that of the young stellar population, corresponding to 
$b$ $\ga$ 10. The SED at lower $b$ is not sensitive to the presence of the old
stellar population. On the other hand, the stellar mass of the young stellar
population $M_{\rm y}$ is reliable because it is primarily determined by the
H$\beta$ luminosity. We also note that the SFR derived from the H$\beta$
emission line is similar to the value $M_{\rm y}$/$t_b$ obtained from our 
modelled SEDs (see Table~\ref{tab5}).
However, the derived SFRs and sSFRs are just an indication of the galaxy
properties. They are not used for SED fitting and thus have no impact on the 
determination of the LyC escape fraction.

\begin{figure*}
\hbox{
\includegraphics[angle=-90,width=0.48\linewidth]{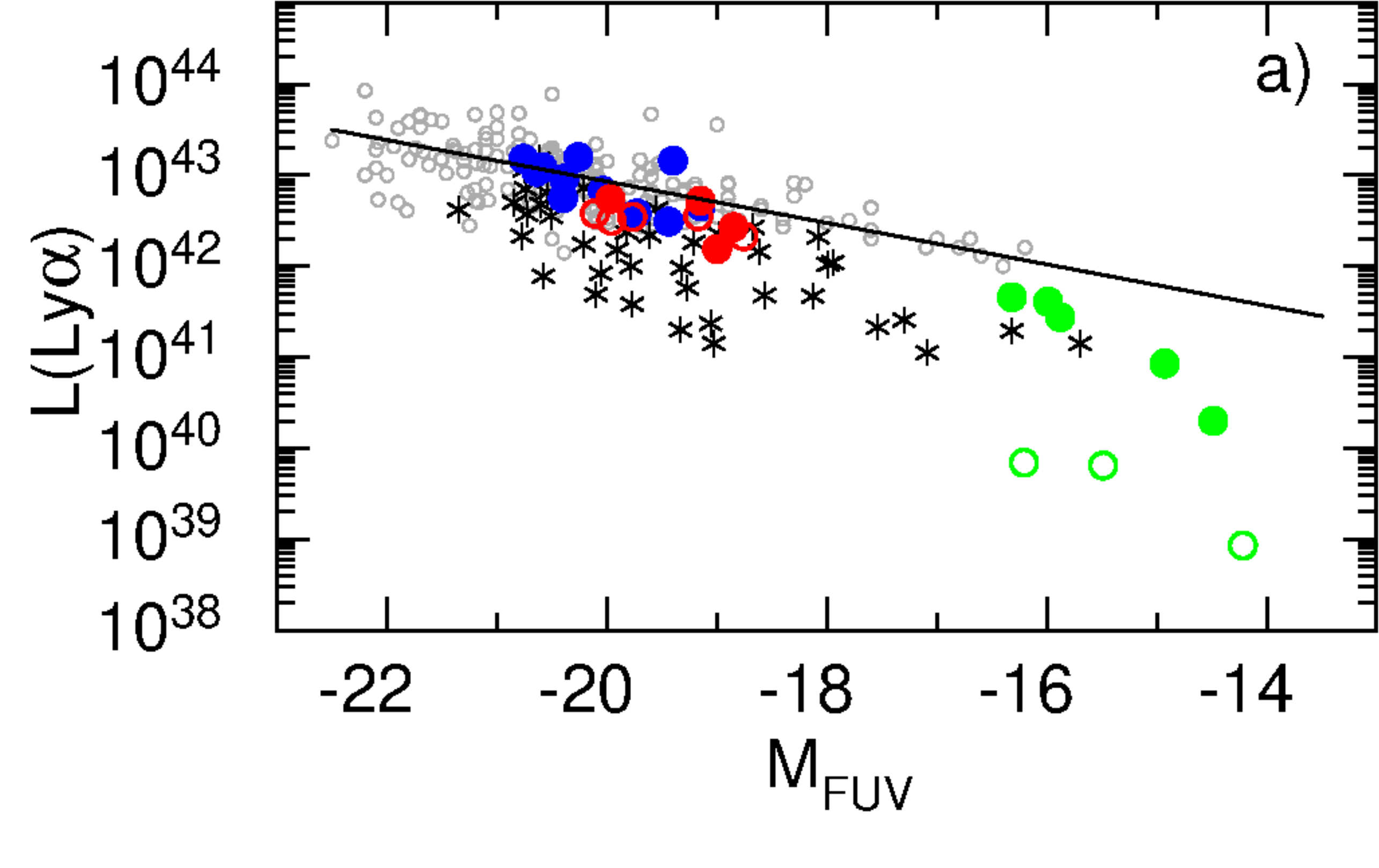}
\includegraphics[angle=-90,width=0.48\linewidth]{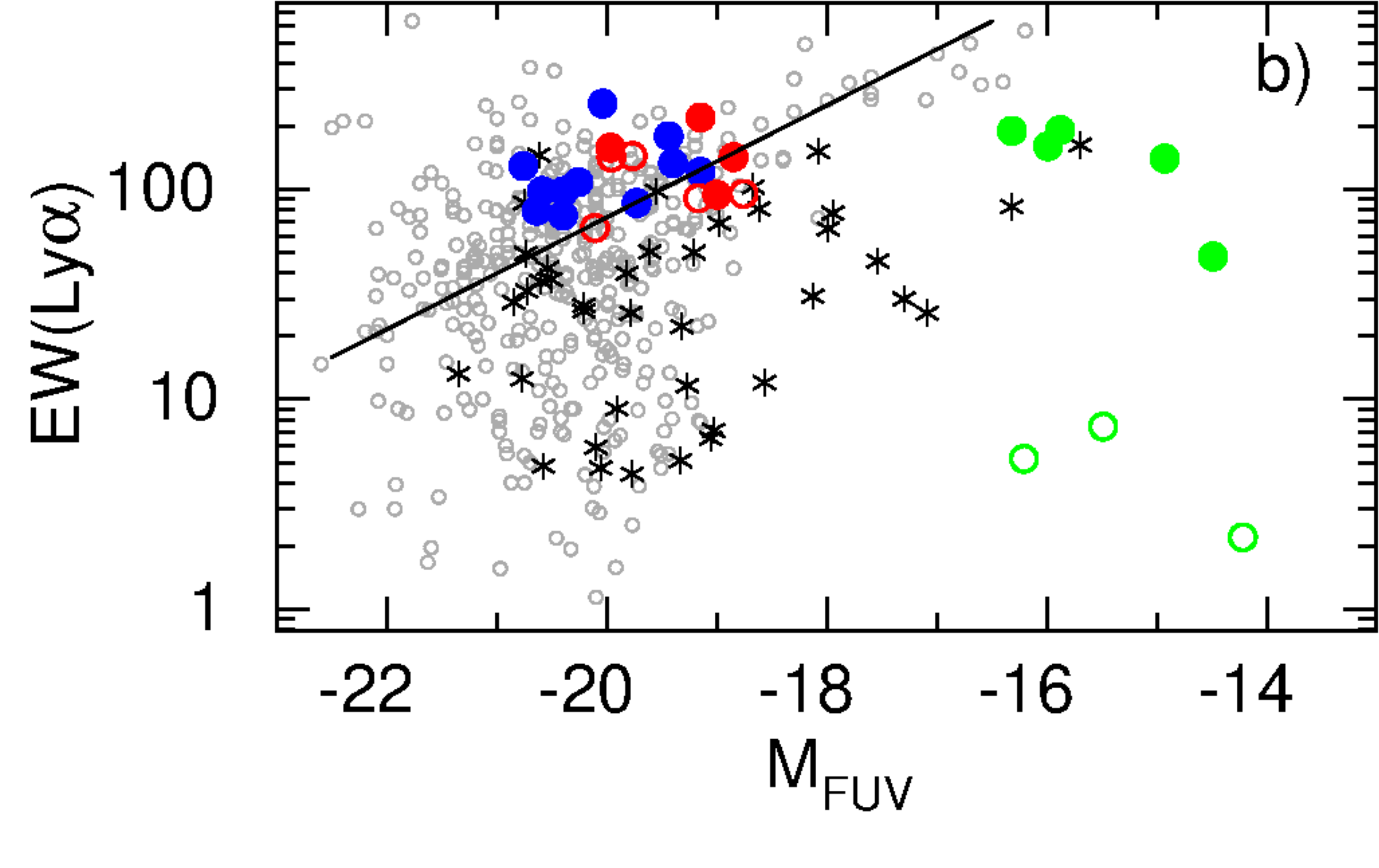}
}
\hbox{
\includegraphics[angle=-90,width=0.48\linewidth]{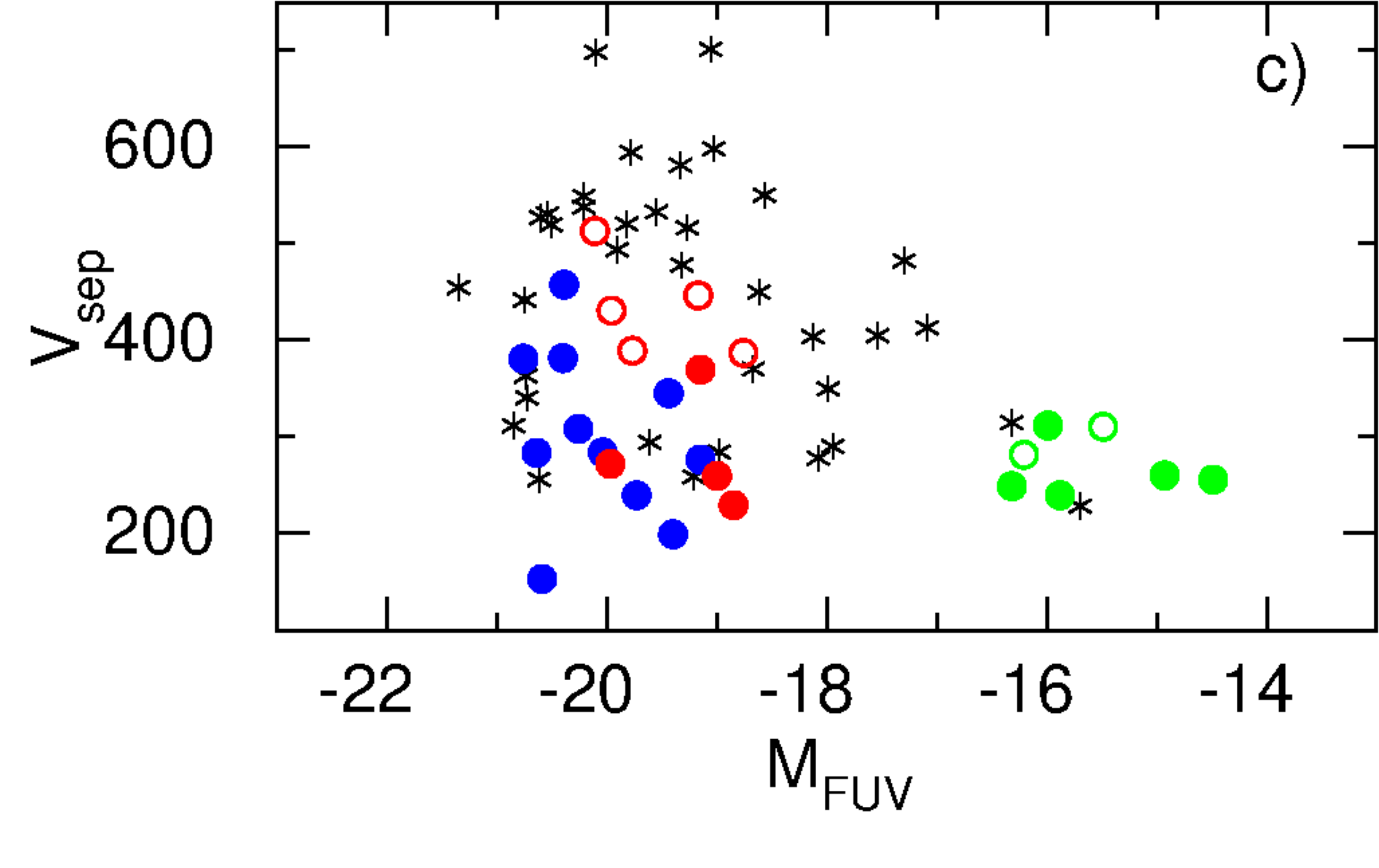}
\includegraphics[angle=-90,width=0.48\linewidth]{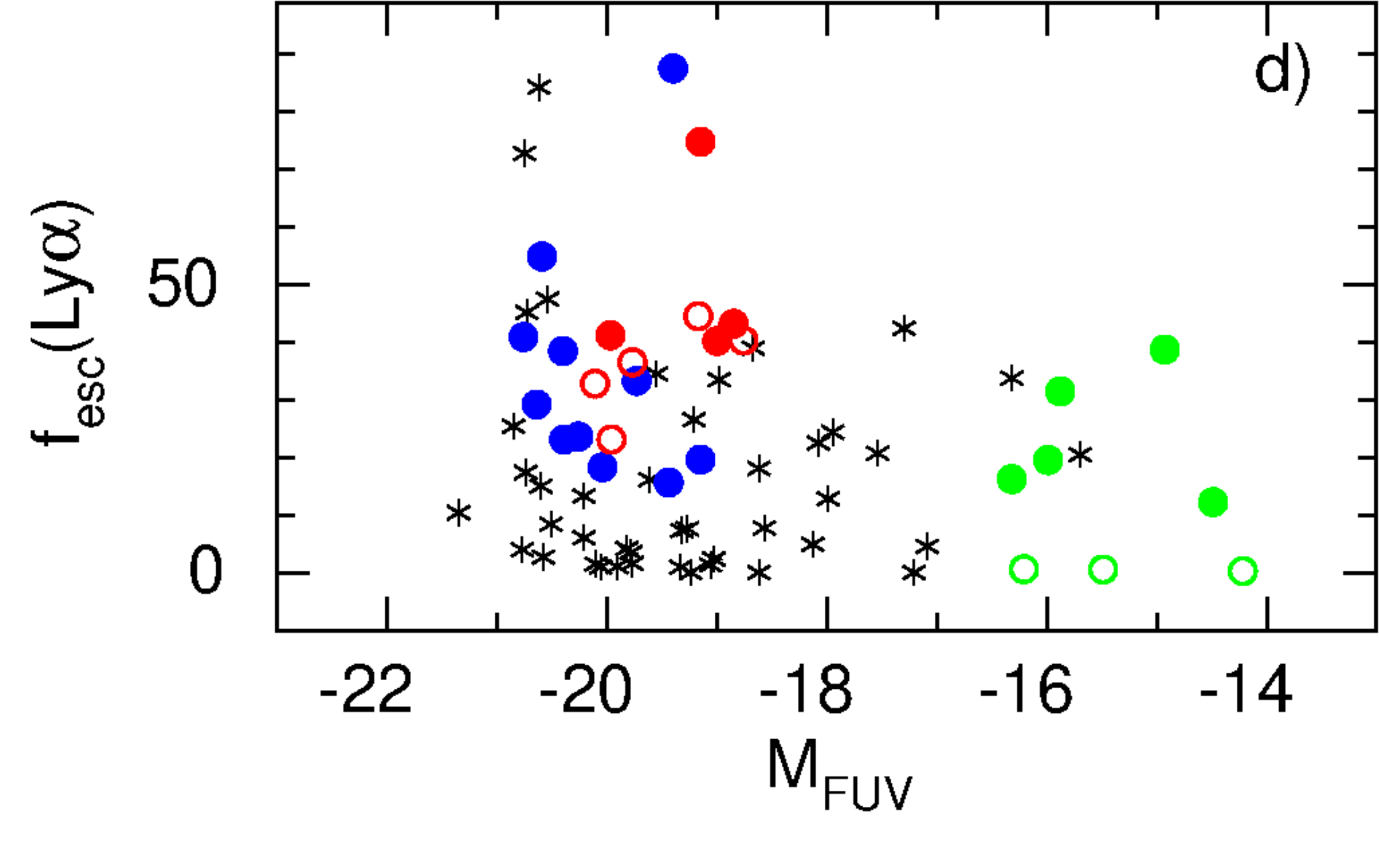}
}
\caption{Dependence of absolute magnitude $M_{\rm FUV}$ at 1500\AA\ in mags on 
{\bf (a)} Ly$\alpha$ luminosity $L$(Ly$\alpha$) in erg s$^{-1}$,
{\bf (b)} equivalent  width EW(Ly$\alpha$) in \AA, {\bf (c)} separation between 
Ly$\alpha$ peaks in km s$^{-1}$, and {\bf (d)} Ly$\alpha$ escape fraction 
$f_{\rm esc}$(Ly$\alpha$) in per cent. Galaxies from this paper with detected
LyC emission and with upper limits of LyC emission are shown by red filled and 
open circles, 
respectively. The confirmed LyC leakers from \citet{I16a,I16b,I18a,I18b} are 
shown by blue filled circles. GPs \citep{H15,JO14,J17,MK19,Y17}  are  
shown  by black asterisks.  High-redshift galaxies 
from \citet{O08}, \citet{H17}, \citet{J18}, \citet{Mat17,Mat18}, \citet{So18} 
in {\bf (a)} and from \citet{O08}, \citet{H17}, \citet{J18}, \citet{Ha18}, 
\citet{Ca18}, \citet{Pe18}, \citet{Mat17,Mat18}, \citet{So18} in {\bf (b)} are 
represented by small grey open circles. Relations for these galaxies are
shown by black solid lines. For comparison, the green symbols
show $z$ $<$ 0.07 compact SFGs with extreme O$_{32}$~$>$~20 \citep{I20}, where
filled symbols are for galaxies with high EW(Ly$\alpha$) without evidence of
the Ly$\alpha$ absorption profiles, whereas open symbols are for galaxies
with low-EW Ly$\alpha$ emission on the top of broad absorption.
\label{fig6}}
\end{figure*}

  \begin{table*}
  \caption{LyC escape fraction \label{tab8}}
\begin{tabular}{lcccrrrr} \hline
Name&$\lambda_0$$^{\rm a}$&$A$(LyC)$_{\rm MW}$$^{\rm b}$&$I_{\rm mod}$$^{\rm c,d}$&$I_{\rm obs}$(total)$^{\rm c,e}$&$I_{\rm esc}$(total)$^{\rm c,f}$&\multicolumn{1}{c}{$f_{\rm esc}$$^{\rm g}$}&\multicolumn{1}{c}{$f_{\rm esc}$$^{\rm h}$} \\
    &(\AA)&(mag)&&&&\multicolumn{1}{c}{(per cent)}&\multicolumn{1}{c}{(per cent)} \\
\hline
J0232$-$0426&870-890&0.203&16.83$\pm$1.92&    $<$0.58$^{\rm i}$ &     $<$0.70        &   $<$4.2          & $<$1.6 \\
J0919$+$4906&890-910&0.144&44.35$\pm$3.70& 6.30$^{+1.67}_{-1.51}$& 7.19$^{+1.92}_{-1.73}$&16.2$^{+5.9}_{-5.3}$&15.1$^{+4.5}_{-4.2}$ \\
J1046$+$5827&890-910&0.080&46.65$\pm$4.87&    $<$0.80$^{\rm i}$ &     $<$0.86        &   $<$1.8          & $<$1.0 \\
J1121$+$3806&880-900&0.210&60.69$\pm$5.30&17.53$^{+2.16}_{-2.09}$&21.25$^{+2.61}_{-2.52}$&35.0$^{+5.6}_{-5.4}$&25.1$^{+3.8}_{-3.7}$ \\
J1127$+$4610&870-900&0.165&29.00$\pm$3.02& 2.76$^{+0.89}_{-0.83}$& 3.22$^{+1.06}_{-0.99}$&11.1$^{+4.0}_{-3.7}$& 6.1$^{+2.4}_{-2.3}$ \\ 
J1233$+$4959&890-910&0.158&66.81$\pm$6.51& 7.00$^{+1.69}_{-1.61}$& 8.09$^{+1.95}_{-1.86}$&12.1$^{+3.4}_{-3.3}$& 9.5$^{+2.9}_{-2.7}$ \\ 
J1349$+$5631&850-870&0.098&17.22$\pm$1.87&    $<$1.05$^{\rm i}$ &     $<$1.15        &   $<$6.7          & $<$2.2 \\
J1355$+$1457&850-870&0.208&86.01$\pm$8.77&    $<$0.85$^{\rm i}$ &     $<$1.03        &   $<$1.2          & $<$0.9 \\
J1455$+$6107&850-870&0.133&69.42$\pm$6.97&    $<$1.36$^{\rm i}$ &     $<$1.54        &   $<$2.2          & $<$1.9 \\
\hline
  \end{tabular}

\hbox{$^{\rm a}$Restframe wavelength range in \AA\ used to determine the LyC flux.}

\hbox{$^{\rm b}$Milky Way extinction at the mean observed wavelengths of the 
range used to determine the LyC flux.} 

\hbox{\, The \citet{C89} reddening law with $R(V)$ = 3.1 is adopted.} 

\hbox{$^{\rm c}$LyC flux in 10$^{-18}$ erg s$^{-1}$cm$^{-2}$\AA$^{-1}$.}

\hbox{$^{\rm d}$LyC flux derived from the modelled SED.}

\hbox{$^{\rm e}$Observed LyC flux derived from the spectrum with shadow exposure, excluding J1046$+$5827, where measurements are from}

\hbox{\, the spectrum with total exposure.}


\hbox{$^{\rm f}$LyC flux which is corrected for the Milky Way extinction.}

\hbox{$^{\rm g}$$f_{\rm esc}$(LyC) = $I_{\rm esc}$(total)/$I_{\rm mod}$, where $I_{\rm mod}$ is derived from SED (first method).}

\hbox{$^{\rm h}$$f_{\rm esc}$(LyC) = $I_{\rm esc}$(total)/$I_{\rm mod}$, where $I_{\rm mod}$ is derived from H$\beta$ flux (second method).}

\hbox{$^{\rm i}$1$\sigma$ confidence upper limit.}

  \end{table*}

\section{{\sl HST}/COS observations and data 
reduction}\label{sec:obs}

{\sl HST}/COS spectroscopy of the nine LyC leaker candidates was obtained
in program GO~15639 (PI: Y.\ I.\ Izotov) during the period September 2019 --
May 2020. The observational details
are presented in Table \ref{tab6}. As in our previous programs
\citep{I16a,I16b,I18a,I18b}, the galaxies were directly acquired by COS near
ultraviolet (NUV) imaging. 
The NUV-brightest region of each target was centered in the
$\sim$~2.5\,arcsec diameter spectroscopic aperture (Fig.~\ref{fig2}).
Although the galaxies show generally a compact star-forming region 
superimposed upon an extended
low-surface-brightness (LSB) component and, in the case of
J0232$-$0426, several star-forming knots, their sizes are smaller
than the central unvignetted $0.8$\,arcsec diameter region of
the spectroscopic aperture \citep{F18}, except for J0232$-$0426. However, even
for this galaxy more than 90 per cent of the emission is concentrated in the
$0.8$\,arcsec diameter region. Hence, the galaxy quantities 
derived from the COS spectra do not require corrections for vignetting.
We note, however, that the acquisition exposure failed for J1046$+$5827. 
Therefore, no acquisition image is available for this galaxy (Fig.~\ref{fig2}).
Furthermore, the location of this galaxy inside the spectroscopic aperture
is not known, introducing uncertainties in the fluxes of the COS spectra of 
J1046$+$5827. The only way to estimate these uncertainties is to compare the 
observed COS spectrum with the extrapolation of the SED, obtained from fitting 
the SDSS optical spectrum, to the UV range (see Sect.~\ref{sec:global}).

The spectra were obtained at COS Lifetime Position~4 with the 
low-resolution grating G140L and medium-resolution grating G160M, applying all 
four focal-plane offset positions.
The 800\AA\ setup was used for the G140L grating 
(COS Lifetime Position 4: wavelength range 770--1950\,\AA, resolving
power $R\simeq 1050$ at 1150\,\AA) to include the redshifted LyC
emission for all targets. We obtained resolved spectra of the galaxies'
Ly$\alpha$ emission lines with the G160M grating ($R\sim 16000$ at
1600\,\AA), varying the G160M central wavelength with galaxy redshift
to cover the emission line and the nearby continuum on a single detector 
segment. The G140L observations of J1046$+$5827 and J1121$+$3806 were partially
executed due to technical problems (Table~\ref{tab6}).

The individual exposures were reduced with the \textsc{calcos} pipeline v3.2.1,
followed by accurate background subtraction and co-addition with custom
software \citep{W16}. We used the same methods and extraction
aperture sizes as in \citet{I18a,I18b} to achieve a homogeneous
reduction of the galaxy sample observed in multiple programs. We checked the
accuracy of our custom correction for scattered light in COS G140L data
by comparing the LyC fluxes obtained in the total exposure and in 
orbital night, respectively. An exception is the galaxy J1046$+$5827, for which
no orbital night exposures were obtained.

\begin{figure*}
\hbox{
\includegraphics[angle=-90,width=0.325\linewidth]{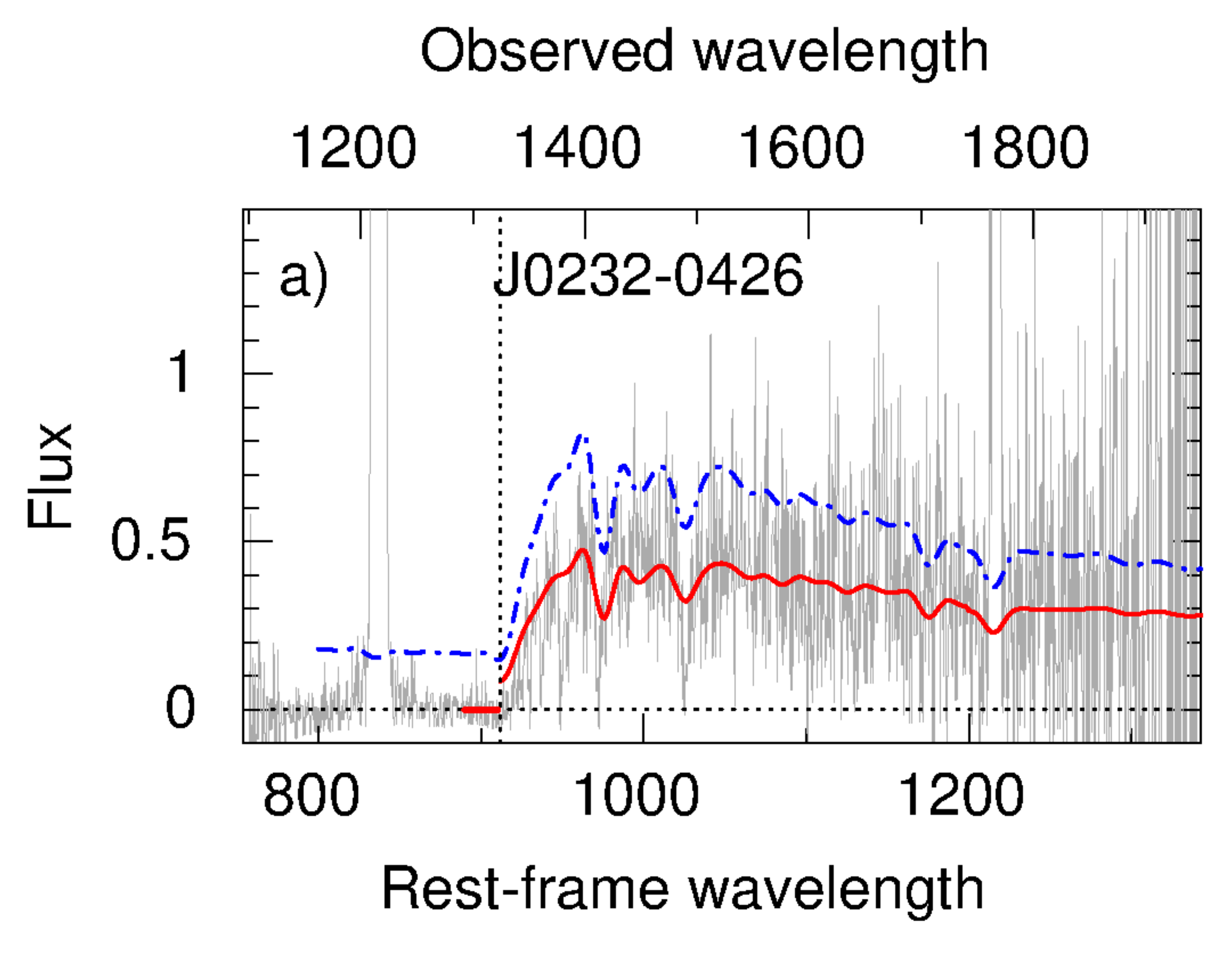}
\includegraphics[angle=-90,width=0.325\linewidth]{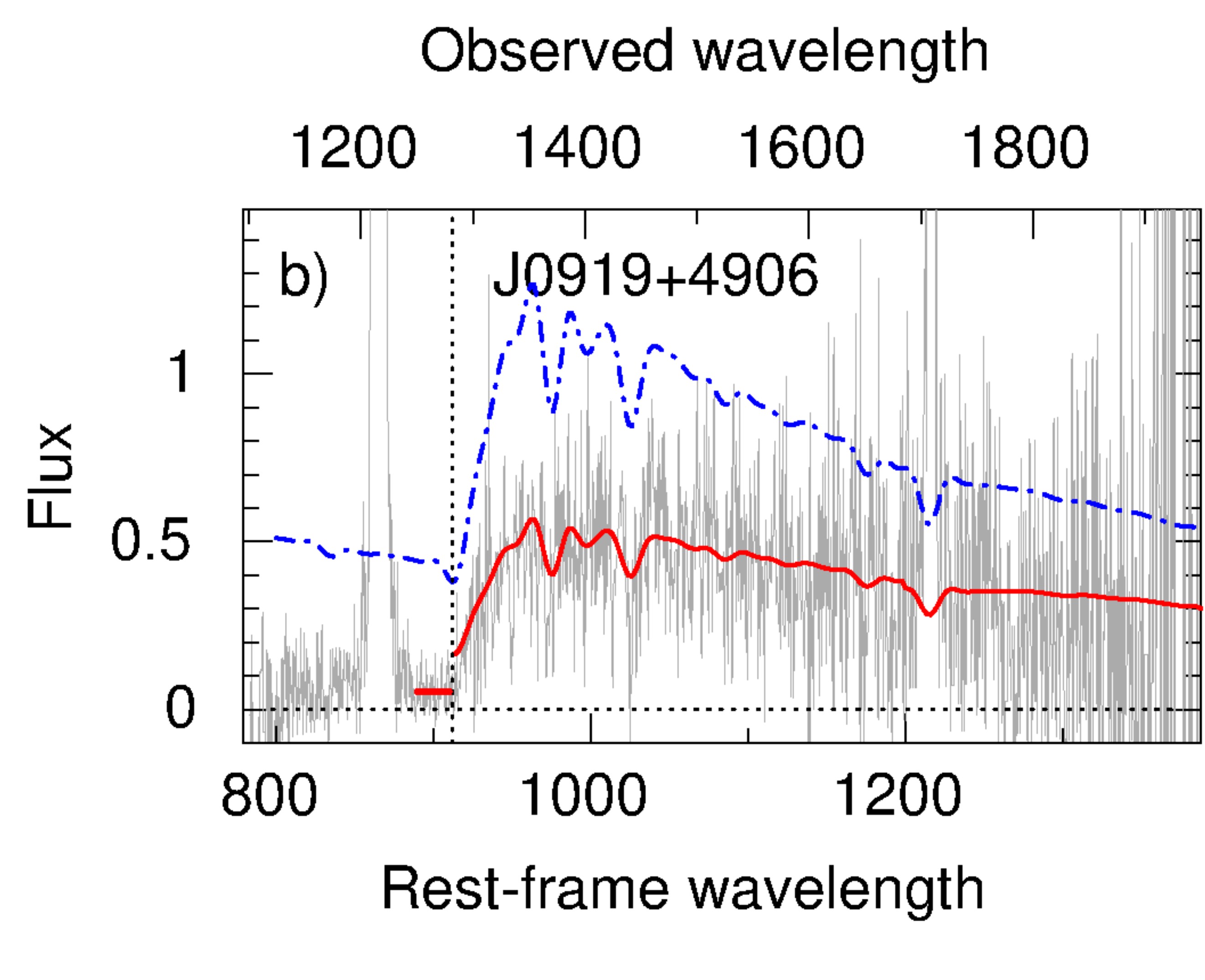}
\includegraphics[angle=-90,width=0.325\linewidth]{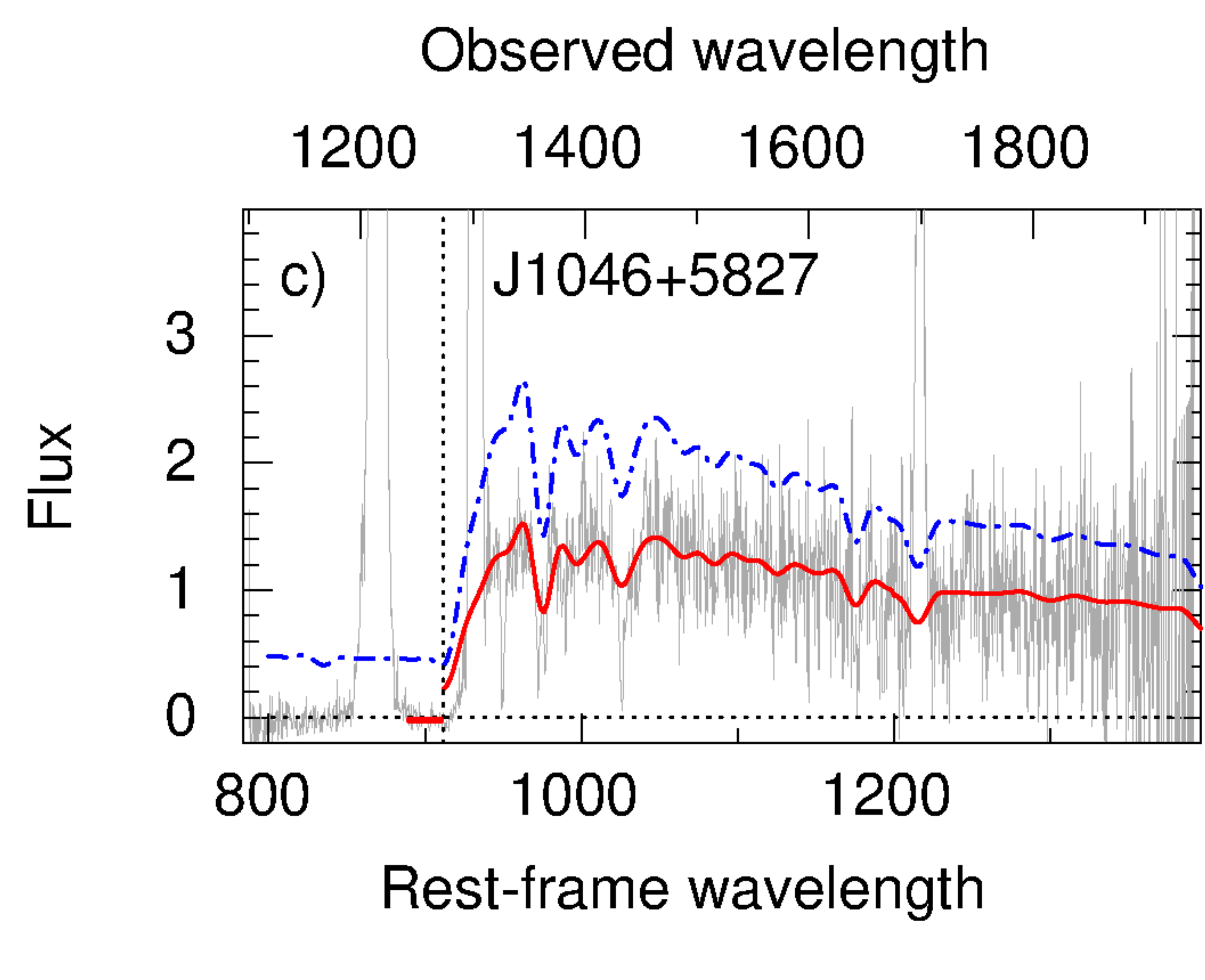}
}
\hbox{
\includegraphics[angle=-90,width=0.325\linewidth]{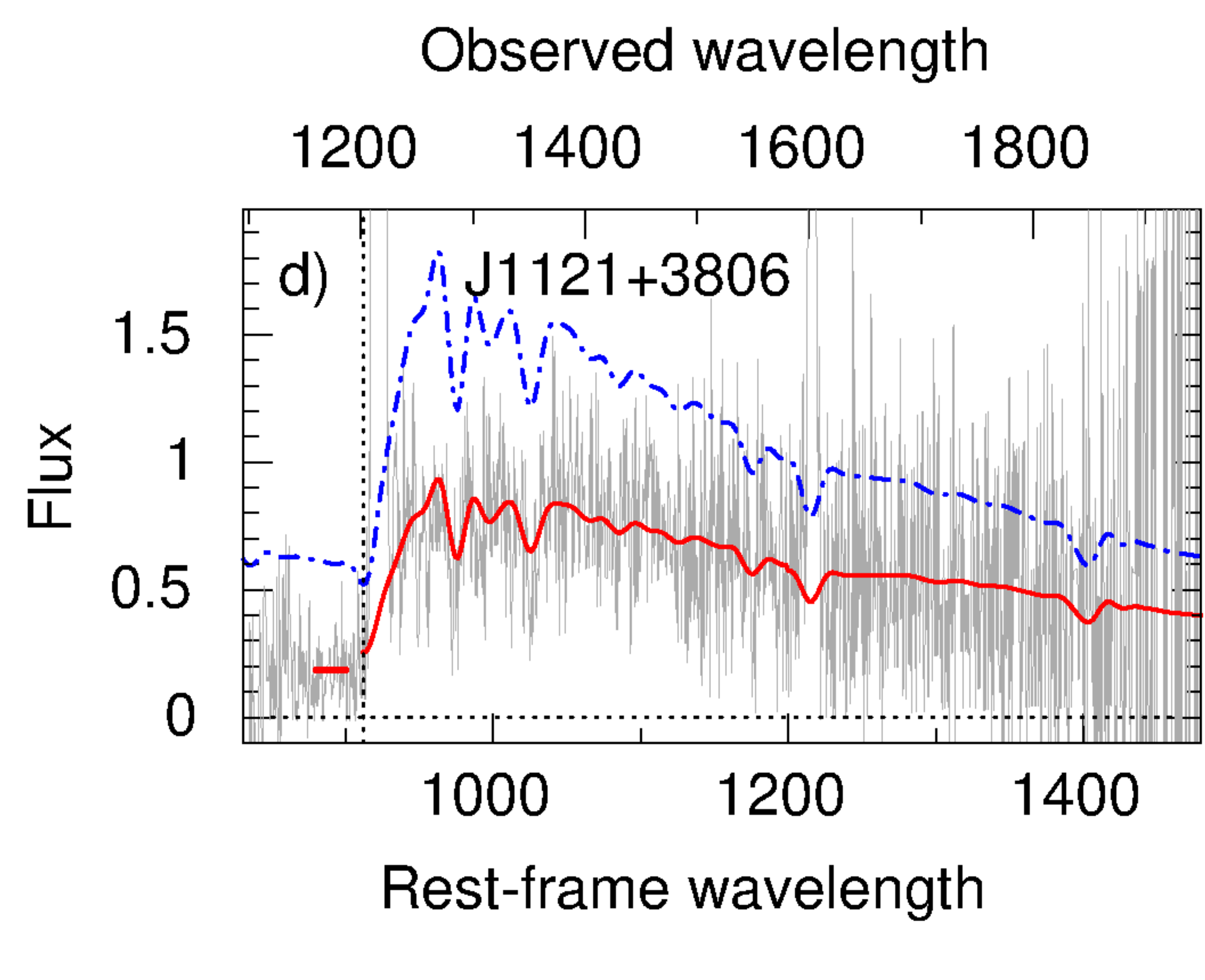}
\includegraphics[angle=-90,width=0.325\linewidth]{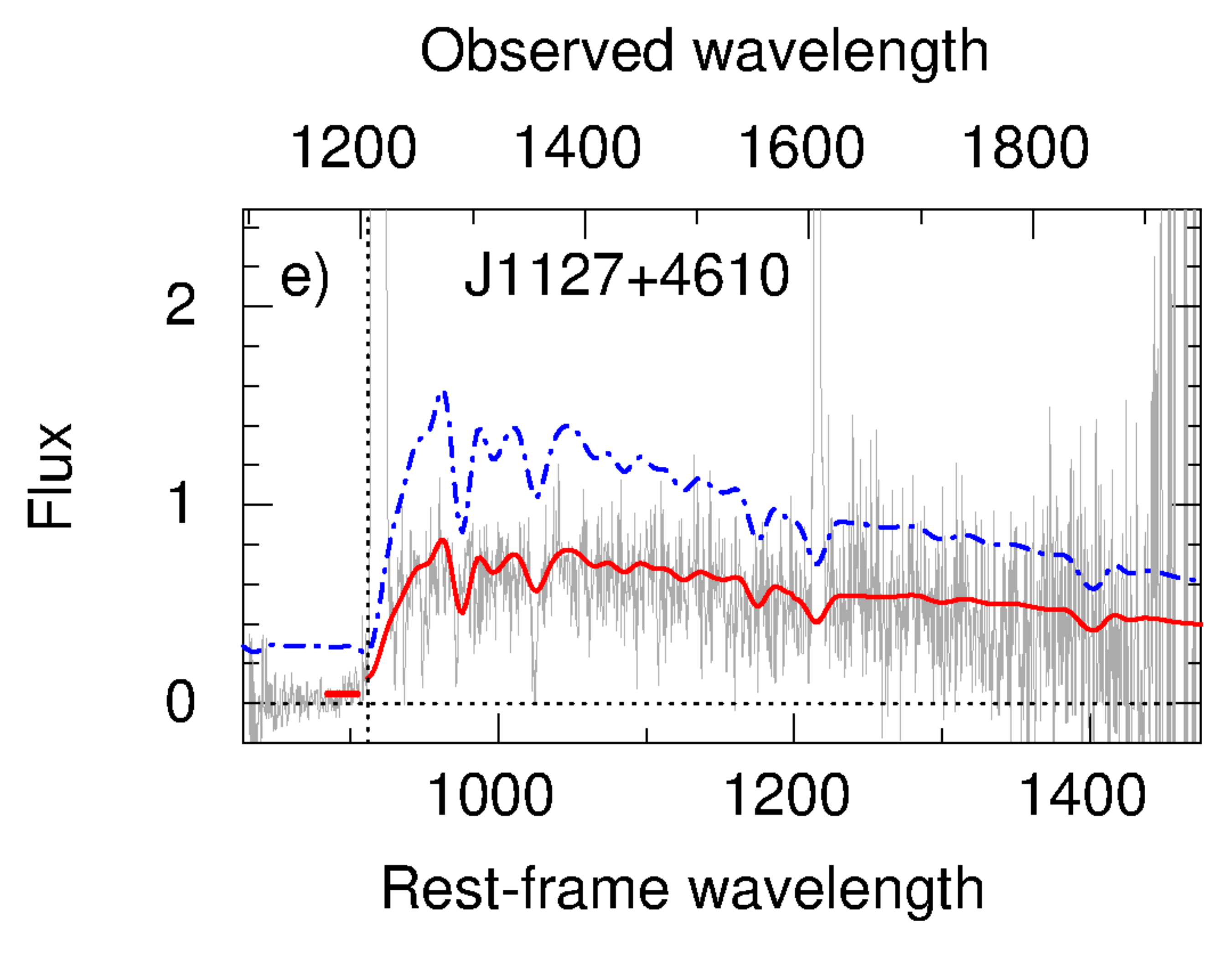}
\includegraphics[angle=-90,width=0.325\linewidth]{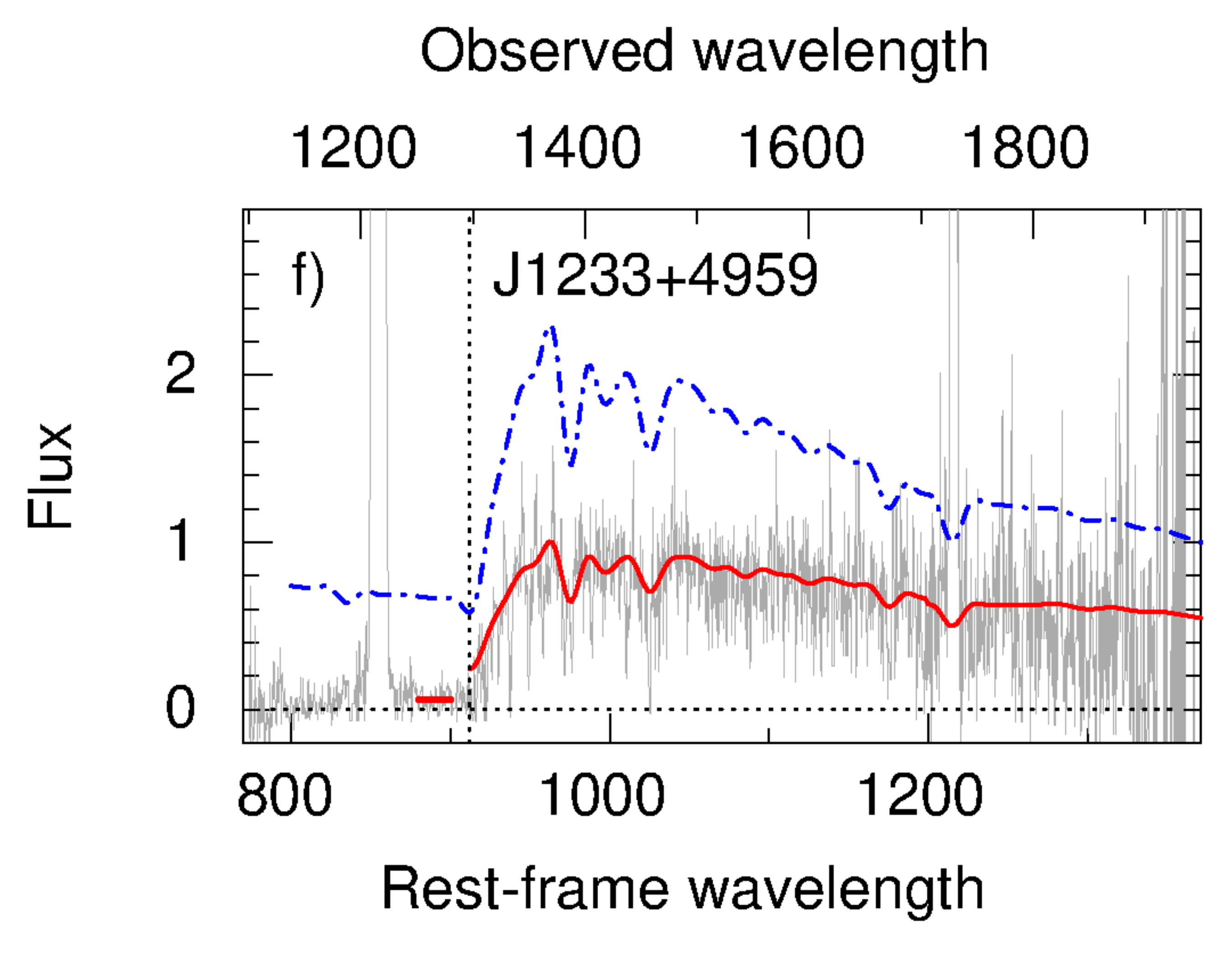}
}
\hbox{
\includegraphics[angle=-90,width=0.325\linewidth]{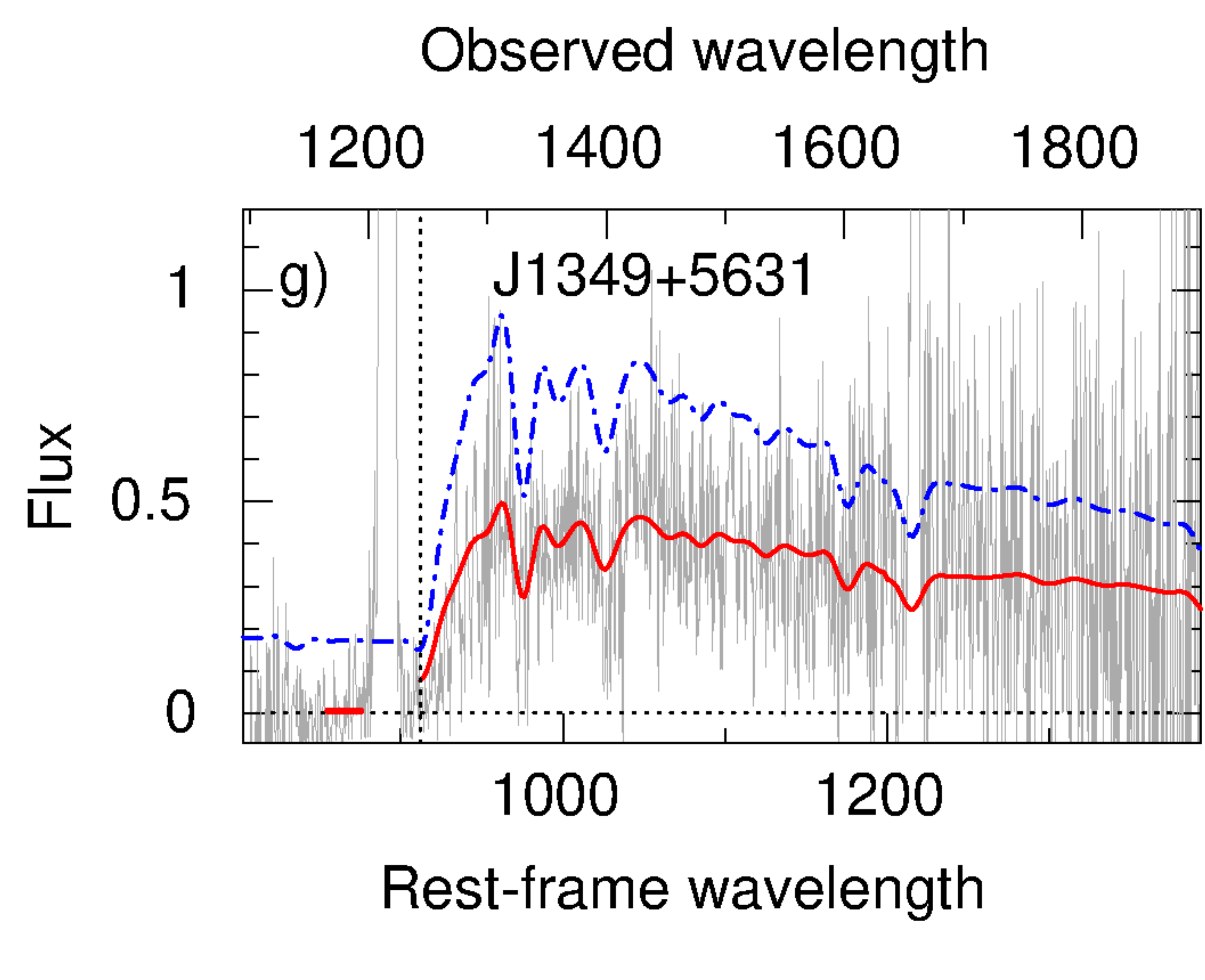}
\includegraphics[angle=-90,width=0.325\linewidth]{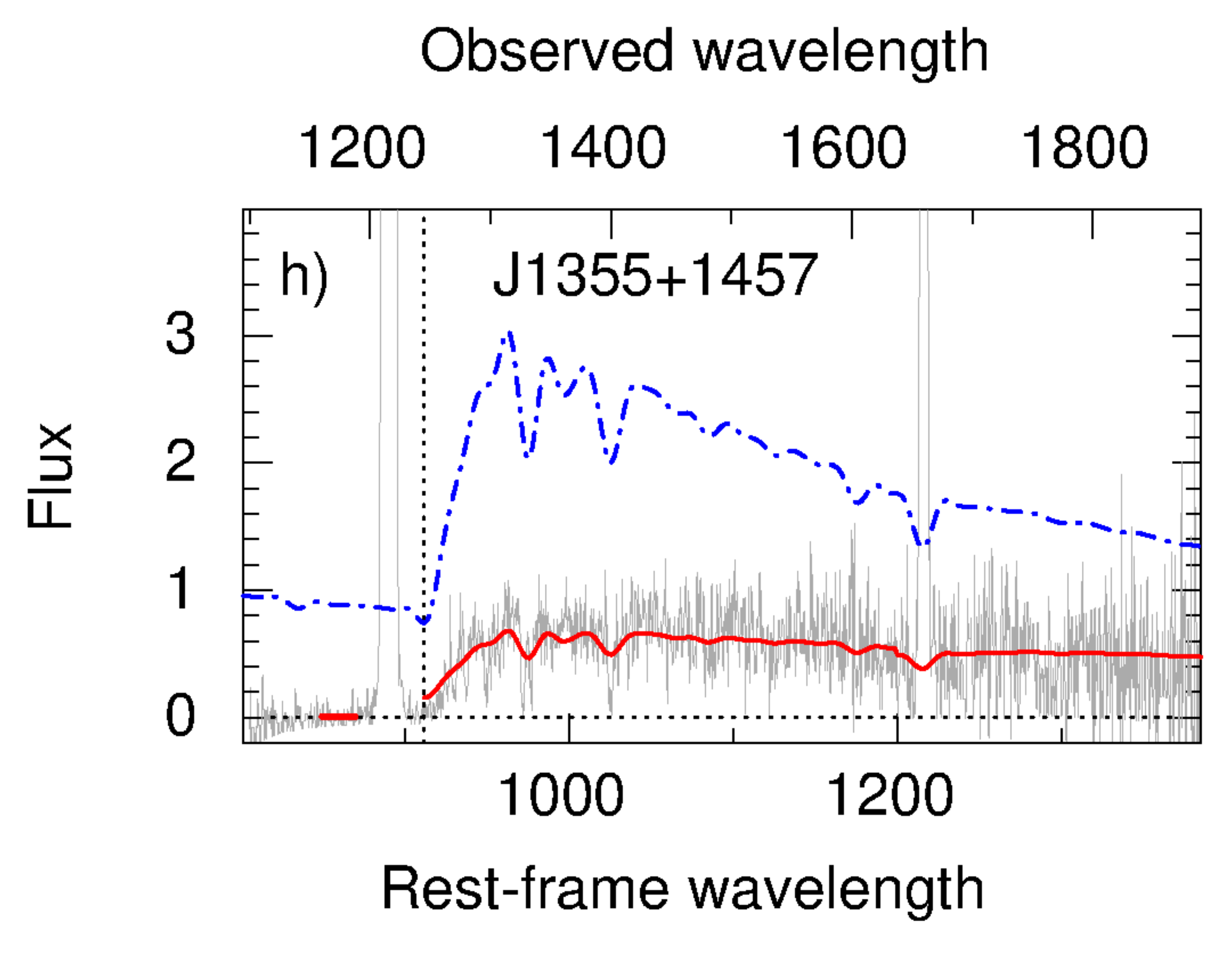}
\includegraphics[angle=-90,width=0.325\linewidth]{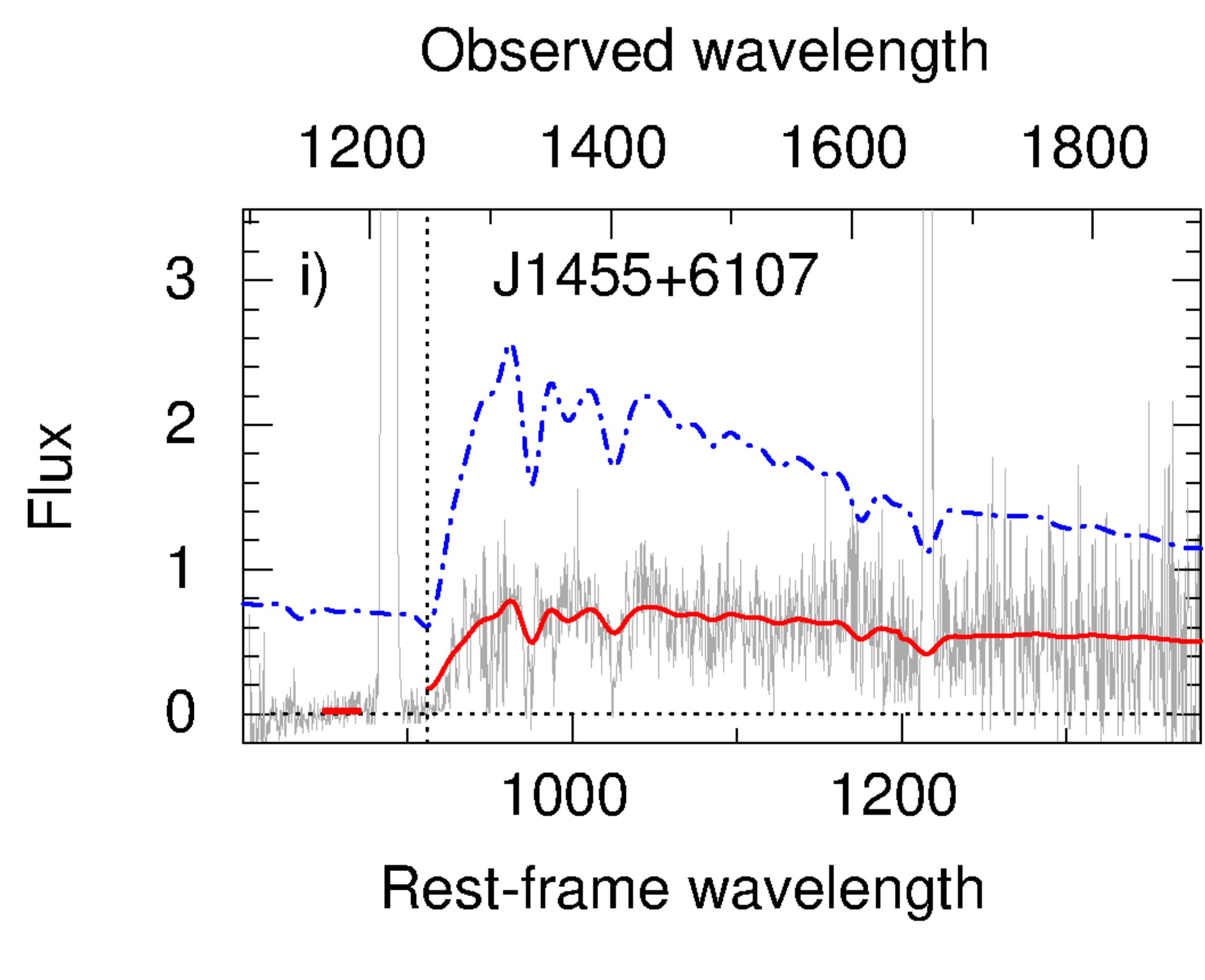}
}
\caption{COS G140L spectra of our sources (grey lines)
superposed by the modelled SEDs (thick red solid lines), reddened by 
both the internal and Milky Way extinctions.  
The unreddened (intrinsic) SEDs are shown by the thick blue dash-dotted lines.
$R(V)$$_{\rm int}$~=~2.7 is adopted in all panels.
The Lyman limit at the rest-frame wavelength 912\AA\ is indicated by dotted
vertical lines. Zero flux is represented by dotted horizontal lines.
Fluxes are in 10$^{-16}$ erg s$^{-1}$ cm$^{-2}$\AA$^{-1}$, wavelengths 
are in \AA. \label{fig7}}
\end{figure*}

\section{Surface brightness distribution in the NUV range}\label{sec:sbp}

The surface brightess (SB) profiles of our galaxies, in accord with previous 
studies by \citet{I16a,I16b,I18a,I18b}, are derived using the COS 
NUV acquisition images and the routine {\it ellipse} in 
{\sc iraf}\footnote{{\sc iraf} is distributed by 
the National Optical Astronomy Observatories, which are operated by the 
Association of Universities for Research in Astronomy, Inc., under cooperative 
agreement with the National Science Foundation.}/{\sc stsdas}\footnote{{\sc stsdas} is a product of 
the Space Telescope Science Institute, which is operated by AURA for NASA.}.
However, we note that the images of our galaxies are not as deep as
those in \citet{I16a,I16b,I18a,I18b} because of lower exposure times.
The profiles were scaled to magnitudes per square arcsec using the ratios
of the uncalibrated galaxy fluxes in the COS NUV images, measured with 
the routine {\it apphot} in {\sc iraf}, to the fluxes corresponding to 
the respective apparent {\sl GALEX} NUV magnitudes. 
An exception is the galaxy J1121$+$3806, for which no {\sl GALEX}
data are available. For this galaxy, we adopt a magnitude derived
from extrapolation of the attenuated SED obtained from the SDSS spectrum. There is no
SB profile for the galaxy J1046$+$5827 because its acquisition exposure failed, as noted before. 

As found before by \citet{I16a,I16b,I18a,I18b}, the outer parts of our galaxies are characterised by a linear decrease in SB
(in mag per square arcsec scale), characteristic of a disc structure, 
while the central part with the bright star-forming region(s) shows a sharp increase (Fig. \ref{fig3}).

The scale lengths $\alpha$ of our galaxies, defined in Eq.~1 of  \citet{I16b},
are in the range $\sim$ 0.3 -- 0.6 kpc (Fig.~\ref{fig3}), lower than 
$\alpha$ = 0.6 -- 1.8 kpc in other LyC leakers \citep{I16a,I16b,I18a,I18b} and
indicating a lower mass and a more compact structure for our galaxies. The 
corresponding surface densities of star-formation rate in the studied galaxies 
$\Sigma$ = SFR/($\pi \alpha^2$) are similar to those of other LyC leakers.
Because the bright star-forming regions are very compact with sharply rising  
brightness profiles, the half-light radii $r_{50}$ of our galaxies in the NUV 
are similar but
considerably smaller than $\alpha$ (see Table \ref{tab5}).
Adopting $r_{50}$ as a measure of the size of these galaxies, 
the corresponding $\Sigma$s are
typically two orders of magnitude larger, and are comparable to those found for 
SFGs at high redshifts \citep{CL16,PA18,Bo17}. 
The values $\Sigma_1$ in Table \ref{tab5} are similar to those found by \citet{Ki20}
for GPs and Lyman Break Analogs.

\section{Comparison of the {\sl HST}/COS spectra with the modelled
SEDs in the UV range}\label{sec:global}

To derive the fraction of the escaping ionising radiation, the two methods, 
which we use \citep[e.g. ][]{I18a}, are based on the comparison
between the observed flux in the Lyman continuum range and the intrinsic
flux produced by stellar populations in the galaxy. 
The intrinsic LyC flux can be obtained from SED fitting of the SDSS spectra or 
from the flux of the H$\beta$ emission line.
To verify the quality of our SED fitting, we extrapolate the attenuated SEDs 
to the UV range and compare them with the observed COS spectra in 
Fig.~\ref{fig4}. For comparison, we also show with blue filled circles the 
{\sl GALEX} FUV and NUV fluxes and the fluxes in the SDSS $u,g,r,i,z$ 
filters. We find that the spectroscopic 
and photometric data in the optical range are consistent, indicating that almost
all the emission of our galaxies is inside the spectroscopic aperture. 
Therefore, aperture corrections are not needed. 
On the other hand, considerable deviations of {\sl GALEX} FUV and NUV fluxes 
to brighter values from both the observed COS spectra and attenuated SED 
extrapolations are found for many of galaxies, most strikingly for J0232$-$0426.
These deviations in the FUV could in part be explained by the redshifted
Ly$\alpha$. However, no such deviation is seen in the spectrum of J0919$+$4906
with the highest EW(Ly$\alpha$). Furthermore, no bright emission lines fall
into the NUV range despite large deviations from the observed spectra and 
modelled SEDs. Therefore, 
these systematic deviations indicate high uncertainties of the {\sl GALEX} FUV 
and NUV fluxes, at least for faint objects.

The modelled intrinsic SEDs in Fig.~\ref{fig4} are attenuated by adopting the 
extinction coefficients $C$(H$\beta$)$_{\rm MW}$ and $C$(H$\beta$)$_{\rm int}$ 
(Table~\ref{tab3}) and the reddening law by 
\citet{C89} with $R(V)_{\rm MW}$ = 3.1 and $R(V)_{\rm int}$ = 2.7 (black solid 
lines and red solid lines, respectively). 

It is seen in Fig.~\ref{fig4} that the models reproduce the SDSS 
spectra quite well and do not depend on the adopted $R(V)_{\rm int}$ 
because of low extinction. There is a stronger dependence of the attenuated 
SEDs on $R(V)_{\rm int}$ in the UV range, but differences of the attenuated SEDs 
with $R(V)_{\rm int}$ = 3.1 and $R(V)_{\rm int}$ = 2.7 are still small 
(black and red lines in Fig.~\ref{fig4}), because of 
small extinction coefficients $C$(H$\beta$)$_{\rm int}$. However, the 
attenuated SEDs with $R(V)_{\rm int}$ = 2.7 reproduce 
on average somewhat better the observed COS spectra.

\begin{figure*}
\hbox{
\includegraphics[angle=-90,width=0.45\linewidth]{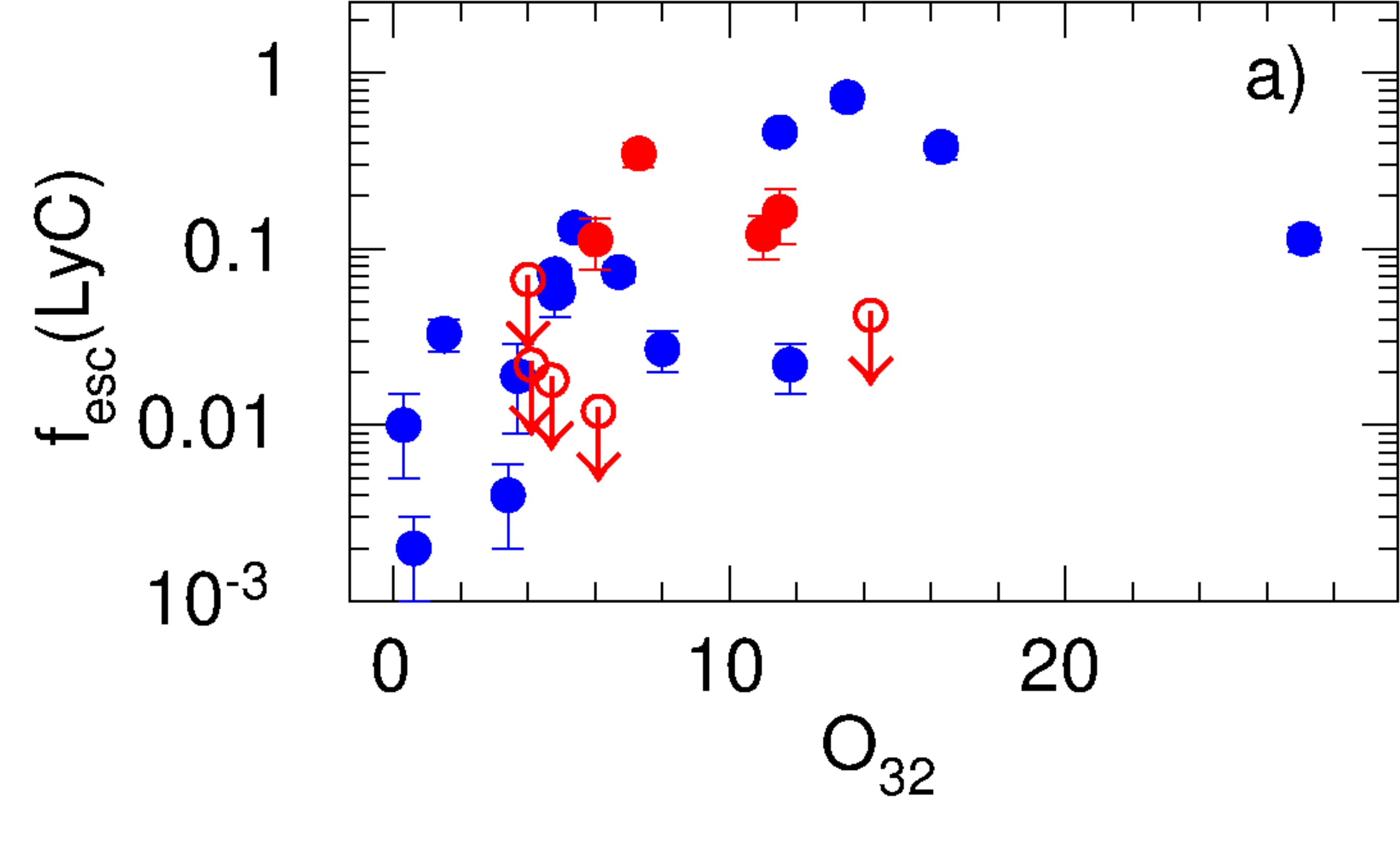}
\includegraphics[angle=-90,width=0.45\linewidth]{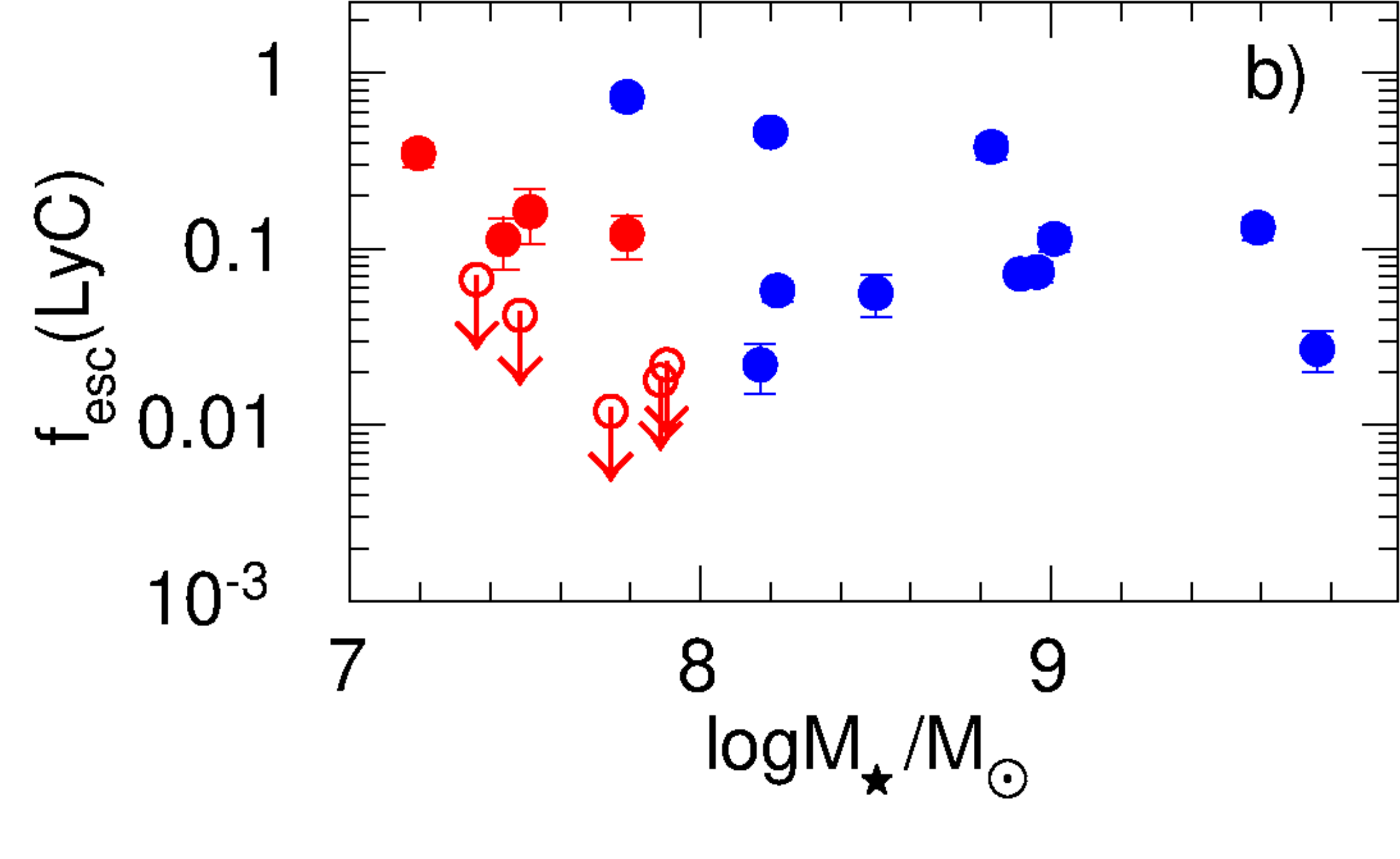}
}
\hbox{
\includegraphics[angle=-90,width=0.45\linewidth]{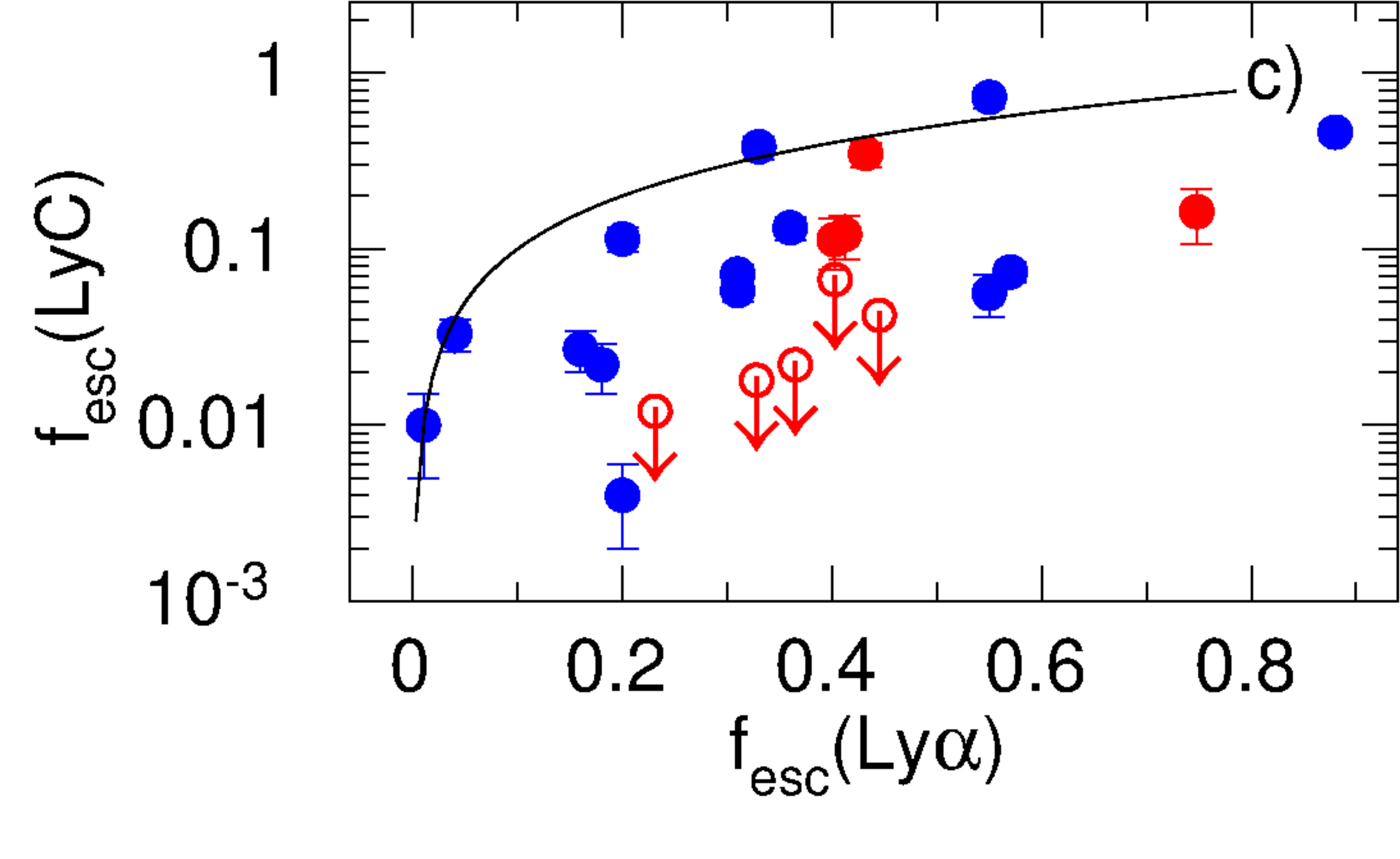}
\includegraphics[angle=-90,width=0.45\linewidth]{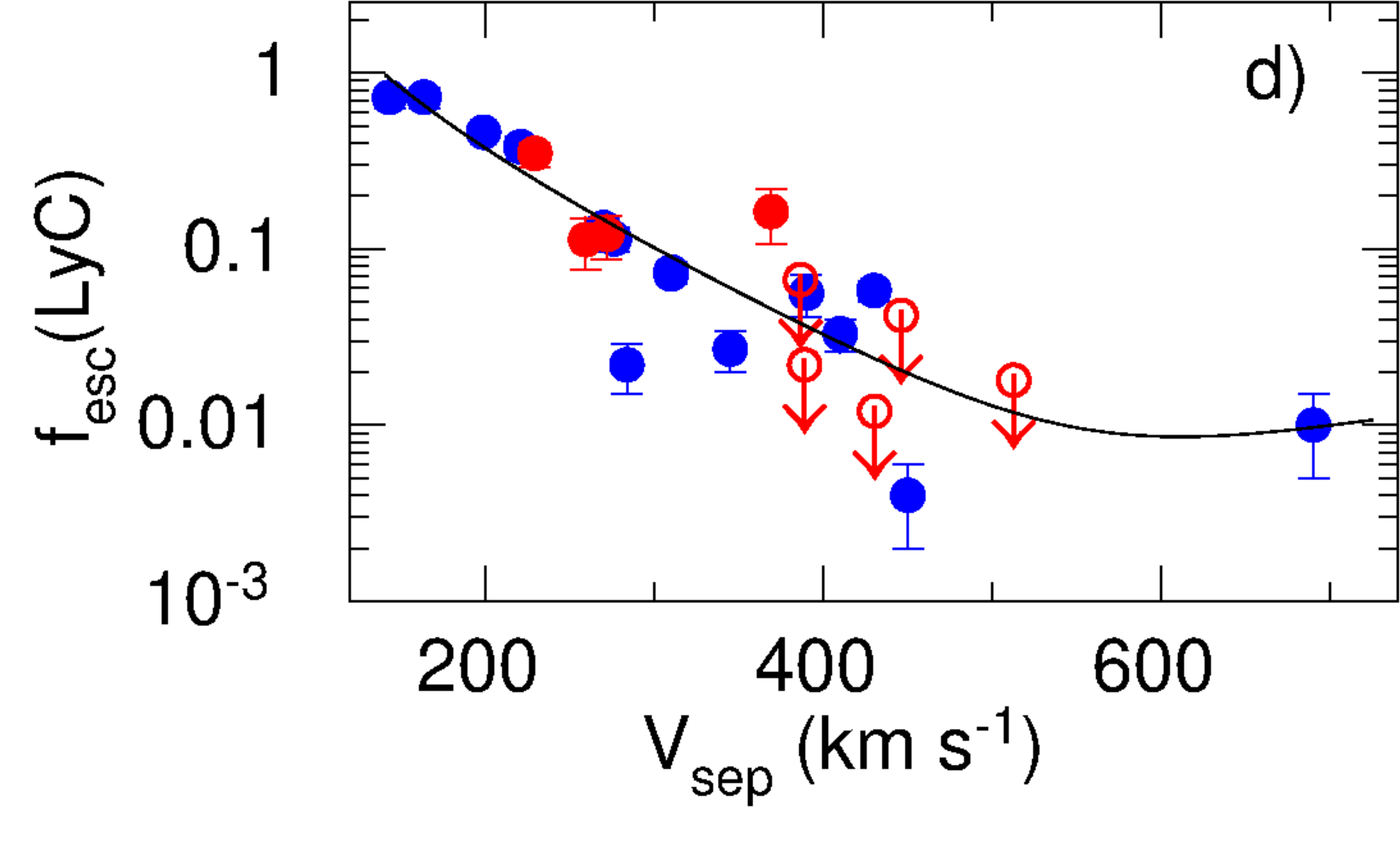}
}
\caption{Relations between the Lyman continuum escape fraction
$f_{\rm esc}$(LyC) in low-redshift LyC leaking galaxies derived from the SED
fits and {\bf a)} the 
[O~{\sc iii}]$\lambda$5007/[O~{\sc ii}]$\lambda$3727 
emission-line flux ratio, {\bf b)} the stellar mass $M_\star$, {\bf c)} the 
Ly$\alpha$ escape fraction $f_{\rm esc}$(Ly$\alpha$), and
{\bf d)} the separation $V_{\rm sep}$ between the Ly$\alpha$ profile peaks. 
LyC leakers from \citet{I16a,I16b,I18a,I18b}, \citet{L13}, \citet{B14} and 
\citet{L16} are shown by blue filled circles with 1$\sigma$ errors, the galaxies
from this paper with detected LyC emission and upper limits are represented by 
red filled circles with 1$\sigma$ errors and
red open circles and downward arrows, respectively. 
The solid line in {\bf c)} is the equality line 
and the solid line in {\bf d)} represents the relation from \citet{I18b}.
\label{fig8}}
\end{figure*}

\section{Ly$\alpha$ emission}\label{sec:lya}

A resolved Ly$\alpha$ $\lambda$1216\,\AA\ emission line is detected
in the medium-resolution spectra of all galaxies (Fig.~\ref{fig5}), whereas 
this line is unresolved in the low-resolution spectra. 
The shape of the resolved Ly$\alpha$ line is similar to that observed in most known 
LyC leakers \citep{I16a,I16b,I18a,I18b} and in some other galaxies 
\citep{JO14,H15,Y17,I20}. In the present sample, profiles with two peaks are detected in
the spectra of seven galaxies. The signal-to-noise ratio
in the spectrum of J0232$-$0426 is too low to definitely derive the profile 
of its Ly$\alpha$ emission line. The Ly$\alpha$ profile of J1127$+$4610 may 
show more than two peaks, but the S/N~$\sim$~5 even in undersampled spectra is 
too low to definitely determine that, similar to the data of J0232$-$0426.
The parameters of Ly$\alpha$ emission are shown in Table~\ref{tab7}. 

We show in Fig.~\ref{fig6} the dependences of some Ly$\alpha$
characteristics on absolute FUV magnitude. Our galaxies with detected LyC
emission are shown by red filled circles and those with upper limits of LyC
emission are represented by red open circles. 
All our low-mass galaxies are characterised by moderate Ly$\alpha$
luminosities in a narrow range between 10$^{42.19}$ and 10$^{42.74}$ erg s$^{-1}$ 
(Fig.~\ref{fig6}a), that are slightly below the values for confirmed LyC leakers
(blue filled circles) by \citet{I16a,I16b,I18a,I18b} and high-redshift galaxies 
(grey open circles) by \citet{O08}, \citet{H17}, \citet{J18}, 
\citet{Mat17,Mat18}, \citet{So18}, but are $\sim$ 1 order
of magnitude higher than those for GPs 
\citep[black asterisks, ][]{H15,JO14,J17,MK19,Y17}. Our low-mass galaxies 
have high EW(Ly$\alpha$) $\sim$ 65 -- 220\AA\ (Table~\ref{tab7}, red symbols in 
Fig.~\ref{fig6}b), similar to those in other LyC leakers (blue symbols in 
Fig.~\ref{fig6}b). They are at the high end 
of the EW(Ly$\alpha$) values for high-$z$ LAEs by 
\citet{O08}, \citet{H17}, \citet{J18}, \citet{Ha18}, \citet{Ca18}, \citet{Pe18},
\citet{Mat17,Mat18}, \citet{So18}
(grey open circles) and GPs (black asterisks). However, contrary to 
expectations for galaxies with lower stellar masses and, likely, lower masses
of the neutral gas, the separation between the 
Ly$\alpha$ peaks is on average similar to that in higher-mass LyC leakers 
(Fig.~\ref{fig6}c), and it is higher in galaxies with upper limits of LyC 
emission ($>$ 400 km s$^{-1}$, red open circles). Ly$\alpha$ escape fractions 
$f_{\rm esc}$(Ly$\alpha$) in low-mass galaxies are also similar to those in 
higher-mass LyC leakers, and they are lower in galaxies with upper limits
of LyC emission (red open circles in Fig.~\ref{fig6}d). We also note that the 
average ratio of blue and red peak fluxes of $\sim$ 25 per cent 
(Table~\ref{tab7}) is somewhat lower than the value of 
$\sim$~30 per cent quoted by \citet{Ha21} for $z$ $\sim$ 0 galaxies.

For comparison, we also show by green filled circles in Fig.~\ref{fig6} the
extreme-O$_{32}$ low-redshift ($z$ $<$ 0.07) compact star-forming galaxies with 
high EW(Ly$\alpha$) and by green open circles the galaxies with weak 
Ly$\alpha$ emission on top of broad Ly$\alpha$ absorption profiles. It is seen 
that low-$z$ SFGs with high EW(Ly$\alpha$) are shifted from relations for 
high-$z$ LAEs (black solid lines) to lower $L$(Ly$\alpha$) (Fig.~\ref{fig6}a) 
and EW(Ly$\alpha$) (Fig.~\ref{fig6}b), despite similar excitation conditions, 
as indicated by comparable O$_{32}$ and EW(H$\beta$) values. 

\citet{I20} discussed 
this feature of low-$z$ CSFGs and suggested that it could be due to the presence
of an extended Ly$\alpha$ halo with an angular diameter considerably larger than
the 2.5 arcses in diameter COS spectroscopic aperture, making some Ly$\alpha$
emission unobservable, although there is no direct evidence for that from the
{\sl HST} observations. Furthermore, the Ly$\alpha$ escape fraction 
$f_{\rm esc}$(Ly$\alpha$) in these galaxies is somewhat lower than that in 
LyC leakers (Figs.~\ref{fig6}d), supporting the idea of the extended Ly$\alpha$
halo, exceeding the size of the COS spectroscopic aperture, corresponding
to a linear radius of $\sim$~0.6~-~1.7 kpc at the redshifts of low-$z$ 
CSFGs. This results in a loss of some emission and, consequently, in a 
reduced derived value of $f_{\rm esc}$(Ly$\alpha$). However, the
linear radius inside the COS spectroscopic aperture at redshift $z$~$>$~0.3
is $\ga$~5~kpc and, likely, is larger than the extent of Ly$\alpha$
emission, implying that most of Ly$\alpha$ emission is inside the COS
aperture, despite the fact that the luminosity of Ly$\alpha$ in LyC leakers 
(red and blue circles in Fig.~\ref{fig6}) is 
several times higher than that in low-$z$ CSFGs with extreme O$_{32}$.

\section{Escaping Lyman continuum radiation}\label{sec:lyc}

The observed G140L total-exposure spectra with the LyC spectral region 
(grey lines) and predicted intrinsic 
SEDs  (blue dash-dotted lines) are shown in Fig. \ref{fig7}, together with the  
attenuated intrinsic SEDs (red solid lines). The predicted 
intrinsic SEDs are
obtained from fitting the optical SDSS spectra, corrected for
the Milky Way extinction at observed wavelengths, and adopting  $A(V)_{\rm MW}$ 
from the NED, and for the internal extinction at restframe wavelengths, 
the UV attenuation law with $R(V)$ = 2.7, and the
extinction coefficients $C$(H$\beta$)$_{\rm int}$ derived from the 
hydrogen Balmer decrement.

The level of the observed LyC continuum is indicated by horizontal red lines 
shortward of the vertical dotted line showing the Lyman series limit. 
The Lyman continuum emission is detected in the 
spectra of four galaxies, J0919$+$4906, J1121$+$3806, J1127+4610 and J1233+4959,
and only upper limits are derived in the spectra of the remaining galaxies. 
The measurements are summarised in Table~\ref{tab8}. Due to the faintness of 
the targets, we adopted the LyC fluxes measured during orbital night (except 
for J1046$+$5827) to minimise residual uncertainties in the scattered light 
correction.

\citet{I16a,I16b,I18a,I18b} used the ratio of the escaping
fluxes $I_{\rm esc}$ to the intrinsic fluxes $I_{\rm mod}$
of the Lyman continuum to derive $f_{\rm esc}$(LyC):
\begin{equation}
f_{\rm esc}({\rm LyC}) =\frac{I_{\rm esc}(\lambda)}{I_{\rm mod}(\lambda)}, 
\label{eq:fesc}
\end{equation}
where $\lambda$ is the mean wavelength of the range used for averaging of
the LyC flux density (see Table~\ref{tab8}).
\citet{I16b} proposed two methods to derive iteratively the intrinsic fluxes 
$I_{\rm mod}$ and, correspondingly, the LyC escape fractions $f_{\rm esc}$(LyC): 
1) from the SED fitting and 2) from the equivalent width of the H$\beta$ 
emission line and its extinction-corrected flux
and adopting relations between $I$(H$\beta$)/$I_{\rm mod}$ 
and EW(H$\beta$) from the models of photoionised H~{\sc ii} regions
\citep[equation 4 for instantaneous burst in ][]{I16b}. 
We use both methods in this paper.

The escape fraction $f_{\rm esc}$(LyC), in the range between 11 and 35 per cent, is 
derived in four out of nine galaxies (Table~\ref{tab8}). For the remaining 
galaxies, only upper limits of $f_{\rm esc}$(LyC) were derived. We note that 
the $f_{\rm esc}$(LyC)s derived by the second method are somewhat lower than those
derived by the first method, but are consistent within the errors for most galaxies. 

\begin{figure*}
\hbox{
\includegraphics[angle=-90,width=0.325\linewidth]{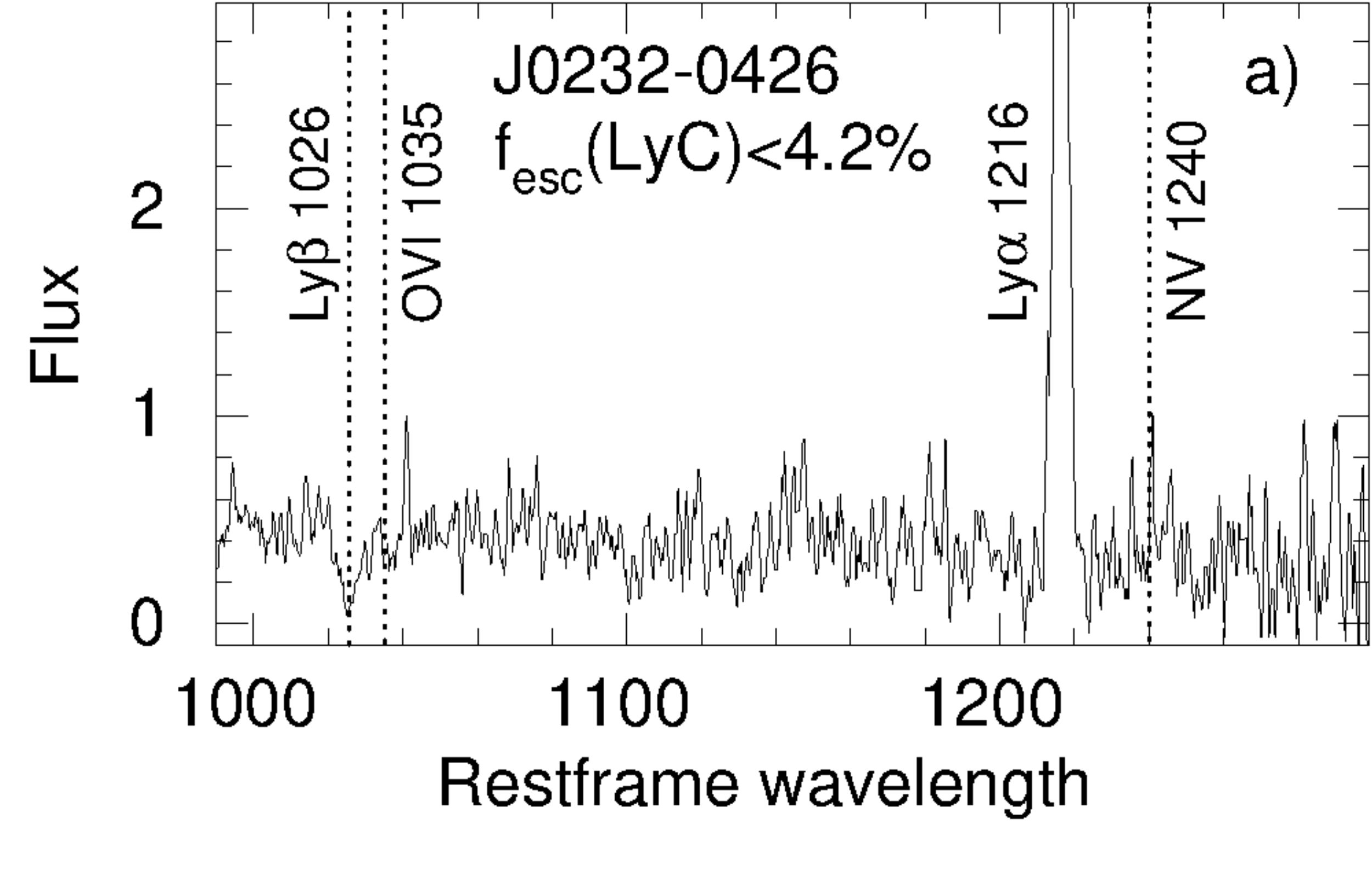}
\includegraphics[angle=-90,width=0.325\linewidth]{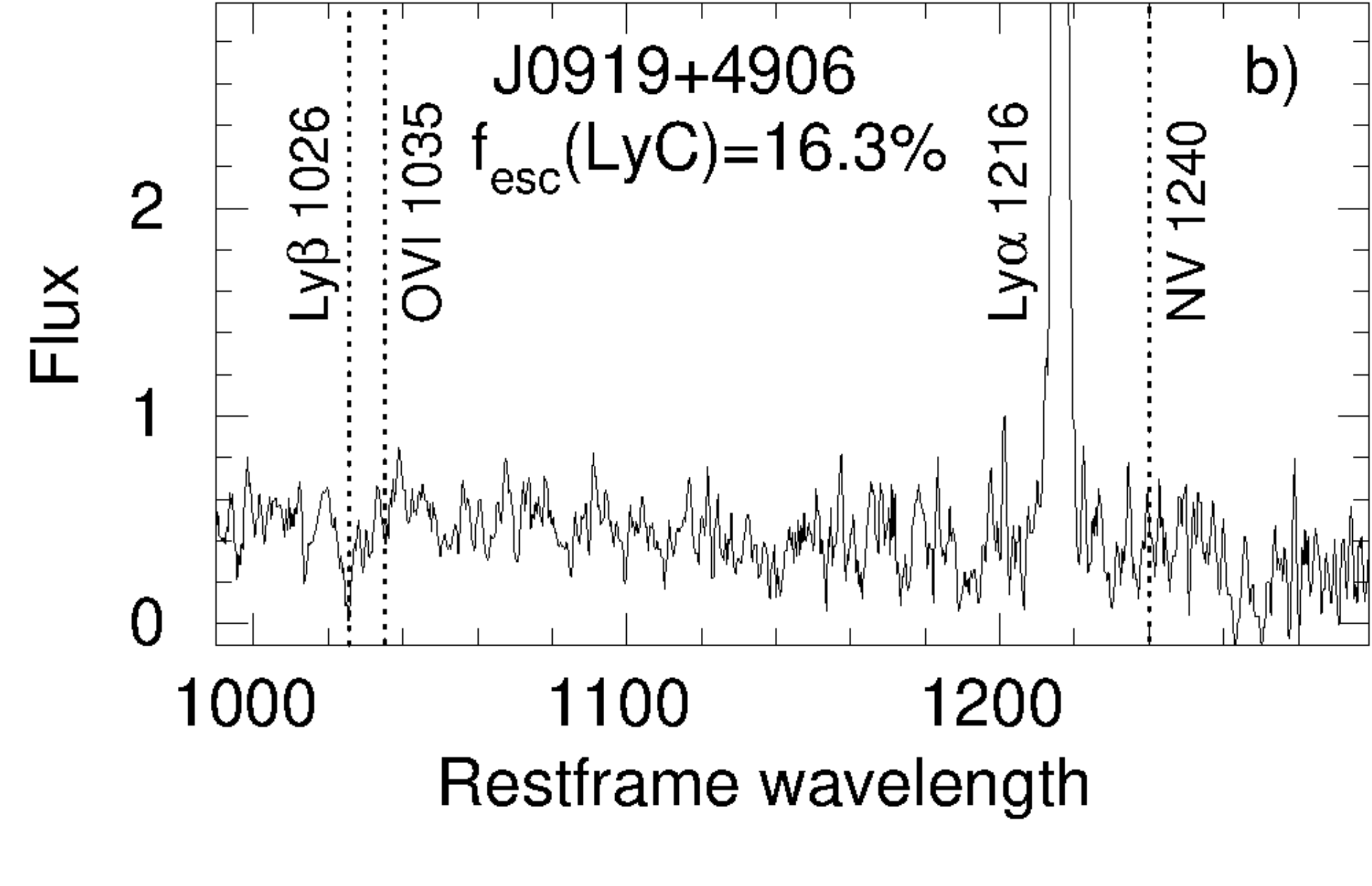}
\includegraphics[angle=-90,width=0.325\linewidth]{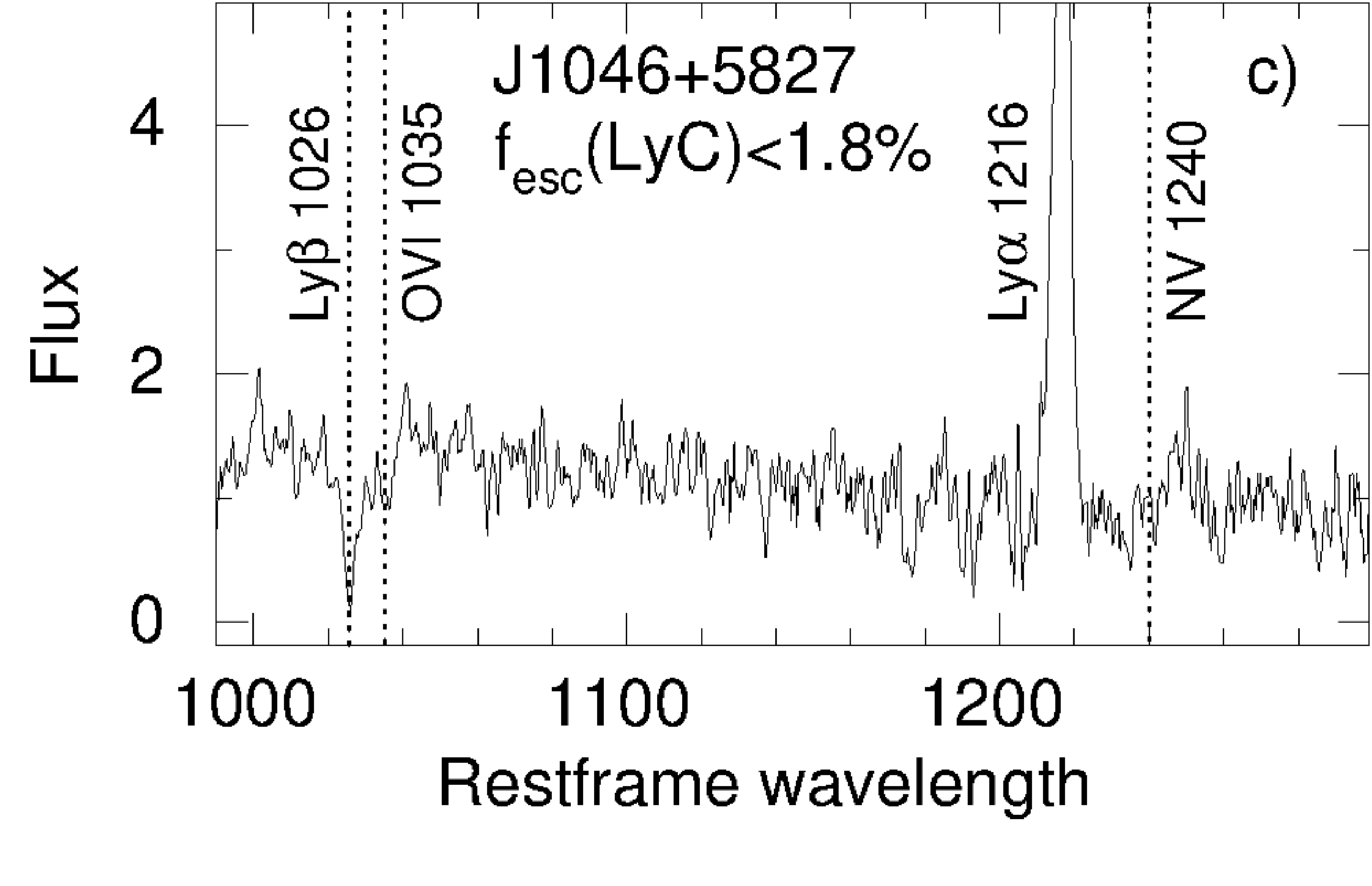}
}
\hbox{
\includegraphics[angle=-90,width=0.325\linewidth]{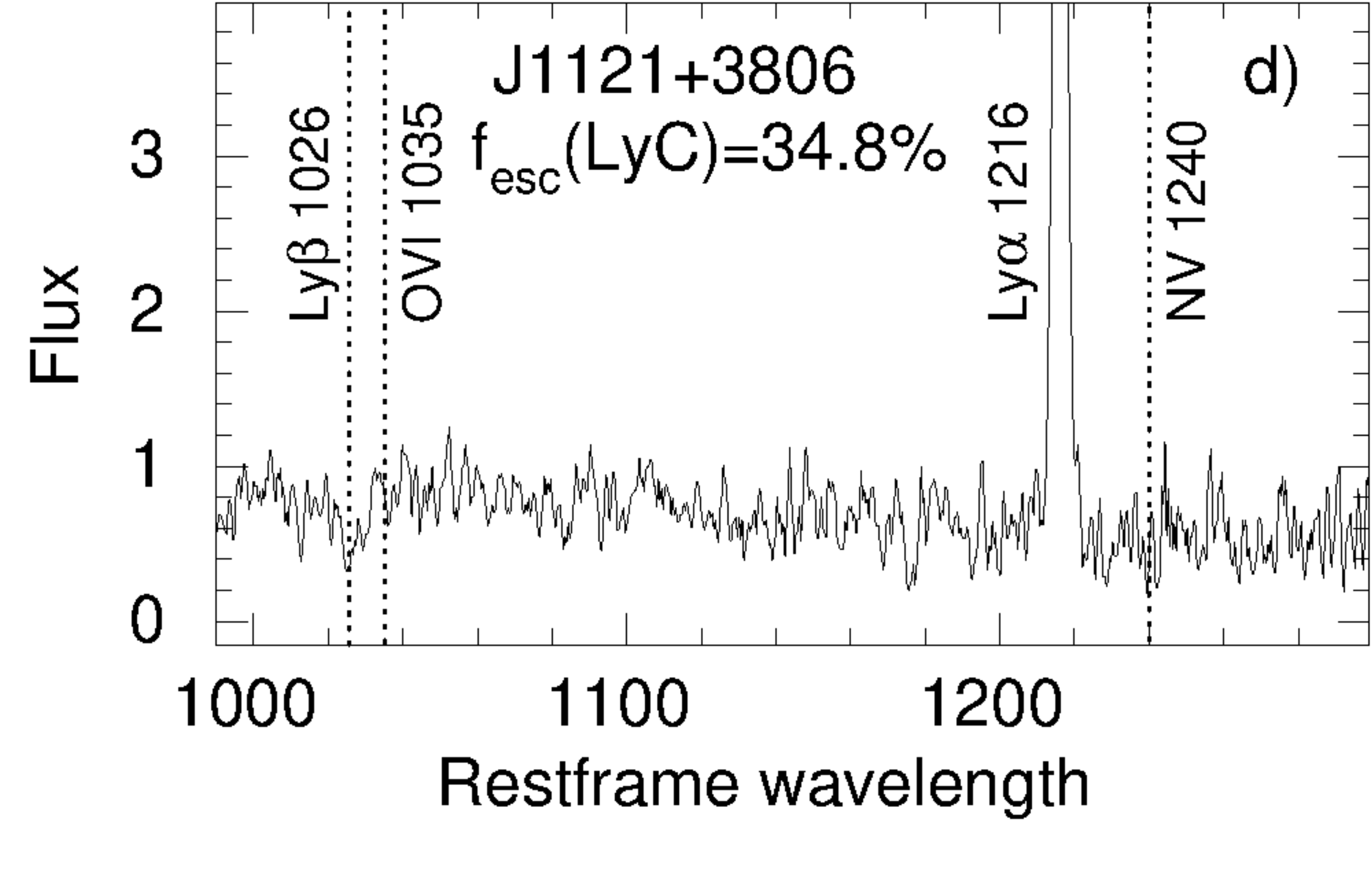}
\includegraphics[angle=-90,width=0.325\linewidth]{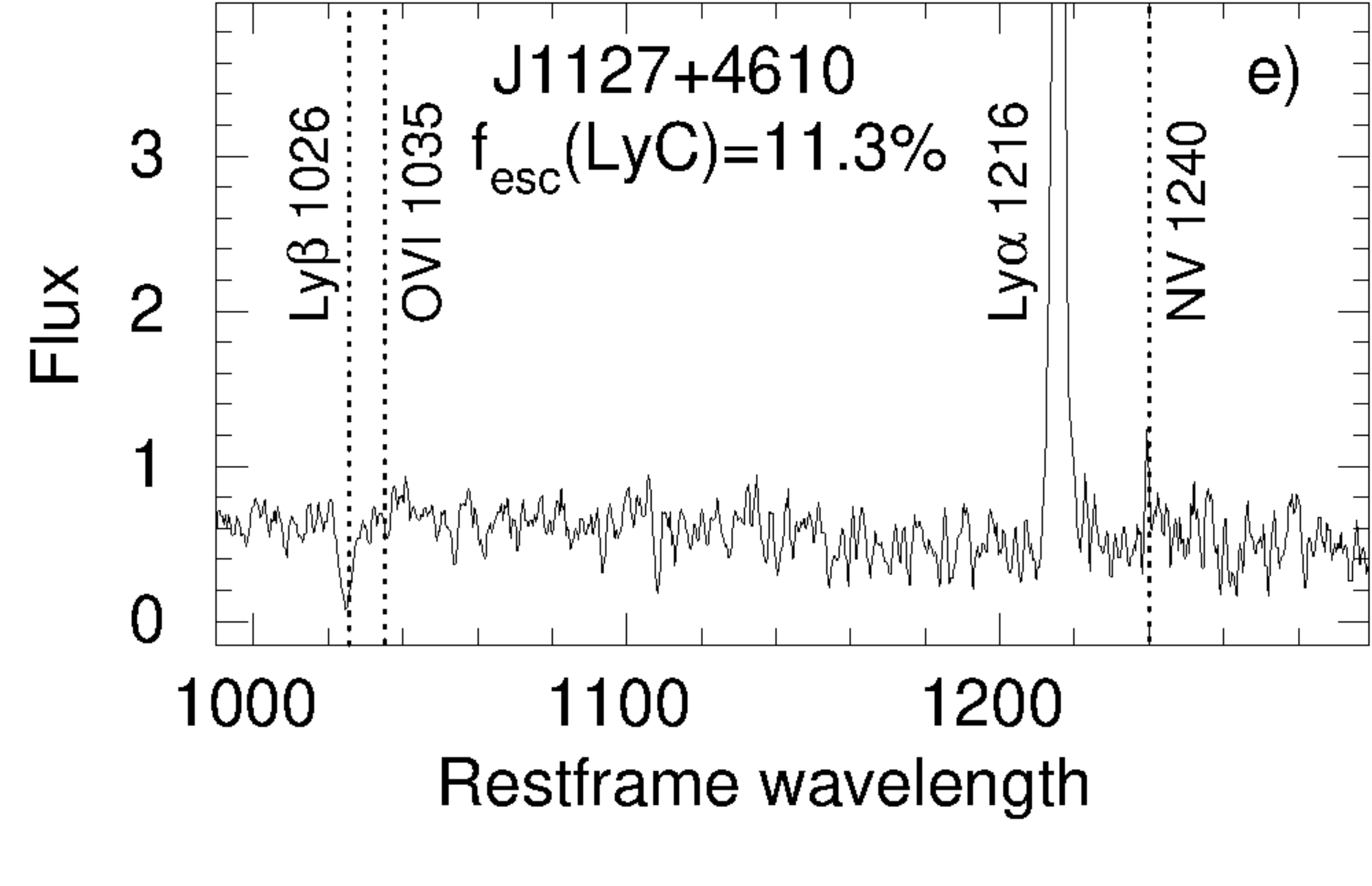}
\includegraphics[angle=-90,width=0.325\linewidth]{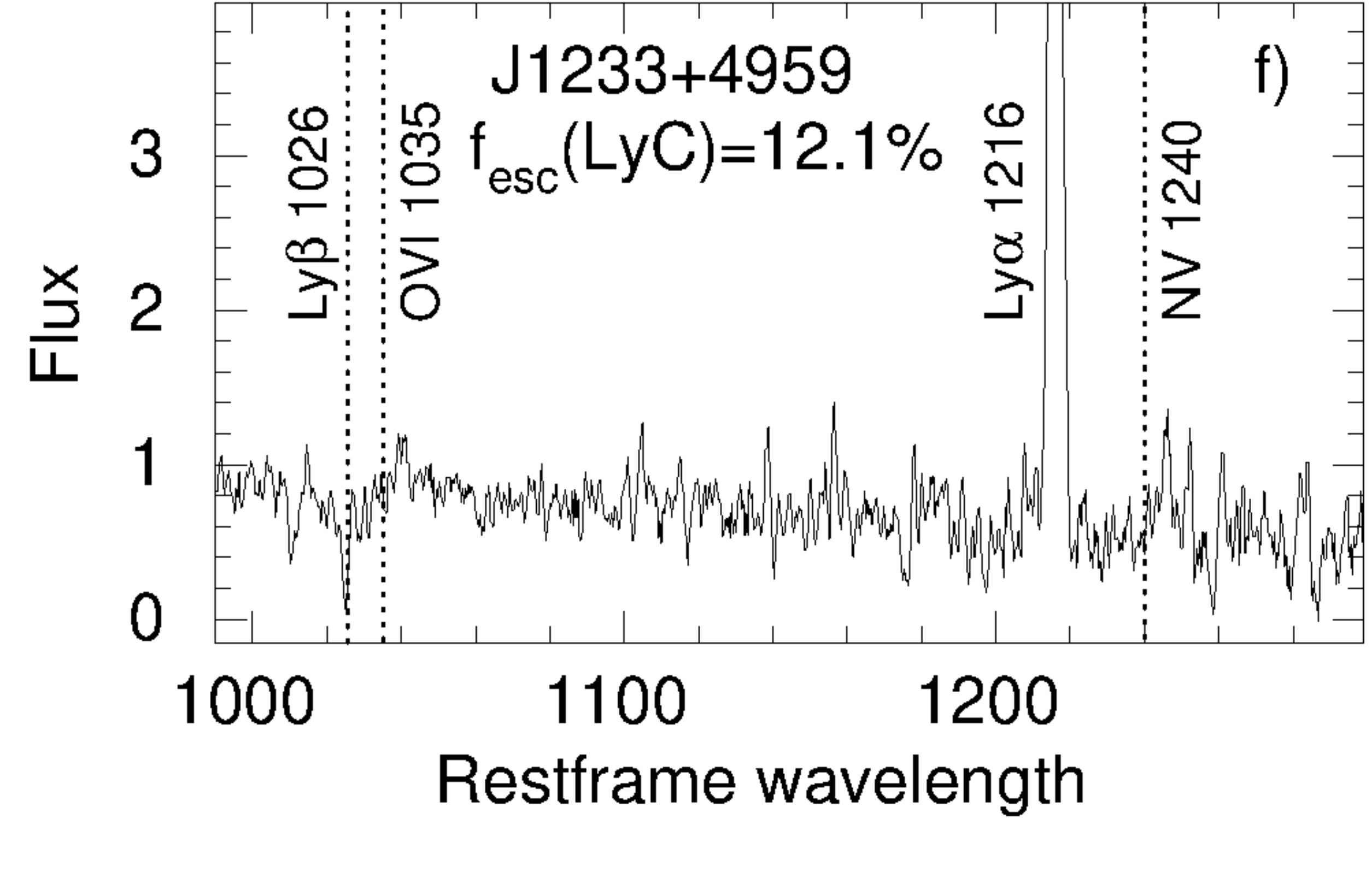}
}
\hbox{
\includegraphics[angle=-90,width=0.325\linewidth]{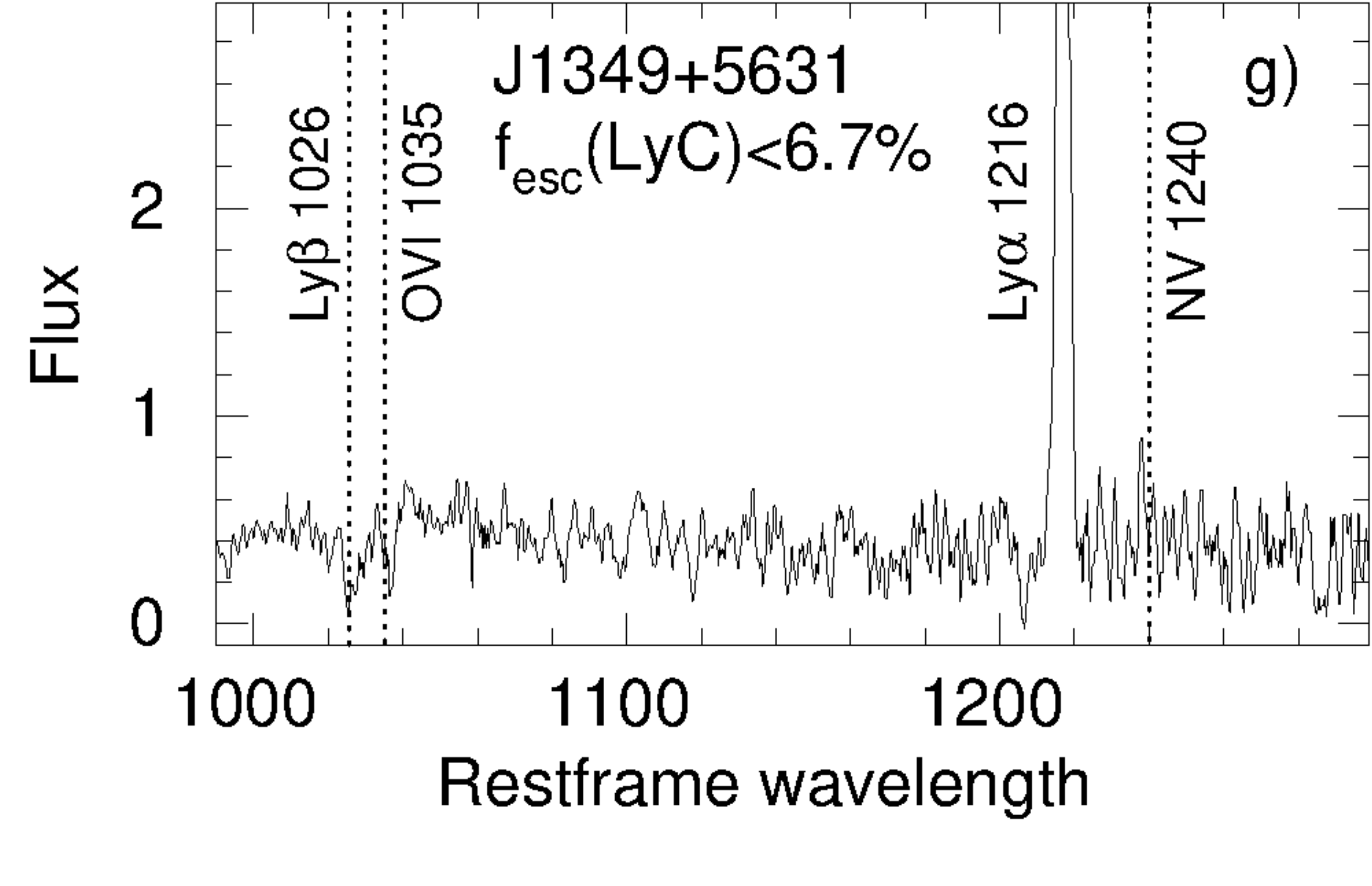}
\includegraphics[angle=-90,width=0.325\linewidth]{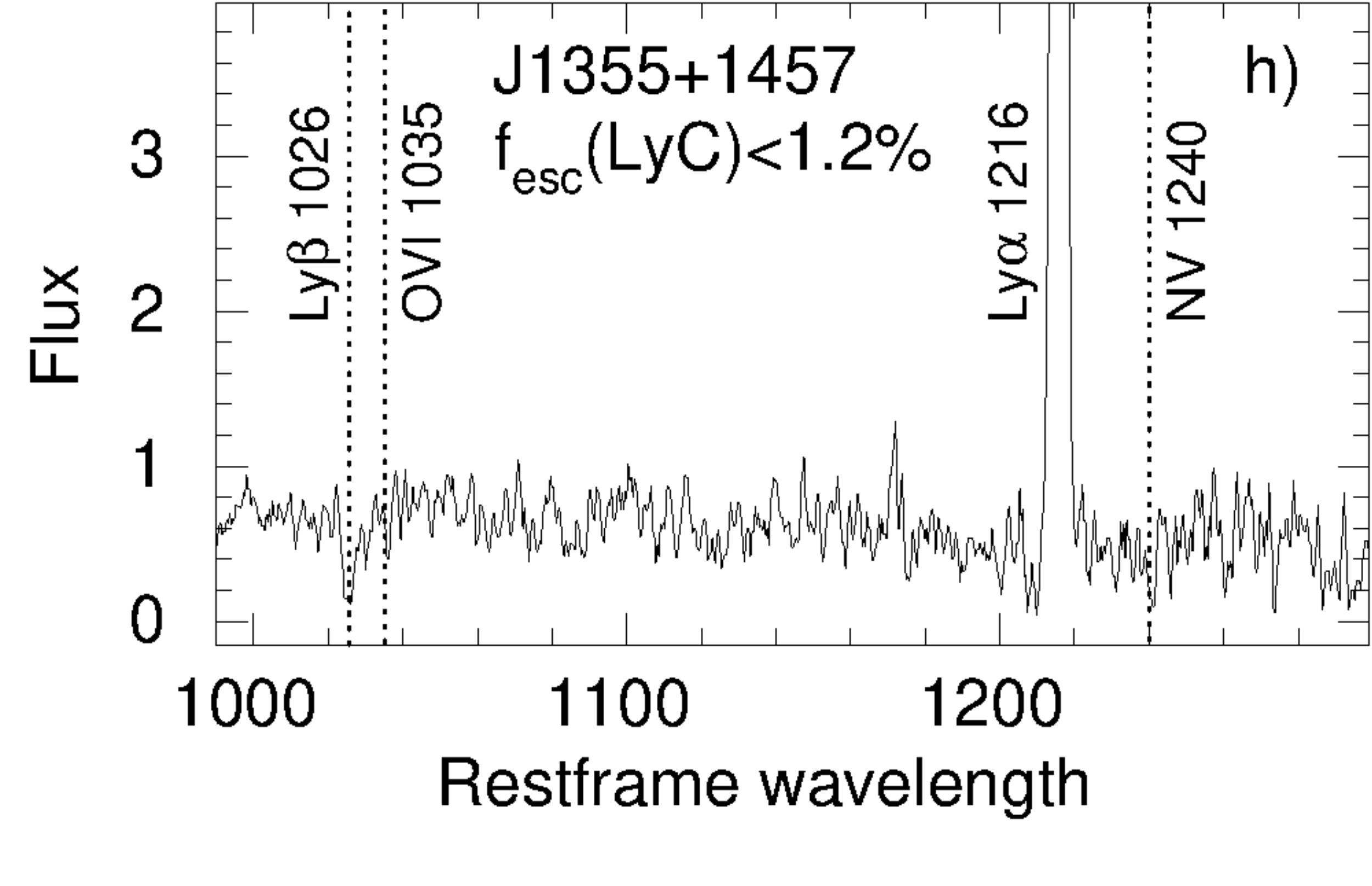}
\includegraphics[angle=-90,width=0.325\linewidth]{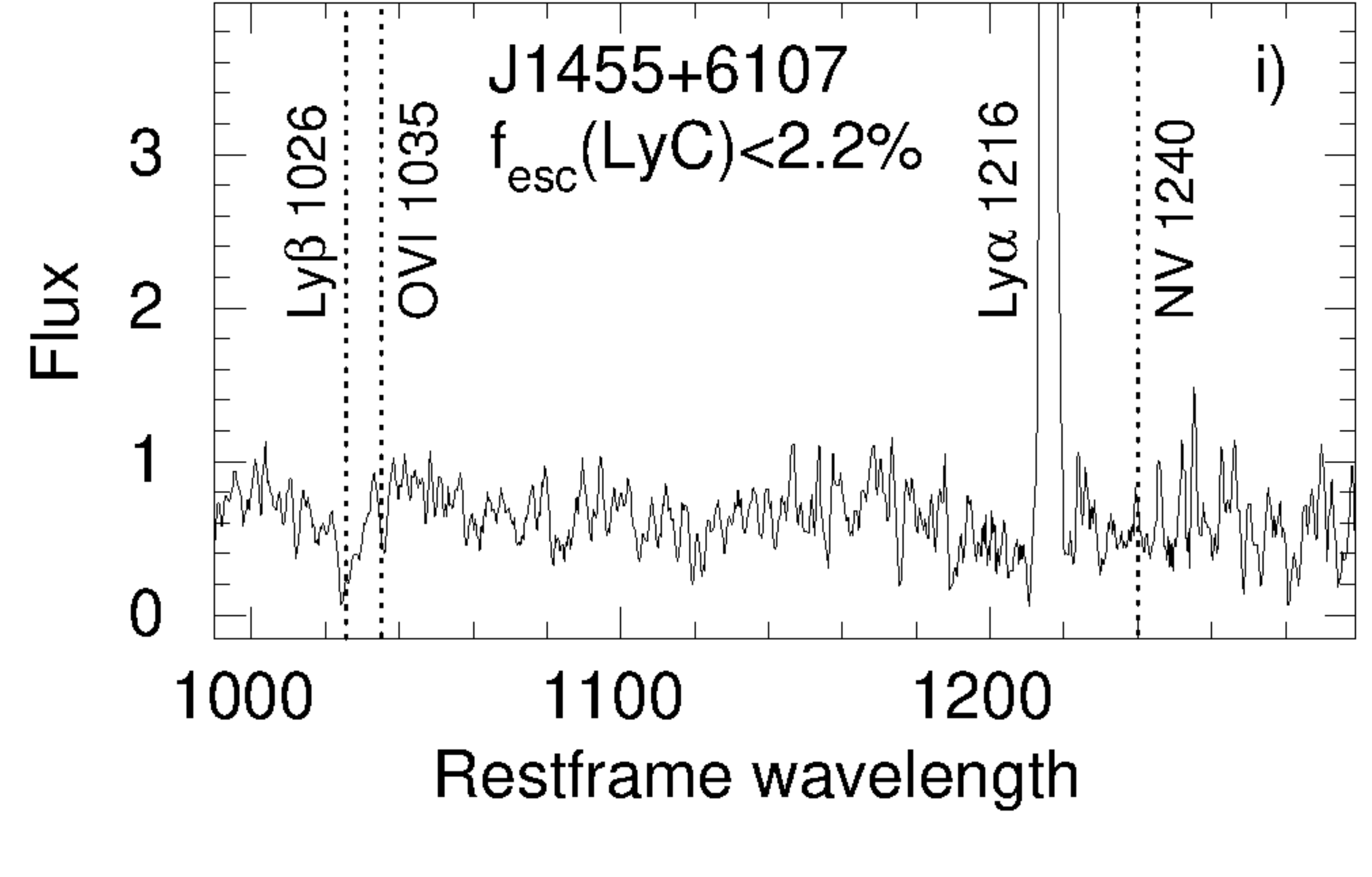}
}
\caption{Segments of COS G140L spectra showing the wavelength regions of the broad stellar lines 
O~{\sc vi}~$\lambda$1035\AA\ and N~{\sc v}~$\lambda$1240\AA\ for all LyC leaker
candidates studied in this paper. The centres of these lines and of the 
Ly$\beta$ line are indicated by vertical dotted lines. Fluxes are in 
10$^{-16}$ erg s$^{-1}$ cm$^{-2}$ \AA$^{-1}$, wavelengths are in \AA.
\label{fig9}}
\end{figure*}

\section{Indicators of high LyC escape fraction} \label{Ind}

The direct detection of LyC emission in low-redshift star-forming galaxies
is a difficult task. At the moment, only {\sl HST} can be used for that 
purpose. Therefore, reasonable indirect indicators of LyC leakage at low
redshift are needed to build a larger sample for statistical studies. 

\citet{JO13} and \citet{NO14} proposed a high
O$_{32}$ ratio as an indication of escaping ionising radiation.
We have in hand, with our new LyC leakers,
a sample of twenty five galaxies with a wide range of O$_{32}$ $\sim$ 0.5 -- 27
and directly derived $f_{\rm esc}$(LyC) \citep{L13,B14,L16,I16a,I16b,I18a,I18b}. 
The $f_{\rm esc}$(LyC) values for several LyC leakers from \citet{L13} and 
\citet{L16} are discrepant. Therefore, for these galaxies, we have used the 
$f_{\rm esc}$(LyC) values re-analysed by \citet{C17}. The relation between 
$f_{\rm esc}$(LyC) and O$_{32}$ is presented in Fig.~\ref{fig8}a. 
It has been discussed before by \citet{F16} and \citet{I18b}.
There is a trend of increasing $f_{\rm esc}$(LyC) with increasing 
of O$_{32}$, but with a substantial scatter, at large O$_{32}$ values.
The large scatter is due to the dependence of O$_{32}$ on other parameters such 
as metallicity, hardness of ionising radiation and ionisation parameter. 
Additionally, the spread of 
$f_{\rm esc}$(LyC) can also be caused by inhomogeneous leakage through channels 
with low optical depth and their orientation relative to the observer. 
Therefore, a high O$_{32}$ is not a very certain indicator of high 
$f_{\rm esc}$(LyC) and it is only a necessary condition for escaping 
radiation \citep{I18b,Na20}. The only definite result from Fig.~\ref{fig8}a is 
that $f_{\rm esc}$(LyC) is very low in objects with O$_{32}$ $\leq$ 4.

\begin{figure*}
\includegraphics[angle=0,width=0.95\linewidth]{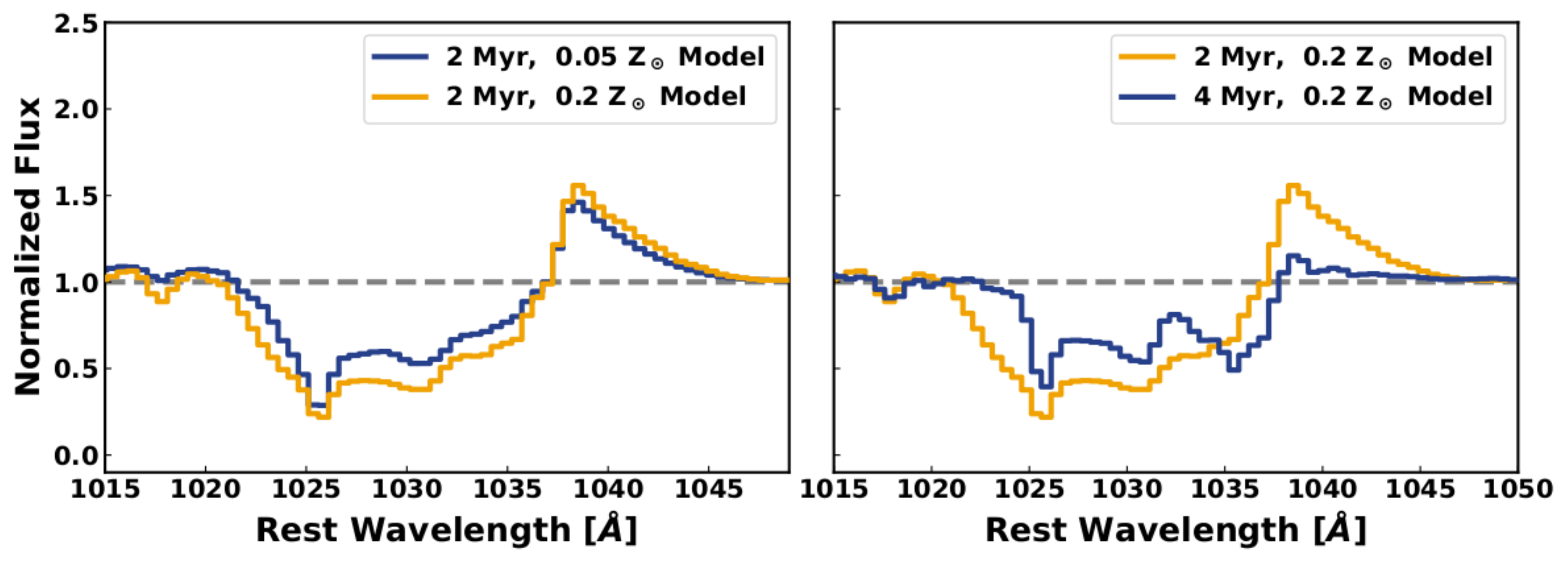}
\caption{Modelled O~{\sc vi}~$\lambda$1035\AA\ line profiles for two 
{\sc starburst}99 stellar populations \citep{L10} with metallicities 
of 0.2~Z$_\odot$ and 0.05~Z$_\odot$ (left) and with two different starburst ages 
(right).
\label{fig10}}
\end{figure*}

It has also been suggested that $f_{\rm esc}$(LyC) tends to be higher in 
low-mass galaxies \citep{W14,T17}. Our new nine galaxies with 
$M_\star$~$<$~10$^8$~M$_\odot$ considerably extend the mass range down to
a stellar mass of $\sim$~10$^7$~M$_\odot$, similar to the lowest-mass known of high-$z$
leakers \citep{Va20}. 
Previous observations have shown a slight tendency for $f_{\rm esc}$(LyC) to 
increase with decreasing stellar mass, albeit with a large intrinsic scatter. 
We present in Fig.~\ref{fig8}b
the relation between $f_{\rm esc}$(LyC) and stellar mass $M_\star$, including our 
new data. The added data 
shows, on the contrary, that there is no clear correlation between 
$f_{\rm esc}$(LyC) and $M_\star$.
This is one of the important findings of this study: contrary to expectation,  
there is no clear trend of increasing $f_{\rm esc}$(LyC) with decreasing 
$M_\star$. However, we caution that $M_\star$ might be higher than those 
derived from SED fitting because of uncertainties in the determination of the
mass of the old stellar population.

In Fig.~\ref{fig8}c, we show the relation $f_{\rm esc}$(LyC) -- 
$f_{\rm esc}$(Ly$\alpha$) for the $z$~$\la$~0.45 LyC leakers observed by 
\citet{I16a,I16b,I18a,I18b}. There is a tendency for $f_{\rm esc}$(LyC)
to increase with increasing $f_{\rm esc}$(Ly$\alpha$). However, the spread of the 
data is large. The only important results are:  
1) $f_{\rm esc}$(LyC) $<$ $f_{\rm esc}$(Ly$\alpha$) for the majority of  
galaxies, in agreement with theoretical predictions \citep*{Di16} and 
2) $f_{\rm esc}$(LyC) is very low for $f_{\rm esc}$(Ly$\alpha$) $\la$ 20 per cent.

The profile of the Ly$\alpha$ emission line can also be used as an indirect
indicator of the LyC leakage. This indicator is most useful because it can be 
applied to nearby low-mass galaxies, for which direct observations of the LyC 
with {\sl HST}/COS are not possible because of their low redshift. 
\citet{V17} and \citet{I18b} found a tight dependence of $f_{\rm esc}$(LyC)
on the separation $V_{\rm sep}$ between the peaks of the Ly$\alpha$
emission line in LyC leakers. The new low-mass LyC leakers discussed in this 
paper also follow the relation discussed by \citet{I18b}, shown by the solid 
line in Fig.~\ref{fig8}d.
There is no new galaxy having a peak separation less than 230~km~s$^{-1}$. 
For comparison, the lowest peak separation in the sample of low-$z$ leakers
\citep{I18b}, is that of J1243$+$4646, $\sim$~150~km~s$^{-1}$. 
We note that \citet{Me21} have recently 
reported the discovery of a double-peak Ly$\alpha$ profile in the galaxy 
A370p\_z1, at the epoch of reionisation ($z$~=~6.803), with the extremely 
low peak separation of 110~km~s$^{-1}$, implying an extremely high LyC escape 
fraction.

The relation between $V_{\rm sep}$
and $f_{\rm esc}$(LyC), as shown in Fig.~\ref{fig8}d, is a consequence of the 
fractions of LyC and Ly$\alpha$ escaping radiation being determined by the
column density of the neutral gas along the line of sight in LyC leaking 
galaxies  \citep[e.g. ][]{V15,V17}.
The majority of the new observations of low-mass galaxies 
(red symbols) support previous findings. However, there is one galaxy,
J0919$+$4906, which deviates significantly from the relation of \citet{I18b}. 
It has a $f_{\rm esc}$(LyC) corresponding to $V_{\rm sep}$ of $\sim$ 250 km s$^{-1}$
derived from the relation, compared to the measured value
of $\sim$ 370 km s$^{-1}$. The cause of this deviation is not clear.

This relation should constitute a strong 
constraint for constructing  radiative transfer and kinematical models which 
simultaneously reproduce $f_{\rm esc}$(LyC) and the Ly$\alpha$ profile.

\citet{I18b} considered also another potential indicator of escaping LyC
radiation. They found that the stellar line O~{\sc vi}~$\lambda$1035\AA\ with
a P-Cygni profile is seen in the spectra of most LyC leaking galaxies observed
by \citet{I16a,I16b,I18a,I18b}, indicative of hot most massive 
stars with masses of $\sim$ 100~M$_\odot$. It is likely that stellar winds from
these hot stars produce channels with ionised gas, through which ionising
radiation escapes the galaxy.

In Fig.~\ref{fig9} we present segments of COS G140L spectra of the LyC leakers 
studied in this paper, in the wavelength range $\sim$1000 -- 1300\AA, 
which includes the stellar lines 
O~{\sc vi}~$\lambda$1035\AA\ and N~{\sc v}~$\lambda$1240\AA. 
At variance with the results of \citet{I16a,I16b,I18a,I18b}, the 
O~{\sc vi}~$\lambda$1035\AA\ lines are not seen in spectra of our galaxies,
including the galaxy J1121$+$3806 with the highest $f_{\rm esc}$(LyC) = 35.0
per cent. Hints of this line are likely seen in the spectra of J1046$+$5827,
J1349$+$5632, and J1233$+$4959 with $f_{\rm esc}$(LyC) $<$ 1.8 per cent, 
$<$ 6.7 per cent, and 12.1 per cent, respectively.

The spectral differences between the galaxies studied 
by \citet{I16a,I16b,I18a,I18b} and in this paper could be caused by the lower 
stellar masses of the present sample. Stellar winds are less likely
in low-metallicity massive stars with more transparent interiors. Furthermore,
it is possible that there is a fundamental difference in the formation of the 
most massive stars in galaxies with different stellar masses. 
The O~{\sc vi} 1035\AA\ spectral feature arises from the stellar winds in the 
most massive stars. The left panel of Fig.~\ref{fig10} shows the O~{\sc vi} 
profile for two {\sc starburst}99 stellar populations \citep{L10} with 
metallicities of 0.2~Z$_\odot$ (similar to the median gas-phase metallicity of 
this sample) and 0.05~Z$_\odot$. This O~{\sc vi} 1035\AA\ profile takes on a 
classic P-Cygni feature with blueshifted absorption and redshifted emission for 
very young stellar populations. While the O~{\sc vi} absorption is slightly 
metallicity dependent, the redshifted emission profile is nearly constant at 
these metallicities and fixed ages. 

However, the right panel of 
Fig.~\ref{fig10} shows the extreme sensitivity of the O~{\sc vi} profile to 
stellar population age (or equivalently the upper end of the IMF), where 
O~{\sc vi} emission disappears (equal to the grey dashed line) for stellar 
populations with ages greater than 4 Myr, once stars with masses greater 
than 60~M$_\odot$ move off the main-sequence \citep{M94}, or for
younger stellar populations with upper stellar mass truncated at 
$\leq$60~M$_\odot$. The stark lack of O~{\sc vi} emission in the observed stellar
populations implies that these low $M_\star$ galaxies do not likely have the 
most massive stars ($>$60M$_\odot$). This could arise from slightly older 
stellar populations than the previous sample, the IMF of the lower mass sample 
may not be sufficiently populated, or the O~{\sc vi} profile may not be fully 
formed in these stellar populations.
Thus, high LyC leakage is not necessarily linked to the presence of 
O~{\sc vi}~$\lambda$1035\AA\ emission, as in the case of J1121$+$3806.

\section{Conclusions}\label{summary}

We present new {\sl Hubble Space Telescope} ({\sl HST}) Cosmic
Origins Spectrograph (COS) observations of nine 
compact star-forming galaxies (SFG), with low stellar masses 
$M_\star$~$\la$~10$^8$~M$_\odot$ and in the redshift range $z$~=~0.3179~--~0.4524.
We use these data to study the Ly$\alpha$ emission and the escaping Lyman 
continuum (LyC) radiation of these SFGs. This study is an extension of the work
reported earlier in \citet{I16a,I16b,I18a,I18b}.
Our main results are summarised as follows:

1. Emission of Lyman continuum radiation is detected in four out of the nine
galaxies with the escape fraction $f_{\rm esc}$(LyC) in the range between
11~$\pm$~4 per cent (J1127$+$4610) and 35~$\pm$~6 per cent (J1121$+$3806). 
Only upper limits of $f_{\rm esc}$(LyC) between $\sim$ 1 and $\sim$ 7 per cent
are obtained for the remaining five galaxies.

2. A Ly$\alpha$ emission line with two peaks is observed in the
spectra of seven galaxies, whereas the Ly$\alpha$ profile in one galaxy,
J1127$+$4610, may show more than two peaks, but the S/N~$\sim$~5 even in 
undersampled spectra is too low to definitely determine that.
The signal-to-noise ratio of the J0232$-$0426 spectrum is also too low to make 
definite conclusions about the shape of Ly$\alpha$.
The flux minimum between the two peaks approaches the zero 
value in eight galaxies. However, the flux between the peaks of one galaxy, 
J1121$+$3806, with the highest $f_{\rm esc}$(LyC), is far above the zero value 
indicating that a considerable fraction of its Ly$\alpha$ emission escaped through
low-density ionised channels (holes) in the neutral ISM along the line of sight.
A strong anti-correlation between $f_{\rm esc}$(LyC) and the peak velocity 
separation  $V_{\rm sep}$ of the Ly$\alpha$ profile is found, confirming
the finding of \citet{I18b} and making $V_{\rm sep}$ the most 
robust indirect indicator of Lyman continuum radiation leakage.

3. Other characteristics such as O$_{32}$ ratio, $f_{\rm esc}$(Ly$\alpha$) and
the stellar mass $M_\star$ show weak or no correlations with $f_{\rm esc}$(LyC), 
with a high spread of values. In particular, our new observations of low-mass 
LyC leaker candidates, with $M_\star$ $<$ 10$^8$~M$_\odot$, do not confirm 
the slight trend of increasing $f_{\rm esc}$(LyC) with decreasing $M_\star$ found
by \citet{I18b} for LyC leakers with higher stellar masses.

4. A bright compact star-forming region (with the exception of J0232$-$0426, 
which shows several knots of star formation) superimposed on a 
low-surface-brightness component, is seen
in the COS near ultraviolet (NUV) acquisition images of all eight galaxies 
(one image is missing due to technical problems). The surface brightness
at the outskirts of our galaxies can be approximated by an exponential disc, 
with a scale length of $\sim$~0.34~--~0.59~kpc. This is $\sim$~4 times lower 
than the scale lengths of the LyC leakers observed by 
\citet{I16a,I16b,I18a,I18b}, indicating 
that our new LyC candidates have much lower masses. However, part of this
difference may be explained by acquisition exposure times that are $\sim$~1.5 
times lower for the present sample, resulting in less deep images.

5. The star formation rates in the range SFR $\sim$~4~--~14~M$_\odot$~yr$^{-1}$ 
of our low-mass galaxies, with stellar masses $M_\star$~$<$~10$^8$~M$_\odot$, 
are several times lower than those in the LyC leakers studied by
\citet{I16a,I16b,I18a,I18b}. However, their specific star formation rates
of $>$~150~Gyr$^{-1}$ are higher than those found in low-redshift LyC leakers. 
The metallicities of our new galaxies, ranging from 12 + logO/H = 7.77 to 8.11,
are similar to those for LyC leakers studied by \citet{I16a,I16b,I18a,I18b}.

6. At variance with the LyC leakers studied by \citet{I16a,I16b,I18a,I18b}, 
the O~{\sc vi}~$\lambda$1035\AA\ line is  not seen in the spectra of our 
low-mass galaxies. 
These spectral differences can be caused by the lower stellar masses and lower 
metallicities of the galaxies in the present sample. The absence of the 
O~{\sc vi}~$\lambda$1035\AA\ line implies that these galaxies do not contain 
the most massive stars.

\section*{Acknowledgements}

Based on observations made with the NASA/ESA {\sl Hubble Space Telescope}, 
obtained from the data archive at the Space Telescope Science Institute. 
STScI is operated by the Association of Universities for Research in Astronomy,
Inc. under NASA contract NAS 5-26555. Support for this work was provided by 
NASA through grant number HST-GO-15639.002-A from the Space Telescope Science 
Institute, which is operated by AURA, Inc., under NASA contract NAS 5-26555.
Y.I. and N.G. acknowledge support from the National Academy of Sciences of 
Ukraine by its priority project No. 0120U100935 ``Fundamental properties of 
the matter in the relativistic collisions of nuclei and in the early Universe''.
Funding for SDSS-III has been provided by the Alfred P. Sloan Foundation, 
the Participating Institutions, the National Science Foundation, and the U.S. 
Department of Energy Office of Science. The SDSS-III web site is 
http://www.sdss3.org/. SDSS-III is managed by the Astrophysical Research 
Consortium for the Participating Institutions of the SDSS-III Collaboration. 
GALEX is a NASA mission  managed  by  the  Jet  Propulsion  Laboratory.
This research has made use of the NASA/IPAC Extragalactic Database (NED) which 
is operated by the Jet  Propulsion  Laboratory,  California  Institute  of  
Technology,  under  contract with the National Aeronautics and Space 
Administration.

\section*{Data availability}

The data underlying this article will be shared on reasonable request to the 
corresponding author.







\bsp	
\label{lastpage}
\end{document}